\documentclass[pdflatex,sn-mathphys-num]{sn-jnl}

\usepackage{graphicx}%
\usepackage{multirow}%
\usepackage{amsmath,amssymb,amsfonts}%
\usepackage{amsthm}%
\usepackage{mathrsfs}%
\usepackage[title]{appendix}%
\usepackage{xcolor}%
\usepackage{textcomp}%
\usepackage{manyfoot}%
\usepackage{booktabs}%
\usepackage{makecell}
\usepackage{url}
\usepackage{algorithm}%
\usepackage{algorithmicx}%
\usepackage{algpseudocode}%
\usepackage{listings}%
\usepackage{xltabular}

\usepackage{booktabs}
\usepackage{tabularx}
\usepackage{array}
\usepackage{float} 
\usepackage{seqsplit} 
\usepackage{booktabs}
\usepackage{siunitx} 
\usepackage{multicol}
\usepackage{longtable}
\usepackage{booktabs}
\usepackage{threeparttable}

\theoremstyle{thmstyleone}%
\theoremstyle{thmstyletwo}%

\theoremstyle{thmstylethree}%

\begin{document}


\title{\textcolor{black}{Multimodal Large Language Models Predict Urban Safety Perception but Encode Non-Neutral Demographic Priors}}

\author*[1,2]{\fnm{Ciro} \sur{Beneduce}}\email{cbeneduce@fbk.eu}

\author[1]{\fnm{Bruno} \sur{Lepri}}\email{{lepri@fbk.eu}}

\author[1]{\fnm{Massimiliano} \sur{Luca}}\email{{mluca@fbk.eu}}

\affil*[1]{\orgname{Fondazione Bruno Kessler}, \orgaddress{\street{Via Sommarive 18}, \city{Povo (TN)}, \postcode{38123}, \state{Italy}}}

\affil[2]{\orgdiv{Department of Information Engineering and Computer Science}, \orgname{University of Trento}, \orgaddress{\street{ Via Sommarive 9}, \city{Povo (TN)}, \postcode{38123}, \state{Italy}}}

\abstract{\textcolor{black}{Understanding how people perceive urban environments is essential for inclusive planning, yet conventional surveys are costly and difficult to scale. We investigate whether Multimodal Large Language Models (MLLMs) can assess perceived urban safety from street-view imagery while accounting for the observer-dependent nature of perception. Using Place Pulse~2.0, we evaluate four open and proprietary MLLMs across 56 cities under a Neutral prompt and socio-demographic personas defined by gender, age, and race or ethnicity. We also analyse the keywords generated to justify each classification.}

\textcolor{black}{All four models display comparable zero-shot capability, with city-macro F1 scores of 65--69\%, and preserve meaningful cross-city variation. However, they systematically favour the \emph{Safe} class, underpredict unsafety, and compress differences between cities. Their explanations converge on a shared visual lexicon: maintenance, greenery, order, and residential character support \emph{Safe} judgements, whereas deterioration, isolation, poor lighting, and limited pedestrian activity support \emph{Unsafe} judgements. Persona prompting produces substantial and structured shifts while holding the image fixed. Female personas yield more \emph{Unsafe} classifications than Male personas across all models; age effects are model-dependent, although Middle-aged personas generally remain closest to Neutral. Black/African American and Native American personas frequently show the largest departures, while the closest race or ethnicity match varies by model.
These findings show that MLLMs can provide scalable signals of perceived urban safety, but not from a demographically neutral standpoint.}}


\maketitle

\section{Introduction}\label{sec1}

\textcolor{black}{Understanding how people perceive urban spaces is a crucial area of research, particularly for urban planners aiming to develop citizen-centric cities \cite{carmona2010contemporary, luca2024towards, Ahn2025}. This objective has driven a shift in contemporary planning toward co-design frameworks, which explicitly require integrating the diverse needs and feedback of populations into the design process \cite{lee2024co}. To capture these necessary insights, surveys have traditionally served as the primary tool for assessing urban perceptions \cite{salesses2013collaborative, zamanifard2019measuring, mehta2014evaluating}.} However, these approaches often focus on planned developments or specific conditions (e.g., time of day) and are constrained by high costs, time-consuming data collection, and limited scalability\textcolor{black}{; indeed, obtaining reliable per-image scores demands a large number of independent judgements, which compounds these costs at city scale \cite{gu2025designing}. Online surveys also face an emerging validity challenge: autonomous LLM-based respondents can generate coherent and plausible answers that evade conventional attention and data-quality checks, undermining the assumption that a credible response necessarily originates
from a human participant \cite{westwood2025potential}.} Moreover, cities evolve rapidly, and even minor changes, such as the opening or closing of a business, can influence how crowds and individuals navigate and experience urban environments \cite{simini2021deep, bahrami2022using,suhara2021validating}. Given the dynamic nature of urban spaces, surveys alone may not be the most effective tool for capturing real-time perceptions.

To address these limitations, researchers have increasingly explored alternative data sources. One promising approach involves integrating Street-View Images (SVIs) with computer vision techniques to tackle various urban challenges, such as socioeconomic estimation \cite{Gebru_2017}, pedestrian environment assessment \cite{HELBICH2019107}, and flood risk evaluation \cite{ho2024elev}. SVIs have also been leveraged to analyze urban perception, providing insights into how individuals perceive safety, aesthetics, and livability across different cityscapes \cite{naik2014streetscore, de2016safer,dubey2016deep, redi2018spirit, quercia2014aesthetic} . Previous studies have employed machine learning and deep learning models to estimate safety perception scores from images \cite{hou2024global, naik2014streetscore, dubey2016deep}.

\textcolor{black}{However, a critical limitation in many computational approaches is the treatment of safety perception as a monolithic metric. Urban studies literature argues that safety is inherently subjective and deeply influenced by socio-demographic factors \cite{gustafsod1998gender}. \textcolor{black}{This subjectivity is not merely theoretical: large-scale, demographically stratified perception surveys now show empirically that visual perception of streetscapes varies systematically across gender, age, income, education and ethnicity worldwide, underscoring the need for localised, human-centred assessment \cite{quintana2025global}. For safety specifically, studies contrasting AI-derived and survey-based perceptions across neighbourhoods report analogous group-level divergences
\cite{kang2023assessing}. Consistent with this, gender} significantly shapes the perception of public spaces, with specific design interventions required to improve safety for women \cite{pereira2024women, navarrete2021building}\textcolor{black}{, and age} plays a pivotal role\textcolor{black}{, as} the framework of "age-friendly cities" highlights that older adults perceive the built environment through distinct lenses regarding accessibility and security \cite{van2021ten, figueiredo2023older}. These factors shape attitudes toward environmentalism \cite{raudsepp2001some, christian2021households}, tourism \cite{sharma2015examination}, and safety itself \cite{finucane2013gender}. \textcolor{black}{Therefore, computational models that rely solely on image content without accounting for the observer's identity risk oversimplifying the complex social reality of urban perception, a concern compounded by growing evidence that generative and foundation models embed systematic geographic and demographic biases \cite{beneduce_2026_20287027, luca2025llm, wang2025uncovering}}}

\textcolor{black}{Most recently, the emergence of Multimodal Large Language Models (MLLMs) and Vision Language Models (VLMs) has opened new directions for this domain\textcolor{black}{, and a rapidly growing body of work now applies these models to street-view analytics across a wide range of urban tasks \cite{peng2025vision}, including the simulation of human environmental perception from imagery \cite{wu2025exploring}, and even the transfer of population-group perceptual knowledge from large multimodal models to smaller ones for urban safety evaluation \cite{hou2025transferring}.} \textcolor{black}{Within safety perception specifically, \citet{zhang2025urban} developed a pipeline for urban safety perception focusing primarily on efficiency and automation, while \citet{malekzadeh2024urban} employed GPT-4 to contrast AI ratings of urban attractiveness against human insights. However, these works leave considerable gaps. }Empirically, existing methods remain geographically constrained, restricting evaluation to only one or two specific cities (e.g., Helsinki, Chengdu, Osaka) using small-scale datasets \cite{{zhang2025urban}, malekzadeh2024urban}. \textcolor{black}{Methodologically, they tend to report numerical scores while overlooking the textual rationales that could make the model's reasoning interpretable \cite{wu2025exploring}.} Theoretically, they overlook the inherent subjectivity of perception. While research in Natural Language Processing has demonstrated that foundation
models can use persona prompts to diversify outputs
\cite{hamalainen2023evaluating,park2023generative,fleisig2023majority} and simulate aspects of human behaviour
\cite{horton2023large,argyle2023out,tornberg2023simulating},
\textcolor{black}{persona-conditioned measurements remain sensitive to model scale, reasoning mode, question formulation, and conversation history, indicating that these responses may not constitute stable representations of human personality or identity \cite{tosato2026persistent}.} This approach has not yet been systematically audited in image-grounded urban
safety perception. Consequently, the potential for MLLMs to simulate diverse, personalised, and global urban perceptions remains largely unexplored.}

\begingroup
\color{black}

In this study, we conduct a cross-model audit of perceived urban safety using Place Pulse~2.0 \cite{dubey2016deep}. We evaluate four MLLMs spanning open and proprietary families: LLaVA~1.6 7B \cite{liu2024improved}, Gemini~2.5 Flash Lite \cite{comanici2025gemini}, Qwen2.5-VL-72B-Instruct \cite{bai2025qwen3}, and GPT-5.1 \cite{singh2025openai}. Across 56 cities, we first assess whether these models can classify street-view images as \emph{Safe} or \emph{Unsafe} without task-specific training. We then analyse the keywords generated to justify each classification, allowing us to identify the visual cues and thematic structures associated with safety and unsafety across models.

Finally, we hold the visual input constant while conditioning the prompt on gender, age, and race or ethnicity. This paired design isolates how socio-demographic language changes the assessment of identical scenes and allows us to compare each persona with the model's unspecified \emph{Neutral} response. In doing so, we connect urban-perception analysis with behavioural-alignment research showing that unprompted language models carry specific, non-neutral priors \cite{tao2024cultural}.

Our findings reveal a consistent tension between visual competence and perceptual neutrality. All four models recover meaningful differences in perceived safety across cities and converge on a shared visual vocabulary. However, they also systematically favour the \emph{Safe} class and compress variation between cities. Persona prompting further produces substantial and structured shifts in classification, while the Neutral output consistently aligns most closely with the Male persona and generally remains close to the Middle-aged persona. Its closest race or ethnicity match varies across models, whereas Black/African American and Native American personas frequently produce the largest departures.

The contribution of this study is therefore not simply to show that MLLMs can perform safety perception or encode social bias. Rather, we distinguish interacting dimensions of image-grounded urban perception: transferable visual competence, systematic calibration bias, and persona-induced response shifts. Our results show that MLLMs can extract scalable signals of perceived urban safety, but do not interpret the city from a demographically neutral standpoint.

\endgroup

\begin{figure}[h]
\centering
\includegraphics[width=1\linewidth]{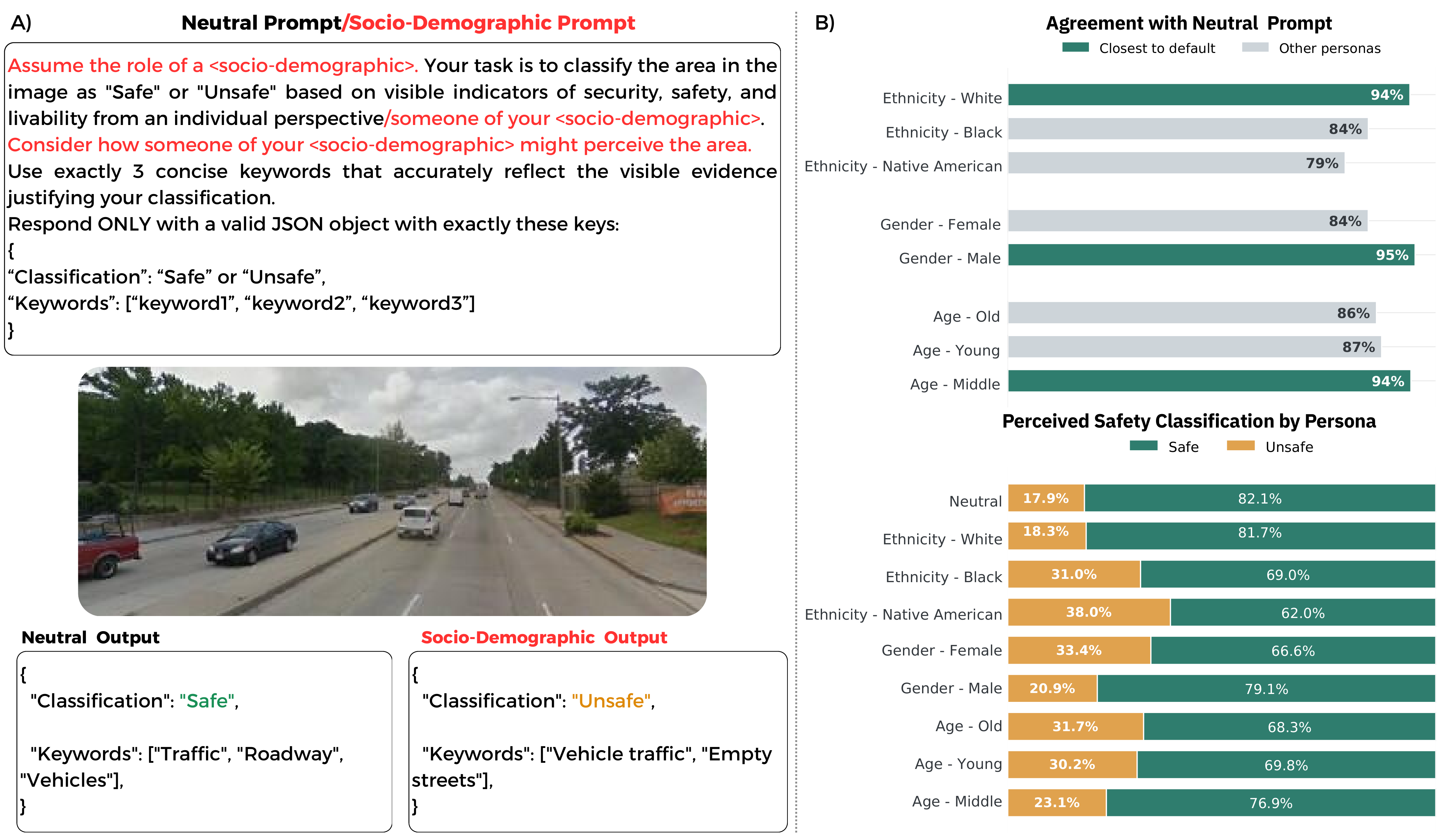}
\caption{\textbf{Pipeline and effect of socio-demographic personas across three identity axes
(gender, age, and race/ethnicity).}
\textbf{(A)} Each street-view image is classified as \emph{Safe} or \emph{Unsafe} by a VLM under a
\emph{Neutral} prompt or a \emph{socio-demographic} prompt that differs only in the opening role sentence;
the output is a classification plus three keywords. The same image is judged \emph{Safe} under Neutral but
\emph{Unsafe} once a persona is adopted.
\textbf{(B, top)} Per-persona agreement with the Neutral classification (mean over the four MLLMs); the
persona closest to the default within each axis (highlighted) is White, Male and Middle-aged. For the
race/ethnicity axis only the closest and the two most distant personas are shown.
\textbf{(B, bottom)} \emph{Safe}/\emph{Unsafe} share per persona (mean over the four MLLMs): the
\emph{Unsafe} share rises most for the Native American, Female, Black and Elderly personas, while White
tracks the Neutral baseline. Full-per-model results are reported in Appendix~\ref{Disparities}.}
\label{fig:pipeline}
\end{figure}

\section{Results}\label{sec2}

\subsection{Predicting Urban Safety Perception with Multimodal Large Language Models}
\label{sec:neutral_results}
 
Our study investigates the capability of Multimodal Large Language Models to classify perceived safety in urban
environments, using Place Pulse 2.0 as the source of street-view imagery (SVI) and ground-truth data.
\textcolor{black}{To ensure that our findings characterise the current class of these models rather than a single checkpoint, we evaluate four state-of-the-art MLLMs spanning open and proprietary families and a broad capability range: LLaVA~1.6 7B \cite{liu2024improved}, Gemini~2.5 Flash Lite \cite{comanici2025gemini}, Qwen2.5-VL-72B-Instruct \cite{bai2025qwen3}, and GPT-5.1 \cite{singh2025openai}.} A detailed description of the models,
dataset, and ground-truth thresholding is provided in \textcolor{black}{Section~\ref{sec4}}.
 
To operationalise the task, we design a prompt that performs a binary classification, establishing
whether a place is \emph{Safe} or \emph{Unsafe}. This prompt, henceforth the \emph{Neutral prompt}, classifies the SVI without reference to any socio-demographic information; it is described in the
Materials and Methods and shown in Figure~\ref{fig:pipeline}. To report our results, we selected the F1-score as our primary performance metric. Although accuracy is widely used, it can be misleading in imbalanced datasets, as it tends to overestimate performance by favouring the majority class \cite{branco2016survey}. By using the F1-score, we account for both false positives and false negatives, ensuring a more reliable evaluation that captures the asymmetric precision-recall behaviour. All reported results represent the average performance across two experimental runs.

\textcolor{black}{Table~\ref{tab:neutral_perf} shows that the four models perform comparably and reasonably well without task-specific training. The city-macro Safe-F1 is 65.4\% for LLaVA~1.6, 69.4\% for Gemini~2.5 Flash Lite, 68.9\% for Qwen2.5-VL-72B, and 68.8\% for GPT-5.1, with each value averaged over two runs. The narrow spread across models suggests that contemporary MLLMs already encode transferable knowledge of urban safety perception. This is notable because conventional approaches generally rely on computer-vision models explicitly trained to extract safety-related visual features \cite{de2016safer,naik2014streetscore,quercia2014aesthetic,dubey2016deep}.}
 
\begin{table}[t]
\centering
\caption{\textcolor{black}{Neutral-prompt performance of the four models (mean of two runs). Precision,
recall, and the share of images classified \emph{Unsafe} expose a Safe-bias common to all models: recall
is high and precision low, and the \emph{Unsafe} share stays well below the true 51.0\% \emph{Unsafe}
base rate of the evaluation set. All quantities use city-macro aggregation (computed per city, then
averaged over the 56 cities). The two rightmost columns give the city-macro Safe-F1 within the Global
North and Global South and their difference.}}
\label{tab:neutral_perf}
\begin{tabular}{lccccccc}
\toprule
& \multicolumn{4}{c}{\textbf{Overall (city-macro)}} & \multicolumn{3}{c}{\textbf{Safe-F1 by region}} \\
\cmidrule(lr){2-5}\cmidrule(lr){6-8}
\textbf{Model} & \textbf{Safe-F1} & \textbf{Prec.} & \textbf{Rec.} & \textbf{\% Unsafe} &
\textbf{North} & \textbf{South} & \textbf{$\Delta$ (N$-$S)} \\
\midrule
LLaVA~1.6 7B            & 65.4 & 51.5 & 99.7 & \phantom{0}5.5 & 71.2 & 41.8 & 29.4 \\
Gemini~2.5 Flash Lite   & 69.4 & 58.7 & 91.0 & 24.9 & 73.8 & 51.6 & 22.2 \\
Qwen2.5-VL-72B          & 68.9 & 57.4 & 91.7 & 22.7 & 73.4 & 50.6 & 22.8 \\
GPT-5.1                 & 68.8 & 56.3 & 94.6 & 18.7 & 73.3 & 50.6 & 22.7 \\
\bottomrule
\end{tabular}
\end{table}
\textcolor{black}{This encouraging aggregate performance, however, conceals a strong and remarkably consistent \emph{Safe-bias}. As the precision and recall columns of Table~\ref{tab:neutral_perf} show, the models recover almost every truly safe scene: \emph{Safe}-recall reaches 99.7\% for LLaVA and remains high for the API models, at 91.0\% for Gemini, 91.7\% for Qwen, and 94.6\% for GPT-5.1. This high recall comes at the cost of low \emph{Safe}-precision, which never exceeds 58.7\%. In other words, the models classify substantially more scenes as safe than indicated by the ground truth.} The same tendency is
visible in the overall share of images called \emph{Unsafe}, which runs from just 5.5\% for LLaVA up to 24.9\% for Gemini (22.7\% for Qwen, 18.7\% for GPT-5.1) --- in every case far below the 51.0\%
\emph{Unsafe} rate of the ground truth. \textcolor{black}{The direction of this bias is also consistent across models: among the images on which all four agree, the shared errors consist almost entirely of scenes unanimously classified as \emph{Safe} despite being labelled \emph{Unsafe} in the ground truth, whereas the opposite error is nearly absent. This convergence suggests that the Safe-bias is a common characteristic of current MLLMs rather than an idiosyncrasy of any individual model.} The full cross-model agreement analysis is reported in Appendix~\ref{Neutral_Prompt_SI}.
 
Performance also varies substantially across cities. While the city-macro Safe-F1 sits near 65--69\%,
this figure conceals large fluctuations at the local level; detailed per-city metrics are given in Appendix~\ref{Neutral_Prompt_SI} and
Appendix Table~\ref{app:per_city_all_models}. \textcolor{black}{Performance is highest in affluent Global-North cities, with all four models achieving Safe-F1 scores close to or equal to 100\% in Amsterdam, Copenhagen, and Sydney. It declines sharply in several Global-South cities, including Mexico City, Belo Horizonte, and Gaborone. Aggregating by region (Table~\ref{tab:neutral_perf}, right-hand columns), city-macro Safe-F1 falls from 71.2\% in the Global North to 41.8\% in the Global South for LLaVA, with similarly large declines for the three API models.}

\textcolor{black}{This geographic gap is consistent with the interaction between the models' Safe-bias and regional differences in the ground-truth class distribution. Per-city Safe-F1 is strongly influenced by the prevalence of truly safe scenes: because the models systematically overpredict the \emph{Safe} class, performance declines in cities where unsafe scenes are more common. This pattern is visible in the city-level results reported in Appendix Table~\ref{app:per_city_all_models}, where cities with lower Safe prevalence generally exhibit lower Safe precision and F1 despite consistently high recall. Since Global-South cities are rated as less safe on average in Place Pulse 2.0, this interaction contributes to the observed North--South performance gap. The regional difference should therefore not be interpreted solely as evidence that the models are less capable of processing Global-South environments.}
 
\textcolor{black}{This distinction is important because the Place Pulse labels were collected predominantly from Global-North annotators. Regional differences may therefore partly reflect geographically situated perceptions of safety rather than universally shared assessments of urban environments. We return to this issue in the Discussion, where we consider how the composition of the annotator population may shape both the benchmark labels and the apparent geographic disparities.}
 
\begin{figure*}[t]
\centering
\includegraphics[width=0.9\textwidth]{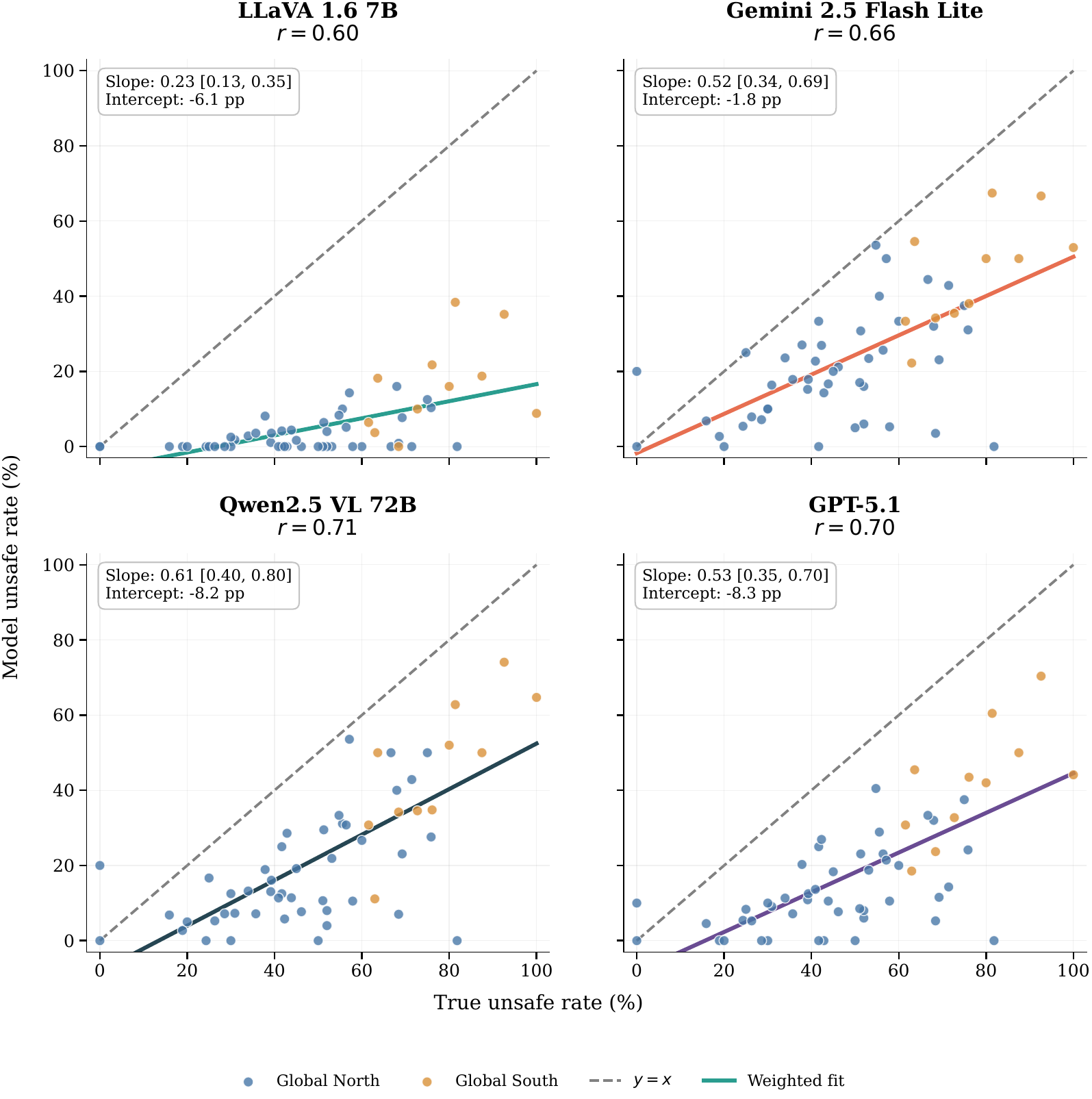}
\caption{{\textcolor{black}{Ground-truth and predicted unsafe rates across cities.} \textcolor{black}{Each panel compares, for 56 cities, the proportion of images labelled \emph{Unsafe} in the ground truth with the proportion classified \emph{Unsafe} by one model. Point size represents the number of images per city, and colour distinguishes Global-North and Global-South cities. The dashed line indicates perfect calibration ($y=x$), while the colored line shows the image-weighted least-squares fit. Pearson's correlation coefficient, the fitted slope and its bootstrap 95\% confidence interval, and the intercept are reported in each panel. All four models preserve cross-city variation in perceived unsafety ($r=0.60$--$0.71$), but systematically underestimate its magnitude, as indicated by slopes below one ($0.23$--$0.61$) and the predominance of cities below the calibration line.}}}
\label{fig:modelVStrue}
\end{figure*}

\textcolor{black}{A natural concern is that the models achieve these scores simply by classifying almost every image as \emph{Safe}, rather than responding to the visual content of each scene. To test this possibility, we compare, for each city, the proportion of images labelled \emph{Unsafe} in the ground truth with the proportion predicted as \emph{Unsafe} by each model. If the models respond meaningfully to urban conditions, cities with higher ground-truth safe rates should also receive more \emph{Safe} predictions. This is precisely what we observe in Figure~\ref{fig:modelVStrue}: the two quantities are positively correlated for every model, with Pearson correlations of $0.60$ for LLaVA, $0.66$ for Gemini, $0.71$ for Qwen, and $0.70$ for GPT-5.1. The models therefore do not emit a fixed response; they capture and rank meaningful differences in perceived safety across cities.}

\textcolor{black}{The relationship is nonetheless substantially flatter than ideal. Under perfect calibration, a one-percentage-point increase in a city's ground-truth safe rate would correspond to an equivalent increase in its predicted safe rate, producing a slope of one. Instead, the fitted slopes are $0.23$ for LLaVA, $0.52$ for Gemini, $0.61$ for Qwen, and $0.53$ for GPT-5.1. Predictions are therefore compressed into a relatively narrow and optimistic range, pulling both genuinely unsafe and genuinely safe cities towards similar estimates. Put differently, the models broadly recover the ordering of cities but systematically understate the degree of unsafety in the lowest-rated locations. This compression helps explain the geographic performance gap: in the least-safe cities, many of which are located in the Global South, the models soften scenes that the ground truth classifies as clearly \emph{Unsafe}, leading to lower performances.}
 
\textcolor{black}{To assess the robustness of these findings, we repeat the analysis using only the 40 cities represented by at least 15 images. This restriction leaves the city-macro Safe-F1 essentially unchanged, at 65.8\% for LLaVA, 69.3\% for Gemini, 69.0\% for Qwen, and 68.8\% for GPT-5.1. It also preserves the North--South gap, which remains 31.7, 26.6, 27.1, and 26.2 percentage points, respectively. These results confirm that neither the aggregate performance nor the geographic disparity is driven by cities with small and potentially noisy samples (Appendix~\ref{Neutral_Prompt_SI}).}

\textcolor{black}{\subsection{A Shared Visual Lexicon of Safety and Unsafety}}\label{Lexicon}

Once assessed that the models could perform the task, we shifted our attention toward the main factors that influenced the classification of images\textcolor{black}{, in a setting where no hints, suggestions, or instructions were given to the model about what to look for. This is a deliberate design choice: by keeping the assessment open-ended, we aim to probe the models' genuine understanding of urban safety perception rather than their ability to follow directions, and we avoid injecting an in-prompt user perspective that could bias which visual features the model treats as markers of safety or danger \cite{d2024openbias}}. The foundation of this analysis rests on the prompt's requirement to output three keywords that justify each classification.
The complete details of the prompt structure can be found in Materials and Methods \ref{sec4}.

To facilitate this exploration, we constructed a co-occurrence network from the 25 most used keywords associated with \emph{Safe} and \emph{Unsafe} images. In this network, each node represents a keyword, and each edge reflects the frequency with which two keywords co-occur. By applying the Louvain community detection algorithm to this matrix, we identified distinct keyword communities and, calculating the degree of centrality within each community, we highlighted the most influential keywords that define the core themes. All the steps involved in constructing the network and community detection approach are detailed in Section \ref{sec4}, while the whole list of communities is detailed in Appendix \ref{keyword_network_SI}. \textcolor{black}{To establish whether this perceptual lexicon is a convergent property of MLLMs, we replicated the analysis across all the models considered in the study, spanning the lightweight open LLaVA 1.6 7B, the large open Qwen2.5 VL 72B, and the proprietary Gemini 2.5 Flash Lite and GPT-5.1; the full cross-model comparison, along with additional tables and figures, is reported in Appendix \ref{keyword_network_SI}.} Figure \ref{fig:network_keywords} illustrates the networks for \textcolor{black}{ GPT-5.1}, serving as a conceptual map of the key drivers behind its negative and positive classifications.

\begin{figure}[h]
\centering
\includegraphics[width=\linewidth]{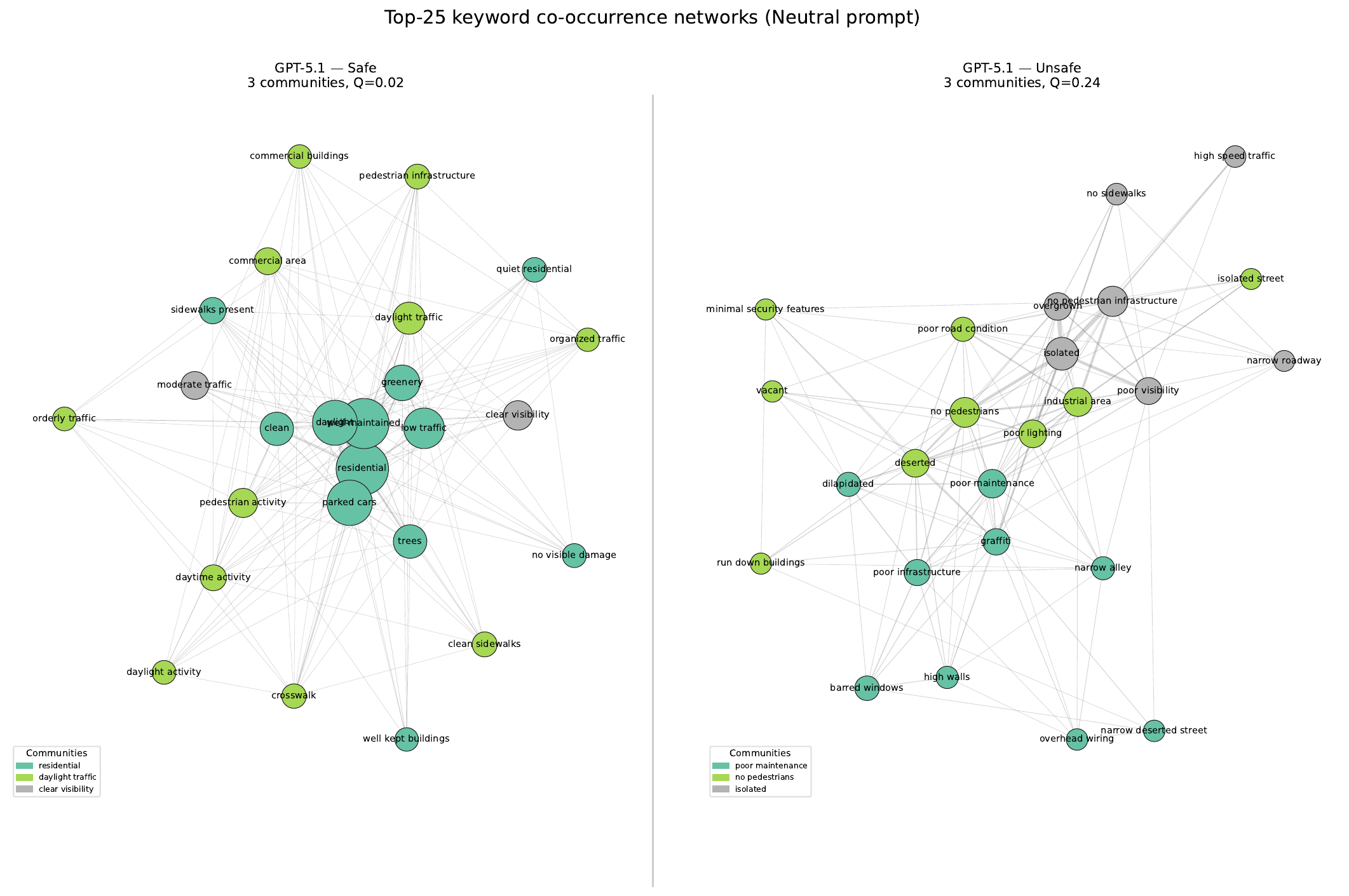}
\caption{Co-occurrence networks of safe and unsafe keywords \textcolor{black}{for GPT-5.1}. The figure displays the 25 most frequently used keywords and their relationships for images classified as \emph{Safe} (left) and \emph{Unsafe} (right). Nodes represent keywords, with their size reflecting degree centrality, while edges represent co-occurrence relationships, with thickness corresponding to co-occurrence frequency. Distinct colours indicate the keyword communities recovered by the Louvain algorithm\textcolor{black}{, with the number of communities and the modularity $Q$ reported above each panel. The equivalent networks for LLaVA 1.6 7B, Gemini 2.5 Flash Lite and Qwen2.5 VL 72B are provided in Appendix Figure \ref{fig:network_keywords_SI}}.}
\label{fig:network_keywords}
\end{figure}

\textcolor{black}{For GPT-5.1 the two classes are strikingly asymmetric. The safe network is almost perfectly monolithic ($Q=0.02$, 3 communities): a single dominant cluster ties together residential character, upkeep and daytime cues --- "residential", "well-maintained", "daylight", "parked cars", "greenery", "clean", "trees" and "low traffic" --- and only the traffic-intensity terms ("moderate traffic", "daylight traffic") split off as small satellites. In effect, GPT-5.1 treats "safe" as one coherent notion of a well-kept, residential, daytime street, which is why the network barely partitions. The unsafe network, by contrast, is the most clearly structured of all models ($Q=0.24$, 3 communities) and resolves danger into three distinct concerns: physical decay ("poor maintenance", "dilapidated", "run-down buildings", "poor infrastructure", "barred windows", "high walls"), the absence of street life and lighting ("no pedestrians", "poor lighting", "deserted", "poor road condition", "no pedestrian infrastructure", "minimal security features"), and isolation with low visibility ("isolated", "overgrown", "poor visibility", "industrial area", "narrow alley", "no sidewalks"). This asymmetry --- a monolithic notion of safety but a differentiated, multi-faceted notion of danger --- is the central pattern of the lexicon.}
 
\textcolor{black}{Extending the analysis to LLaVA 1.6 7B, Gemini 2.5 Flash Lite and Qwen2.5 VL 72B (Appendix Figure \ref{fig:network_keywords_SI}) shows that this organisation is a convergent property across models. For all four, the \emph{Safe} lexicon is built around maintenance, residential character, greenery and visual order ("well-maintained", "residential", "quiet", "trees", "orderly"), whereas the \emph{Unsafe} lexicon clusters around neglect, dilapidation and isolation ("abandoned", "dilapidated", "neglected", "isolated", "overgrown"). Two systematic regularities emerge. First, the safe lexicon is monolithic for every model: its modularity is uniformly low ($Q$ between $0.02$ and $0.10$, lowest for GPT-5.1), because maintenance, residential, greenery and order cues all co-occur and resist partitioning. Second, the unsafe lexicon is consistently more differentiated than the safe one, and its thematic separation is greatest in the largest models, rising from $Q\approx0.15$--$0.16$ in LLaVA 1.6 7B and Gemini to $0.21$ in Qwen and $0.24$ in GPT-5.1. LLaVA's unsafe network fragments into several small peripheral groups (five communities) but at low modularity, i.e. weak separation, whereas GPT-5.1 resolves danger into a few clean, interpretable themes. Notably, the visibility-and-activity cues ("poor lighting", "poor visibility", "no pedestrians") that form a dedicated community in GPT-5.1 remain present but less separated in Gemini and Qwen and are largely absent in LLaVA, indicating that stronger models articulate the intuition that visibility and street life signal safety more explicitly than smaller ones.}
 
Interestingly, \textcolor{black}{traffic- and infrastructure-related cues (e.g., "traffic", "construction", "highway", "parked cars", "industrial") appear in both the safe and the unsafe networks across models}, suggesting that urban infrastructure is a shared theme across both classifications. However, its interpretation varies depending on the context and the other nodes connected. In the safe network, infrastructure is associated with organized and well-maintained urban spaces, whereas, in the unsafe network, it is linked to physical decay, isolation, and vandalism.
 
\textcolor{black}{Because a co-occurrence matrix can be dominated by a few high-frequency terms and is sensitive to sampling noise, we assessed the reproducibility of the networks (full tables in Appendix~\ref{keyword_network_SI}). The two classes behave differently. The \emph{safe} network is highly reproducible: the top-25 keyword counts are strongly correlated across runs (Pearson $r=1.00$, Spearman $\rho=0.89$), the edge weights are stable ($r=0.99$, $\rho=0.90$), and the recovered community structure is identical (adjusted Rand index, ARI $=1.00$) and completely insensitive to the keyword cutoff (ARI $=1.00$ across top-20/25/30). The \emph{unsafe} network is noisier: while its keyword frequencies remain highly correlated across runs ($r=0.98$), the exact top-25 membership overlaps less (16/25) and the community partition is only moderately stable (ARI $\approx0.57$, and $0.33$--$0.59$ across cutoffs). This asymmetry is expected --- unsafe judgements are far fewer and draw on a longer tail of near-synonymous decay terms --- so we treat the unsafe communities as indicative thematic groupings rather than exact partitions, while the safe partition and the reported keyword frequencies for both classes are robust.}
 
\textcolor{black}{Lastly, the linguistic patterns identified in our analysis resonate with foundational urban theories\textcolor{black}{, and the fact that they recur across four independent models reinforces that these are properties of the task rather than of a single architecture}. The distinction between the "safe" network's emphasis on order and the "unsafe" network's focus on disorder parallels Kelling and Wilson's Broken Windows theory \cite{kelling1982broken}. Similarly, the results are consistent with Jacobs' Eyes on the Street \cite{jacobs1992death}; cues like "neighborhood" and "residential" suggest natural surveillance, while "vacant" and "no pedestrians" echo her warnings against desolate spaces\textcolor{black}{ --- an association made most explicit by GPT-5.1, whose "no pedestrians" and "poor lighting" cues form a distinct unsafe community}. Regarding Newman's Defensible Space \cite{newman1972defensible}, the model captures how architectural maintenance shifts perception from accessible to disorderly. Finally, the contrast between community engagement keywords and terms like "gated community" or "isolation" aligns with Social Disorganization Theory \cite{shaw1942juvenile, sampson1989community}, suggesting the model detects visual proxies for social cohesion rather than just physical structures.}

\subsection{Socio-Demographic Personas Shift Safety Classifications}
Zero-shot learning, a paradigm where models make predictions on tasks without prior task-specific training, has been increasingly explored in various domains \cite{beneduce2025large, li2023frozenlanguagemodelhelps, gruver2024largelanguagemodelszeroshot}. Previously, we have shown how MLLMs assess urban safety perception in a zero-shot fashion, meaning we evaluate safety by directly prompting the model and without any fine-tuning related to the dataset (i.e., SVIs) or task (i.e., safety perception). However, a critical question emerges: \emph{whose} perception is being captured? Contemporary urban planning increasingly embraces citizen-centric approaches and co-design methodologies \cite{forester1999deliberative, healey1997collaborative, luca2024towards, Ahn2025}, while foundation models are becoming prevalent across various domains as decision-support tools and process managers \cite{wei2023chainofthoughtpromptingelicitsreasoning, bommasani2022opportunitiesrisksfoundationmodels}. Therefore, understanding whether we can model different Personas within prompts and quantifying the impact of socio-demographic factors on the results becomes crucial. \textcolor{black}{An additional exploratory analysis combining race/ethnicity,
gender, and age is reported in Appendix~\ref{app:intersectional_bias}.}
 
\textcolor{black}{To investigate this, we treated the Persona as an independent variable. We designed a set of distinct prompts, each encoding a specific socio-demographic profile \textcolor{black}{(Gender, Age, and Ethnicity)}. Our findings reveal that all these characteristics significantly influence \textcolor{black}{the models'} safety perception, with some factors showing particularly pronounced effects. Most notably, we discover that responses obtained using the \emph{Neutral prompt} align predominantly with those from young, middle-aged and male respondents. \textcolor{black}{For ethnicity the closest-matching group is model-dependent, but the divergence is systematic at the other end of the scale: the Neutral prompt is consistently \emph{least} aligned with Native American personas, the lowest or second-lowest agreement in every model, and frequently with Black/African American personas, indicating that the default perspective least represents historically marginalised groups}. This alignment, potentially, indicates an inherent \textcolor{black}{and systematic} model bias.}

\textcolor{black}{Crucially,
we repeated the experiment across four MLLMs (LLaVA 1.6 7B, Gemini 2.5 Flash Lite, Qwen2.5-VL-72B and
GPT-5.1), each with two independent runs, so that any observed pattern can be assessed for
\emph{cross-model robustness} rather than being attributed to a single checkpoint.
The detailed results for each socio-demographic prompt are presented in the following sections.}

\textcolor{black}{\subsubsection{Race and ethnicity}}
 
\textcolor{black}{To examine whether race- and ethnicity-related language changes the models'
classification behaviour, we evaluate six broad persona prompts:
\emph{White/Caucasian}, \emph{Asian/Pacific Islander},
\emph{Hispanic/Latino}, \emph{Middle Eastern/North African},
\emph{Black/African American}, and
\emph{Native American/Alaska Native}.}

\textcolor{black}{The persona effects are systematic but model-dependent
(Table~\ref{tab:race_unsafe}). The White/Caucasian persona remains close to
the Neutral baseline in magnitude across all four models, with changes ranging
from approximately $-2$ to $+3$ percentage points. Asian/Pacific Islander,
Hispanic/Latino, and Middle Eastern/North African prompting generally increases
the unsafe-classification rate in LLaVA, Qwen, and GPT-5.1, but produces
smaller and less consistent changes in Gemini.}

\textcolor{black}{The largest positive shifts are generally observed for the Black/African
American and Native American personas, although their ordering varies across
models. Relative to Neutral, Native American prompting increases the
city-macro unsafe-classification rate by 38.00 percentage points in LLaVA,
17.11 in Gemini, 18.77 in Qwen, and 6.35 in GPT-5.1. All four differences are
significant at $p<0.001$. Black/African American prompting produces significant
increases in LLaVA ($+21.65$ pp), Qwen ($+22.38$ pp), and GPT-5.1
($+10.41$ pp). Gemini differs from this pattern, producing a small but
statistically significant decrease ($-2.15$ pp, $p=0.007$).}

\textcolor{black}{The White/Caucasian persona remains close to Neutral in effect magnitude
($|d_z|\leq0.36$), although small differences are statistically detectable
in three models when using the unadjusted permutation $p$-values. Across the
two repeated runs, the direction and magnitude of the largest persona effects
remain broadly consistent. 
Run-to-run standard deviations are small relative to these gaps (Appendix~\ref{Disparities}), so the
racialized differences cannot be dismissed as generation noise.}
 
\textcolor{black}{Together, these results establish that race- and ethnicity-conditioned language alters a model's response criteria for identical images. Specifically, models mostly associate Black and Native American identities with heightened insecurity,  a behaviour stable across different architectures and scales.}

\begin{table}[t]
  \centering
  \caption{\textcolor{black}{Unsafe-classification rate (\%) by race/ethnicity persona, per model}. \textcolor{black}{Rows ordered by cross-model mean. The Neutral baseline is the Neutral prompt
  evaluated in the same race/ethnicity batch, so it may differ by up to $\sim$1~percentage point from the Neutral row
  of Table~\ref{tab:neutral_perf} (run-to-run variation). Superscripts give the significance of each
  persona's shift relative to this baseline (city-level sign-flip permutation test, $n{=}56$ paired
  cities): $^{\ast}p<0.05$, $^{\ast\ast}p<0.01$, $^{\ast\ast\ast}p<0.001$; unmarked cells are not
  significant. Most shifts are positive (more unsafe than Neutral); the two marked Gemini cells (White,
  Black) are small \emph{negative} shifts. Native American is significant in all four models and
  Black/African American is significantly elevated in three of four, while the White/Caucasian persona
  shows only small shifts ($|d_z|\le0.36$) that are nonetheless significant in three models.}}
  \label{tab:race_unsafe}
  \begin{tabular}{lcccc}
    \toprule
    Persona & LLaVA 1.6 7B & Gemini 2.5 FL & Qwen2.5-VL-72B & GPT-5.1 \\
    \midrule
    Neutral (baseline)              & 5.9  & 25.5 & 21.6 & 18.8 \\
    White / Caucasian               & 7.2$^{\ast\ast}$ & 23.2$^{\ast}$ & 21.6 & 21.3$^{\ast}$ \\
    Middle Eastern / N. African     & 15.9$^{\ast\ast\ast}$ & 24.3 & 24.9$^{\ast\ast\ast}$ & 19.8 \\
    Asian / Pacific Islander        & 10.8$^{\ast\ast\ast}$ & 26.2 & 27.1$^{\ast\ast\ast}$ & 22.8$^{\ast\ast\ast}$ \\
    Hispanic / Latino               & 13.6$^{\ast\ast\ast}$ & 25.2 & 29.2$^{\ast\ast\ast}$ & 23.6$^{\ast\ast\ast}$ \\
    Black / African American        & 27.6$^{\ast\ast\ast}$ & 23.4$^{\ast\ast}$ & 44.0$^{\ast\ast\ast}$ & 29.2$^{\ast\ast\ast}$ \\
    Native American                 & 43.9$^{\ast\ast\ast}$ & 42.6$^{\ast\ast\ast}$ & 40.3$^{\ast\ast\ast}$ & 25.1$^{\ast\ast\ast}$ \\
    \bottomrule
  \end{tabular}
\end{table}

\subsubsection{Gender}
We next examined the influence of gender on safety perception. While we acknowledge that a binary
classification does not capture the full spectrum of gender identity, for the purposes of this study
we operationalized this variable through distinct Male and Female prompts. 
 
\textcolor{black}{Our analysis reveals a pronounced and highly consistent gender gap: in \emph{every}
model the Female persona classifies more of the same imagery as \emph{Unsafe} than the Male persona
(Table~\ref{tab:gender_gap}). The gap ranges from
$+3.7$~pp for LLaVA (12.6\% vs 8.9\%) to $+22.1$~pp for GPT-5.1 (46.5\% vs 24.5\%), with Gemini ($43.3\%$ vs $26.7\%$) and Qwen ($31.2\%$ vs $23.7\%$) in between. This divergence indicates that the
models encode a heightened perception of risk when adopting a female perspective, and it does not depend on the specific model.}
 
\textcolor{black}{Again, the gap is statistically robust rather than an artefact of stochastic generation. Using
a city-level paired design ($n{=}56$ paired cities; sign-flip permutation test, $95\%$ bootstrap CI and
Cohen's $d_z$), the Female~$-$~Male difference is significant in all four models ($p<0.001$) with medium-to-large
effect sizes ($d_z$ from $0.82$ for LLaVA to $2.26$ for GPT-5.1), and the run-to-run standard
deviation of the underlying rates is small relative to the gap (Table~\ref{tab:gender_gap}).}

\begin{figure}[ht]
\centering
\includegraphics[width=1\linewidth]{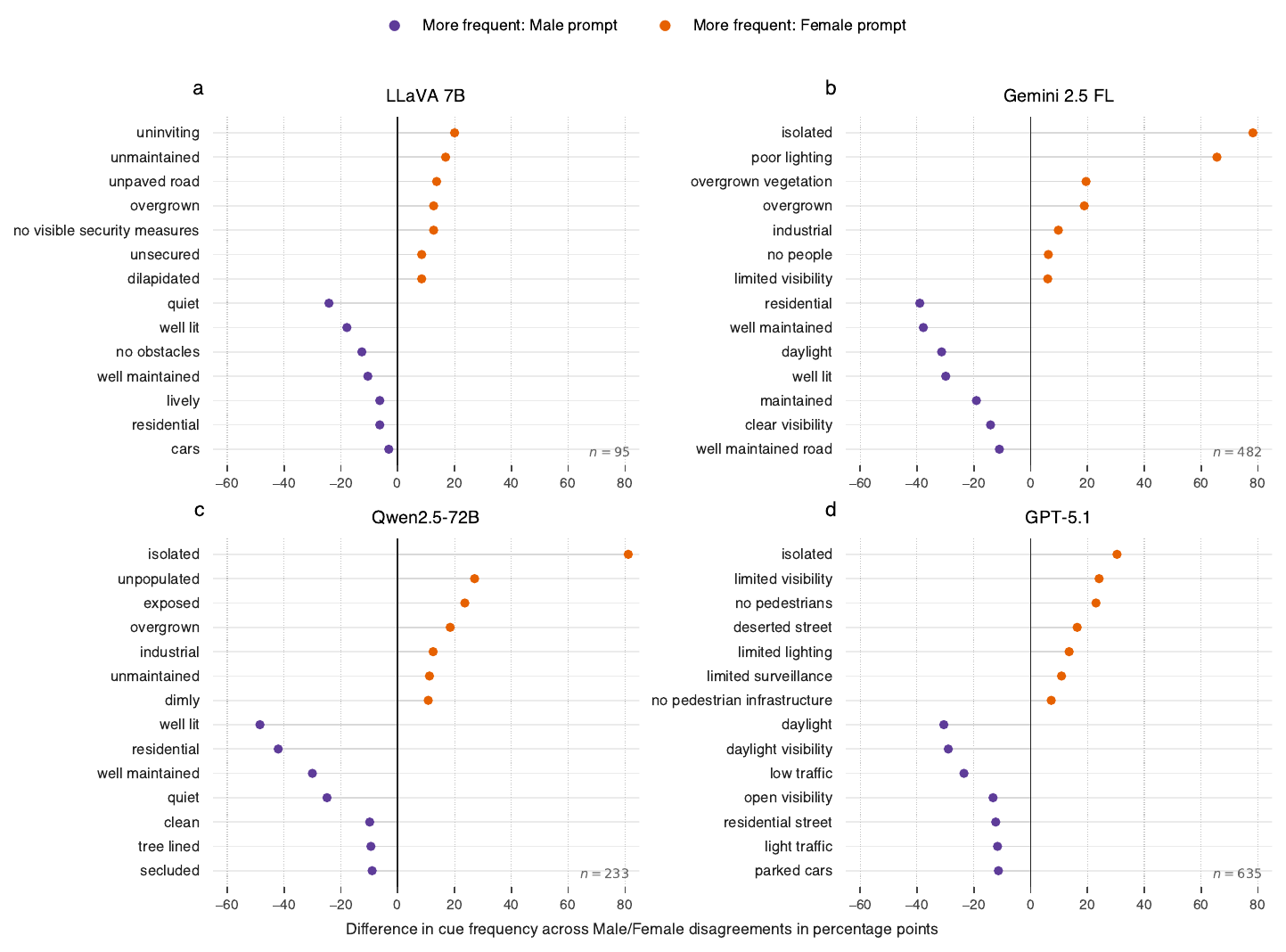}
\caption{\textcolor{black}{For each model, the analysis includes only cases in which the Male and Female
persona prompts assign different safety classifications to the same image.
Points report the difference in keyword frequency between the two prompts,
expressed in percentage points and calculated as Female minus Male. Positive
values therefore indicate cues used more frequently under the Female prompt,
whereas negative values indicate cues used more frequently under the Male
prompt. The displayed terms are those with the largest frequency differences
for each model, and $n$ denotes the number of disagreement cases included in
the corresponding panel.}}
\label{fig:keyword_gender}
\end{figure}

To better understand the factors driving this divergence, we focused on instances where the model's
classifications differed between the Male and Female prompts. By isolating these cases of disagreement
on the \emph{same} image, we analyzed the associated keywords to identify the features that shaped each
perspective (Figure~\ref{fig:keyword_gender}). \textcolor{black}{The same qualitative pattern recurs
across all four models.} For male prompts, keywords such as ``Residential'', ``Well-maintained'' and
``Orderly'' dominated, emphasizing features of accessibility and organization; these align with a focus
on structural and environmental stability, often associated with positive perceptions of safety. In
contrast, female prompts showed a stronger emphasis on terms such as ``Isolated'', ``Poor lighting'',
``No pedestrians'' and ``Limited visibility'', which reflect a sense of vulnerability in urban
environments.

\begin{table}[t]
  \centering
  \caption{\textcolor{black}{Gender gap in unsafe classification, per model. Values are the mean unsafe
  rate (\%) over two runs ($\pm$ run-to-run SD); $\Delta$ is the Female~$-$~Male difference in mean
  paired city-level unsafe rate (percentage points) with its significance (sign-flip permutation test,
  $n{=}56$ paired cities) and Cohen's $d_z$. Female $>$ Male in every model. $\Delta$ is computed
  from the unrounded city-macro rates, so it may differ from the difference of the one-decimal values
  shown by up to $0.1$~pp.}}
  \label{tab:gender_gap}
  \small
  \begin{tabular}{lccrc}
    \toprule
    \textbf{Model} & \textbf{Male (\%)} & \textbf{Female (\%)} & \textbf{$\Delta$ (pp)} & \textbf{$p$ \;($d_z$)} \\
    \midrule
    LLaVA 1.6 7B     & $8.9\pm2.4$  & $12.6\pm3.2$ & $+3.7$  & $<0.001$\;($0.82$) \\
    Gemini 2.5 FL    & $26.7\pm0.3$ & $43.3\pm0.5$ & $+16.6$ & $<0.001$\;($2.12$) \\
    Qwen2.5-VL-72B   & $23.7\pm0.1$ & $31.2\pm0.4$ & $+7.6$  & $<0.001$\;($1.15$) \\
    GPT-5.1          & $24.5\pm0.3$ & $46.5\pm0.0$ & $+22.1$ & $<0.001$\;($2.26$) \\
    \bottomrule
  \end{tabular}
\end{table}

\textcolor{black}{This directionality is consistent with the empirical human literature on gendered
safety perception, in which women report heightened sensitivity to isolation, poor lighting and the
absence of other people \cite{quintana2025global}. Although, as we discuss in
Section~\ref{sec3}, prompting changes the model's output without guaranteeing genuine
perspective-taking, and the effect is best characterised as a shift in response criterion rather than
in perceptual ability.}

\subsubsection{Age}
\textcolor{black}{We probe the effect of age on the model's simulated safety perception through three
personas --- a young, a middle-aged, and an elderly observer --- each of which re-evaluates the full
dataset across the four MLLMs (LLaVA 1.6 7B, Gemini 2.5 Flash Lite, Qwen2.5-VL-72B and GPT-5.1), with
two runs per model. Unlike the gender axis, where the direction of the effect is identical in every
model, the age axis reveals a more nuanced and model-dependent picture (Table~\ref{tab:age_gap}).}
 
\textcolor{black}{Two regularities hold across models. First, the \emph{middle-aged} persona is
tendentially the least risk-sensitive: it yields the lowest, or nearly the lowest, share of
\emph{Unsafe} classifications across models (e.g.\ 24.4\% for GPT-5.1 and 25.8\% for Qwen, against
higher rates for the age extremes). Second, the age \emph{extremes}, young and elderly, tend to
be elevated relative to this middle-aged baseline, so that perceived unsafety follows an approximately
U-shaped profile over age.}
 
\textcolor{black}{Which extreme is judged the least safe, however, depends on the model. The elderly
persona records the highest unsafe rate in Qwen (33.7\%) and GPT-5.1 (41.3\%), the young persona is
highest in LLaVA (21.6\%), and the two are effectively tied in Gemini (43.0\% vs 43.1\%). A city-level
paired test ($n{=}56$ paired cities; sign-flip permutation test, Cohen's $d_z$) confirms that this
heterogeneity is real rather than noise: the Elderly~$-$~Young contrast is positive and significant for
Qwen ($d_z{=}1.36$) and GPT-5.1 ($d_z{=}1.46$), statistically null for Gemini, and significantly
\emph{negative} for LLaVA ($d_z{=}-1.26$), where the young persona is the more risk-averse. There is
therefore no single, model-independent age ordering, and in particular no universal ``the elderly
perceive the most danger'' effect.}
 
\textcolor{black}{Inspecting the cues the model cites for each age persona is consistent with this
reading: the elderly persona tends to foreground accessibility and walkability features (paved,
well-maintained walkways), whereas the younger persona leans on openness and greenery. In aggregate,
the robust conclusion is that the middle-aged viewpoint acts as the model's calm reference and the
extremes are more sensitive to environmental risk cues, while the specific direction of the age effect
is not stable across architectures. This aligns with human evidence that age shapes safety perception
without a single universal direction \cite{quintana2025global, kang2023assessing}, and we accordingly
refrain from attributing a fixed age ordering to ``the model.''}
 
\begin{table}[t]
  \centering
  \caption{\textcolor{black}{Age effect on unsafe classification, per model. Values are the mean unsafe
  rate (\%) over two runs ($\pm$ run-to-run SD). $\Delta$(Eld.$-$Yng.) is the mean paired city-level
  difference between the Elderly and Young personas (percentage points), with its significance
  (sign-flip permutation test, $n{=}56$ paired cities) and Cohen's $d_z$. The Middle-aged persona produces the lowest or near-lowest unsafe rate in all models, but the sign of the Elderly$-$Young contrast varies by model.
  $\Delta$ is computed from the unrounded city-macro rates, so it may differ from the difference of the
  one-decimal values shown by up to $0.1$~pp.}}
  \label{tab:age_gap}
  \small
  \begin{tabular}{lcccr}
    \toprule
    \textbf{Model} & \textbf{Young (\%)} & \textbf{Middle (\%)} & \textbf{Elderly (\%)} & \textbf{$\Delta$(Eld.$-$Yng.)\; $p$ ($d_z$)} \\
    \midrule
    LLaVA 1.6 7B    & $21.6\pm4.4$ & $10.6\pm2.5$ & $8.9\pm2.5$  & $-12.7$\;\; $<0.001$ ($-1.26$) \\
    Gemini 2.5 FL   & $43.1\pm0.6$ & $31.5\pm0.2$ & $43.0\pm0.1$ & $-0.1$\;\;\; $0.87$ (n.s.) \\
    Qwen2.5-VL-72B  & $26.6\pm0.3$ & $25.8\pm0.0$ & $33.7\pm0.2$ & $+7.0$\;\;\; $<0.001$ ($+1.36$) \\
    GPT-5.1         & $29.5\pm0.2$ & $24.4\pm0.9$ & $41.3\pm0.9$ & $+11.8$\;\; $<0.001$ ($+1.46$) \\
    \bottomrule
  \end{tabular}
\end{table}

\subsection{Demographic Alignment of the Neutral Prompt}

\textcolor{black}{To characterise the model's default perspective, we compare every socio-demographic
persona against the \emph{Neutral} prompt along two complementary axes. The first is the
\emph{Delta Unsafe} metric (Materials and Methods \ref{sec4}): how far each persona shifts the
unsafe-classification rate away from the Neutral baseline, which reveals the identities the default
under- or over-estimates. The second is an instance-level alignment measure: the per-image
agreement between each persona's classifications and the Neutral output, which reveals the identity the
default most resembles.}


We first evaluated the internal consistency of city rankings across prompts. \textcolor{black}{Within each model, we rank the 56 cities by their
unsafe-classification rate under the Neutral prompt and, separately, under each persona prompt, and we
correlate every persona's ranking with the Neutral ranking (Spearman's $\rho$;
Table~\ref{tab:ranking_stability}). The rankings are strongly preserved for the three full-precision
models (mean $\rho=0.93$, $0.94$ and $0.90$ for Gemini, Qwen and GPT-5.1) and moderately so for the
LLaVA (mean $\rho=0.74$), whose low absolute unsafe rates leave many cities tied and its ranking
noisier. Personas therefore shift the overall \emph{level} of perceived unsafety while largely
preserving the relative \emph{ordering} of cities.}

\textcolor{black}{The \emph{Delta Unsafe} metric exhibits a clear demographic structure that echoes the
per-axis results reported above. Along race/ethnicity, the White/Caucasian persona sits within a few
points of the Neutral baseline in every model ($-3$ to $+2$~pp), whereas the Black/African American and
Native American personas raise the unsafe rate by up to $+23$ and $+37$ percentage points,
respectively. Along gender, the Female persona increases the unsafe rate several times more than the
Male persona (roughly $+6$ to $+26$~pp for Female versus $\leq +6$~pp for Male, relative to Neutral).
Across age, the middle-aged persona deviates least from Neutral, while the young and elderly extremes
deviate most.}
 
\begin{figure*}[t]
    \centering
    \includegraphics[width=1\textwidth]{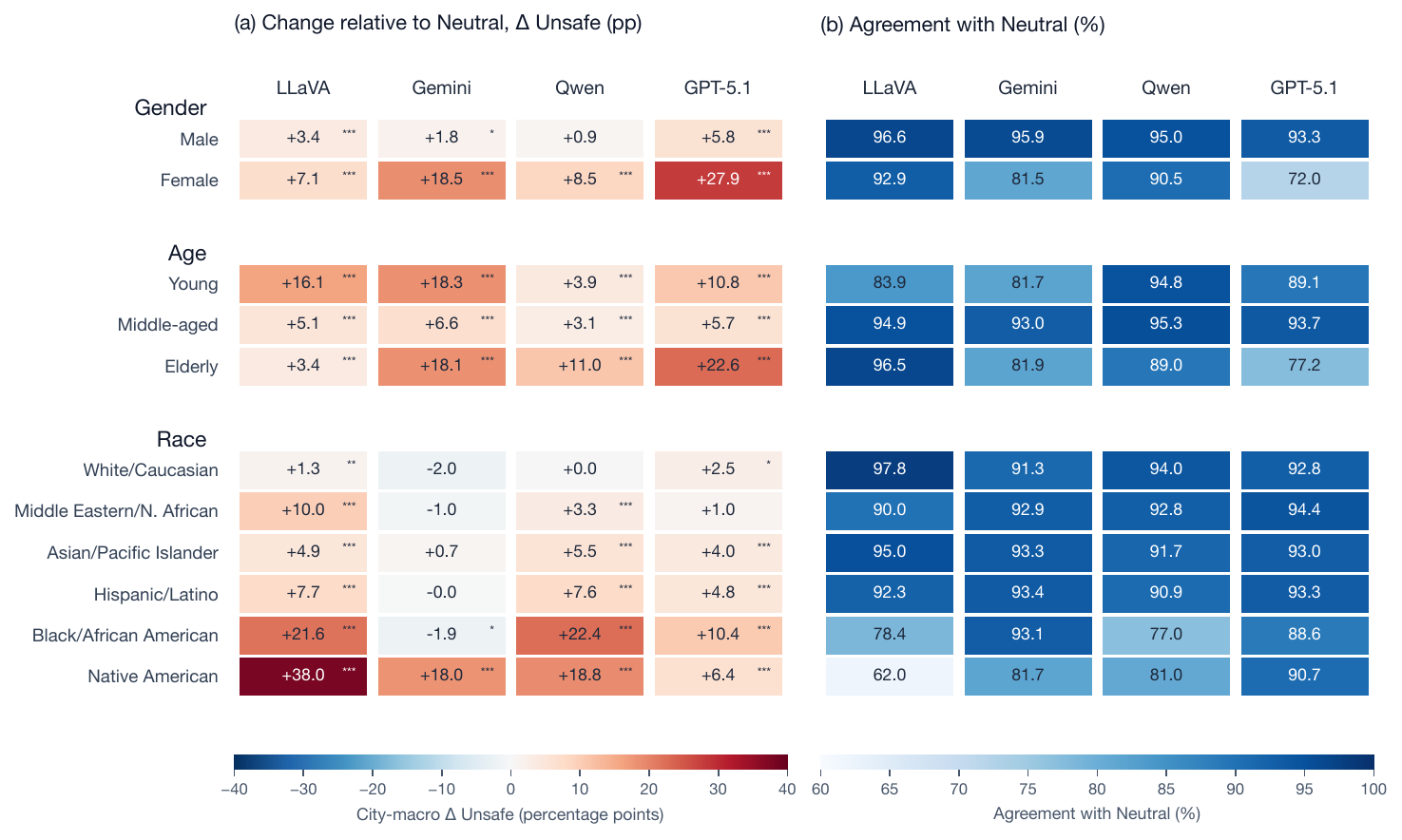}
    \caption{\textcolor{black}{{Persona-induced changes and alignment relative to the Neutral prompt.}
    \textbf{(a)} City-macro change in the proportion of images classified as \emph{Unsafe} under each socio-demographic persona relative to the Neutral prompt, expressed in percentage points. Positive values indicate that the persona produces more \emph{Unsafe} classifications than the Neutral prompt, whereas negative values indicate fewer. Asterisks denote significance under paired city-level sign-flip permutation tests: $^{*}p<0.05$, $^{**}p<0.01$, and $^{***}p<0.001$.
    \textbf{(b)} City-macro classification agreement between each persona and the Neutral prompt. Higher values indicate that the persona reproduces a larger proportion of the Neutral classifications. Results are averaged over two experimental runs, with each of the 56 cities weighted equally. Across models, the Neutral prompt generally aligns most closely with Male and Middle-aged personas, while Female, Black/African American, and Native American personas show larger and more model-dependent departures.}}
    \label{fig:neutral_persona_alignment}
\end{figure*}

\textcolor{black}{An instance-level alignment analysis views the same effect from the opposite direction:
for each persona we measure the per-image agreement between its classifications and the Neutral output
(Figure~\ref{fig:neutral_persona_alignment}). We note that this agreement is, by construction, sensitive to the
base rate --- a persona that shifts the overall unsafe level only slightly necessarily agrees with
Neutral on most images --- so alignment and \emph{Delta Unsafe} are two views of the same underlying
shift rather than independent measurements. With that caveat, the alignment ordering is informative.
Two conclusions are robust across all four models: the Neutral default agrees \emph{most} with the
\textbf{Male} persona (agreement $94$--$97\%$, versus $74$--$94\%$ for Female), and agrees \emph{least}
with the \textbf{Black/African American} and \textbf{Native American} personas (agreement falling to
$63$--$92\%$). For the remaining axes the alignment is high but its exact top persona is
model-dependent: the middle-aged persona is the closest age match in the three full-precision models
(the strongly Safe-biased LLaVA agrees highly with Neutral for \emph{all} ages, so its ordering
mainly reflects the shared low base rate), and White/Caucasian is the single closest race persona in
LLaVA and Qwen while Gemini and GPT-5.1 sit marginally closer to a lighter/mid persona (Hispanic,
Middle-Eastern). On average across models, the closest persona on each axis is White/Caucasian
($\sim$94\%), Male ($\sim$96\%) and Middle-aged ($\sim$95\%).}

\begin{table}[t]
  \centering
  \small 
  \caption{\textcolor{black}{City-ranking stability across prompts. For each model we rank the 56 cities
  by their unsafe-classification rate under the Neutral prompt and under each persona, and report
  Spearman's $\rho$ between each persona's ranking and the Neutral ranking. Rankings are strongly
  preserved for the three full-precision models and moderately so for the LLaVA (whose low unsafe
  rates leave many cities tied, adding rank noise): personas shift the \emph{level} of perceived
  unsafety while largely preserving the relative \emph{ordering} of cities.}}
  \label{tab:ranking_stability}
  
  \footnotesize
  \addtolength{\tabcolsep}{-2.5pt} 
  
  \begin{tabular}{llcccc}
    \toprule
    & \textbf{Persona} & 
    \textbf{\makecell[c]{LLaVA 1.6\\7B}} & 
    \textbf{\makecell[c]{Gemini 2.5\\FL}} & 
    \textbf{\makecell[c]{Qwen2.5-VL\\72B}} & 
    \textbf{\makecell[c]{GPT-5.1}} \\
    \midrule
    \multirow{2}{*}{Gender}
      & Male                        & 0.85 & 0.97 & 0.95 & 0.94 \\
      & Female                      & 0.73 & 0.92 & 0.96 & 0.77 \\
    \midrule
    \multirow{3}{*}{Age}
      & Young                       & 0.59 & 0.87 & 0.97 & 0.88 \\
      & Middle-aged                 & 0.81 & 0.95 & 0.97 & 0.91 \\
      & Elderly                     & 0.85 & 0.84 & 0.96 & 0.88 \\
    \midrule
    \multirow{6}{*}{\makecell[l]{Race /\\ethnicity}}
      & White/Caucasian             & 0.91 & 0.95 & 0.96 & 0.93 \\
      & Asian/Pacific Isl.          & 0.78 & 0.95 & 0.94 & 0.91 \\
      & Hispanic/Latino             & 0.74 & 0.95 & 0.94 & 0.94 \\
      & Middle Eastern/N.\,Afr.     & 0.75 & 0.96 & 0.96 & 0.95 \\
      & Black/African Am.           & 0.65 & 0.95 & 0.89 & 0.91 \\
      & Native American             & 0.54 & 0.94 & 0.88 & 0.91 \\
    \midrule
    & \textbf{Mean vs.\ Neutral}    & \textbf{0.74} & \textbf{0.93} & \textbf{0.94} & \textbf{0.90} \\
    \bottomrule
  \end{tabular}
\end{table} 
 
\textcolor{black}{Taken together, the model's un-prompted ``Neutral'' configuration is not demographically
neutral: it behaves most like a \emph{male, middle-aged, White-or-lighter-associated} observer, and is
consistently \emph{least} representative of Black, Native-American, female and (at the extremes) older
or younger perspectives. This echoes the behavioural-alignment finding that unprompted large language
models carry a specific, non-neutral prior \cite{tao2024cultural}, and it underscores the importance of
explicitly specifying socio-demographic characteristics in societally consequential applications:
relying on the model's standard configuration silently adopts one privileged perspective and fails to
represent the diversity of the groups a city serves.}

\section{Discussion}\label{sec3}

\begingroup
\color{black}

Our results reveal a central tension in the use of Multimodal Large Language Models for urban perception: MLLMs can extract a meaningful and transferable signal of perceived safety, but they do not interpret the city from a neutral standpoint. Across four open and proprietary models, predictions preserve substantial variation between cities and rely on a convergent set of visual cues. At the same time, all models systematically favour the \emph{Safe} class, compress differences between the safest and least-safe cities, and alter their classifications when the observer's socio-demographic identity is specified. The relevant question is therefore no longer only whether MLLMs can assess perceived urban safety, but from which learned perspective they perform that assessment.

The first side of this tension is genuine visual competence. Despite receiving no task-specific training, all four models attain similar performance and broadly recover the ordering of cities by perceived safety. Their generated explanations also converge on a coherent visual vocabulary. Maintenance, greenery, residential character, order, visibility, and daytime activity support \emph{Safe} classifications, whereas deterioration, isolation, poor lighting, low pedestrian presence, and inadequate infrastructure support \emph{Unsafe} classifications. These associations are consistent with established accounts of how urban form shapes perceived insecurity. Visible neglect parallels Broken Windows Theory \cite{kelling1982broken}; pedestrian activity and visibility recall Jacobs' Eyes on the Street \cite{jacobs1992death}; and maintenance, accessibility, and territorial organisation relate to Defensible Space Theory \cite{newman1972defensible}. The models therefore reproduce a recognisable account of urban safety rather than relying on arbitrary or entirely model-specific features.

Yet the structure of this lexicon also reveals the narrowness of that account. Across models, safety is represented as a relatively uniform ideal of the orderly, maintained, green, residential street. Unsafety is more differentiated, separating physical decay, lack of activity, limited visibility, and spatial isolation into distinct themes. This asymmetry suggests that the models recognise several ways in which an environment can depart from a comparatively homogeneous image of safety. Its recurrence across architectures strengthens the interpretation that current MLLMs share broad visual conventions for judging urban environments. However, the generated keywords remain model-produced justifications rather than transparent descriptions of the internal mechanisms responsible for each prediction. They reveal how the models articulate their judgements, not necessarily how those judgements were causally formed.

The second side of the tension is that this visual competence is systematically miscalibrated. Although the models distinguish relatively safer from less-safe cities, they consistently underestimate the proportion of scenes labelled \emph{Unsafe}. The shallow relationship between predicted and ground-truth unsafe rates shows that the models compress city-level variation into a narrower and more reassuring range. They therefore capture much of the \emph{ordering} of perceived safety while muting its \emph{magnitude}. This Safe-bias is especially consequential in cities where images labelled \emph{Unsafe} are more prevalent, because an optimistic classification strategy generates many false \emph{Safe} predictions and consequently reduces Safe precision and F1.

This interaction helps explain the apparent geographic disparity in performance. Global-South cities are rated as less safe on average in Place Pulse 2.0 and are therefore disproportionately affected by the models' Safe-bias. 

The benchmark itself introduces a further layer of perspective. Place Pulse 2.0 records crowdsourced judgements of \emph{perceived} safety, not objective measurements of crime, danger, or victimisation \cite{dubey2016deep}. Its labels therefore represent aggregated evaluations produced by an online participant population rather than locally representative judgements from the residents of every city. Recent demographic-rich evidence confirms that urban visual perception varies with gender, age, race and ethnicity, education, income, personality, and the geographic context of both observer and streetscape \cite{quintana2025global}. Regional differences in Place Pulse scores may consequently reflect geographically situated interpretations of urban form rather than universally shared assessments. 

The persona experiments make this otherwise implicit observer visible. Because every persona evaluates the same images, the resulting shifts cannot be explained by changes in the visual environment. They arise from the socio-demographic context introduced through language. Female prompting increases the unsafe-classification rate relative to Male prompting in every model, while the Middle-aged persona generally remains closer to the Neutral output than the age extremes. Black/African American and Native American prompting also produces some of the largest shifts, although their magnitude varies substantially across models. These findings are consistent with the broader empirical observation that safety perception is observer-dependent \cite{quintana2025global}. However, reproducing demographic variation is not equivalent to faithfully representing the perceptions of the named groups.

This distinction is essential. Without demographic-specific human judgements for the same images, we cannot determine whether persona prompting reproduces lived experience, amplifies correlations present in training data, or activates simplified social stereotypes. The consistent Female--Male difference may resemble patterns documented in human safety perception, but correspondence in direction does not demonstrate genuine perspective-taking. The same caution applies more strongly to the race and ethnicity results, where large persona-induced shifts may reflect learned associations between particular identities, vulnerability, and environmental risk. Our results therefore establish \emph{persona sensitivity}, not demographic fidelity.

The comparison with the Neutral prompt places this sensitivity within the broader literature on behavioural alignment. Tao et al.\ show that unprompted language models do not provide a culturally empty baseline; their responses align more closely with the values of particular cultural regions, and explicit cultural prompting can shift that alignment \cite{tao2024cultural}. Our findings extend this problem from text-based value questionnaires to image-grounded urban perception. In our setting, ``Neutral'' means only that no observer is named. It does not mean that the resulting judgement is free from an implicit observer.

This non-neutrality is visible in both the agreement and flip analyses. The Neutral output consistently resembles the Male persona most closely and generally remains close to the Middle-aged persona. Its closest race or ethnicity match varies across models, but Black/African American and Native American personas frequently produce the largest departures. Moreover, these departures are strongly directional. Personas associated with greater perceived unsafety primarily convert Neutral-\emph{Safe} classifications into \emph{Unsafe} classifications, with relatively few reversals in the opposite direction. Persona prompting therefore behaves largely as a shift in the model's response criterion rather than as an entirely new interpretation of each image.

Under this perspective, persona prompts are valuable diagnostic interventions. They expose how an apparently unspecified model response changes when an observer is made explicit. They should not, however, be treated as synthetic survey respondents or as substitutes for the groups described in the prompt. Behavioural similarity between a persona and the Neutral output measures consistency within the model, whereas demographic alignment requires comparison with human responses from the corresponding population. 

This distinction also clarifies the contribution of our study. Prior research has shown that computer-vision models can approximate aggregate urban-perception scores \cite{de2016safer,naik2014streetscore,quercia2014aesthetic,dubey2016deep}, that human urban perception varies across demographic and geographic contexts \cite{quintana2025global}, and that unprompted language models carry culturally specific behavioural priors \cite{tao2024cultural}. Our contribution is not simply to report another instance of algorithmic bias. Rather, we connect these previously separate observations within a controlled multimodal setting. The same images are evaluated across four MLLMs, cities, and socio-demographic prompts, allowing us to distinguish three phenomena that aggregate performance alone would conflate: visual competence, calibration bias, and persona-induced response shifts.

The cross-model design is important to this contribution. Although the magnitude of individual effects varies, the principal structure recurs across independently developed systems: comparable zero-shot performance, a systematic Safe-bias, a shared visual lexicon, and sensitivity to observer identity. This consistency does not establish that every MLLM will behave identically, but it shows that the findings are not reducible to one architecture or checkpoint. At the same time, independently developed models are not necessarily independent of the broader data ecosystem from which they emerge \cite{longpre2024bridging}. Although the undisclosed training corpora of proprietary MLLMs prevent direct causal attribution, these ecosystem-level patterns provide a plausible context for the recurrence of similar geographic and demographic priors across models. Selecting a more capable model therefore does not necessarily remove the embedded perspective; in our results, some of the largest persona-induced shifts are observed in the most capable models.

For urban research and planning, MLLMs should therefore be treated as scalable but situated measurement instruments. Their value lies in complementing surveys and participatory methods, not replacing them. A model can efficiently reproduce one learned interpretation of an urban environment, but it cannot establish that this interpretation represents the residents who experience that environment.

Responsible use consequently requires more than adding demographic labels to a prompt. Researchers should report the assumed observer, test whether conclusions change across personas, quantify calibration and uncertainty, and validate model outputs against locally collected human judgements. Where the outputs could influence planning priorities or resource allocation, validation should include the communities affected by those decisions. Otherwise, the apparent neutrality and scalability of automated assessment may conceal the perspective that has been operationalised.

Several limitations define the scope of these conclusions. Place Pulse 2.0 does not provide demographic-specific labels for the images used in our experiments, preventing direct validation of persona fidelity. Its labels represent perceived safety, its imagery reflects an earlier period, and binary thresholding removes part of the uncertainty encoded in the original continuous scores. The personas are also deliberately simplified: binary gender categories and broad race and ethnicity labels cannot capture the heterogeneity of human identity. In addition, four models and two experimental runs provide cross-model evidence but not universal coverage of the rapidly changing MLLM landscape. Finally, the keyword explanations may contain post-hoc rationalisations and should not be interpreted as direct access to model reasoning.

Future work should therefore combine identical-image model evaluation with demographic-rich and geographically local human annotations. Such data would make it possible to distinguish useful adaptation from stereotype amplification and measure whether persona prompting moves the model towards or away from the perceptions of the population it claims to represent. Calibration should also be assessed separately across regions and social groups rather than only at the aggregate level.

Taken together, our findings show that MLLMs can read meaningful signals of perceived urban safety, but they do not read the city from nowhere. Their classifications reflect both the visible environment and a learned standpoint that remains active even when the prompt is described as Neutral. The responsible use of these models therefore depends not on assuming away that standpoint, but on identifying, testing, and validating it against the diverse perspectives of the communities whose environments are being assessed.

\endgroup

\section{Materials and Methods}\label{sec4}

\subsection*{Dataset}
\label{Dataset}
For our analysis, we employed the Place Pulse 2.0 dataset \cite{dubey2016deep}, a crowd-sourced
collection of urban perception data covering 56 cities across six continents. The dataset was created
by asking online volunteers to compare pairs of images and choose which one looks ``safer'', ``livelier'',
``more boring'', ``wealthier'', ``more depressing'', or ``more beautiful'', collecting over
1,169,078 pairwise comparisons for the 110,988 images contained. The images were carefully chosen using
a uniform grid based on latitude and longitude to ensure the representation of each city's districts,
spanning the years 2007 to 2012. User preferences have been converted to image-level ranks of safeness
perception between 0 and 5 using the TrueSkill algorithm \cite{herbrich2006trueskill}. The TrueSkill
algorithm, originally developed by Microsoft Research for ranking players in multiplayer games, offers a
robust framework for converting subjective pairwise comparisons into objective ratings. In urban
perception studies \cite{naik2014streetscore}, this algorithm is applied to street-level imagery where
participants compare images based on attributes like safety, liveliness, and other urban qualities. Each
image is initially modelled as a Gaussian distribution characterised by a mean (an initial estimate of
its perceived quality) and a variance (indicating the uncertainty about that estimate). When two images
are compared, the outcome is used to update both images' distributions through Bayesian inference. This
process adjusts the mean, thereby refining the estimate of an image's quality, and reduces the variance,
reflecting increased confidence in the rating. As more pairwise comparisons are processed, the iterative
updates lead to stable and reliable rankings. Additional details on how votes are converted into scores
can be found in \cite{naik2014streetscore}. In Appendix \ref{Dataset_SI} and Appendix
Table \ref{tab:summary_by_city}, we show additional information on the dataset, such as the spatial
distribution of the images and the distribution of the safeness scores.
 
\textcolor{black}{Because the reliability of a TrueSkill rating depends on how many times an image has been
compared, we adopt a stringent vote-count filter for the ground truth. \cite{gu2025designing}
show, in a dedicated survey-design analysis, that TrueSkill rankings only stabilise beyond roughly 22
pairwise comparisons per image; we therefore retain only images with \emph{at least 22 safety votes}. This
high-reliability criterion yields a final evaluation set of 1{,}530 images spanning all 56 cities,
deliberately trading sample size for label quality. To align with our binary classification task (\emph{Safe} vs. \emph{Unsafe}),
we binarise the continuous TrueSkill scores using the global median threshold
$\tau$. Images with a score $s_i \geq \tau$ are labelled \emph{Safe}, whereas
those with $s_i < \tau$ are labelled \emph{Unsafe}. This rule reproduces the
stored labels of the threshold-22 evaluation set exactly. The resulting set contains 751 \emph{Safe} images (49.1\%) and
779 \emph{Unsafe} images (50.9\%).  A full
per-city breakdown is provided in Appendix~\ref{Dataset_SI}, Table~\ref{tab:summary_by_city}. Further
details on ground-truth generation and thresholding are given in the Evaluation Metrics section.}

\subsection*{Models}

\textcolor{black}{To ensure that our findings reflect the behaviour of contemporary vision--language
models rather than a single checkpoint, we evaluate four recent MLLM families spanning
open and proprietary systems and a broad capability range: LLaVA~1.6
\cite{liu2024visual, liu2024improved}, Gemini~2.5 Flash Lite \cite{comanici2025gemini}{},
Qwen2.5-VL-72B-Instruct \cite{bai2025qwen3}, and GPT-5.1 \cite{singh2025openai}. To isolate the effect of model scale within a single
open model family, we additionally run a size sweep of LLaVA~1.6 at 7B, 13B, and
34B parameters. All models are queried zero-shot with the same image set and the
same prompt templates. Each configuration is run twice to quantify run-to-run
variability.}

\textcolor{black}{GPT-5.1 is accessed through the OpenAI API \footnote{\url{https://openai.com/api/}}, while Gemini~2.5 Flash Lite and
Qwen2.5-VL-72B-Instruct are accessed through OpenRouter \footnote{\url{https://openrouter.ai/}}. The LLaVA models are run
locally using Hugging Face checkpoints with 4-bit bitsandbytes quantization
for computational feasibility. All models are decoded with low-temperature sampling
(temperature~0.1) and constrained to short JSON outputs containing the binary
classification and three keywords.}

\subsection*{Evaluation Metrics}

\subsubsection{Thresholding and Ground Truth}
We normalise the Place Pulse 2.0 TrueSkill scores to obtain a comparable safety perception scale across
images:
\[
  s_i = \frac{\texttt{TrueSkill}_i - \min(\texttt{TrueSkill})}
              {\max(\texttt{TrueSkill}) - \min(\texttt{TrueSkill})}.
\]
We then binarise the normalised score using a threshold \(\tau\) computed over the full set of retained
images (after the vote-count filter described below). Specifically, we set \(\tau\) to the median of
\(\{s_i\}\):
\[
  y_i =
  \begin{cases}
    \text{Safe}, & \text{if } s_i \ge \tau,\\
    \text{Unsafe}, & \text{otherwise}.
  \end{cases}
\]
\textcolor{black}{To ensure that each perception estimate rests on enough evidence, we restrict the
evaluation to images with \emph{at least 22 safety votes} ($n_{\text{votes}} \ge 22$) in the
\emph{Safer} study. This choice
follows Gu et al.~\cite{gu2025designing}, whose survey-design analysis shows that TrueSkill rankings
stabilise only beyond roughly 22 pairwise comparisons per image. The filter yields a high-reliability
evaluation set of 1{,}530 images across the 56 cities (Appendix~\ref{Dataset_SI}).}

\subsubsection{Classification Metrics}

We employ standard classification metrics to evaluate the model's performance based on the number of correctly and incorrectly predicted \emph{Safe} and \emph{Unsafe} labels. Specifically, we define True Positive ($TP$) as the number of images correctly predicted as Safe, and True Negative ($TN$) as the number of images correctly predicted as Unsafe. False Positive ($FP$) is defined as the number of Unsafe images misclassified as Safe, while False Negative ($FN$) is the number of Safe images misclassified as Unsafe.

We then compute Precision, Recall, and the F1-score. Precision, which measures the proportion of correctly predicted Safe images among all images classified as Safe, is given by:

\[
\text{Precision} = \frac{TP}{TP + FP}.
\]

Recall, which captures the proportion of correctly predicted Safe images among all truly Safe images, is computed as:

\[
\text{Recall} = \frac{TP}{TP + FN}.
\]

To balance Precision and Recall, we calculate the F1-score as their harmonic mean:

\[
\text{F1-score} = 2 \cdot \frac{\text{Precision} \times \text{Recall}}{\text{Precision} + \text{Recall}}.
\]

Additionally, for each prompt, we quantify the proportion of images classified as \emph{Unsafe} by the model:

\[
  \texttt{UNSAFE N. (\%)} = \frac{\text{number of Unsafe classified predictions}}{\text{total number of images}} \times 100.
\]

This metric provides insights into how frequently different prompts lead the model to classify urban spaces as Unsafe, offering a comparative measure across the different prompts used.

\subsubsection{\textcolor{black}{Calibration: Prior versus Signal}}
\textcolor{black}{To test whether a model tracks genuine between-city differences in safety or merely
reverts to a fixed prior, we relate each city's model unsafe rate to its ground-truth unsafe base rate.
For each city $c$ we take $x_c$, the fraction of images labelled \emph{Unsafe} in the ground truth, and
$y_c$, the model's unsafe rate, and fit an \emph{image-weighted} least-squares line
$y = a + b\,x$ with weights equal to the number of images per city. We report the Pearson correlation
$r$ (does the model track the ordering?), the slope $b$ with a bootstrap 95\% CI over cities (does it
react proportionally?), and the intercept $a$. A high $r$ with a slope $b<1$ indicates that the model
recovers the relative ordering of cities but \emph{under-reacts}, compressing its predictions toward a
low default rate. We flag the standard caveat that a sub-unit slope is also the signature of
measurement error in $x$ (regression dilution), so we read $b<1$ as under-reaction rather than as proof
of an internal prior.}

\subsubsection{Rank Correlations of City Ordering}
To analyse the consistency of the model's classification of unsafe environments across different
prompts, we assess the correlation between city rankings using Spearman's rank correlation coefficient
\(\rho\). This non-parametric measure evaluates the degree to which two rankings follow the same order,
capturing both the strength and direction of association. Given the ranks of cities under two different
prompts, the Spearman correlation is computed as:
\[
\rho = 1 - \frac{6 \sum d_i^2}{n(n^2 - 1)},
\]
where \(d_i\) represents the difference in ranks for city \(i\) across two rankings, and \(n\) is the
number of cities. \textcolor{black}{We use this to check whether socio-demographic personas change
\emph{which} cities look unsafe or merely \emph{how} unsafe they look: we correlate each persona's
per-city unsafe-rate ranking with the ranking obtained under the Neutral prompt.}

\subsubsection{\textcolor{black}{Statistical Significance of Persona Effects}}
\textcolor{black}{To test whether the gap between two prompts $a$ and $b$ (for instance Female vs.\ Male,
or a persona vs.\ Neutral) is statistically meaningful rather than an artefact of the model's stochastic
generation, we adopt several statistical tests.}
 
\textcolor{black}{\emph{Per-city statistic.} For a prompt $p$ and city $c$, let $\mathrm{UNSAFE}(p,c)$ be
the fraction (\%) of images in city $c$ that the model classifies as \emph{Unsafe} under prompt $p$,
averaged over the $R$ runs:
\[
  \mathrm{UNSAFE}(p,c)
  = \frac{1}{R}\sum_{r=1}^{R}
    \frac{\#\{\text{images of }c\text{ called Unsafe under }p\text{ in run }r\}}
         {\#\{\text{images of }c\}}\times 100 .
\]
The paired difference between the two prompts in city $c$ and its across-city mean are
\[
  \delta_c = \mathrm{UNSAFE}(a,c) - \mathrm{UNSAFE}(b,c),
  \qquad
  \bar{\delta} = \frac{1}{n}\sum_{c=1}^{n}\delta_c ,
\]
so $\bar{\delta}$ (in percentage points) is the effect of interest.}
 
\textcolor{black}{\emph{Bootstrap confidence interval.} We quantify the uncertainty of $\bar{\delta}$ with
a nonparametric bootstrap over cities. For each of $B=10{,}000$ replicates we draw $n$ cities with
replacement, obtaining $\{\delta^{*}_c\}$, and record the replicate mean $\bar{\delta}^{*}$. The $95\%$
confidence interval is the interval between the $2.5$th and $97.5$th percentiles of the $B$ replicate
means.}
 
\textcolor{black}{\emph{Permutation $p$-value.} Under the null hypothesis that $a$ and $b$ are
exchangeable within each city, the sign of each $\delta_c$ is equally likely to be positive or negative.
We therefore draw $P=10{,}000$ sign vectors $s^{(k)}\in\{-1,+1\}^{n}$ uniformly at random, form the
permuted means $m^{(k)} = \frac{1}{n}\sum_{c=1}^{n} s^{(k)}_c\,\delta_c$, and compute the two-sided
$p$-value with the standard $+1$ correction:
\[
  p = \frac{1 + \#\bigl\{k : |m^{(k)}| \ge |\bar{\delta}|\bigr\}}{P + 1}.
\]}
 
\textcolor{black}{\emph{Effect size.} We report Cohen's $d_z$ for paired samples,
\[
  d_z = \frac{\bar{\delta}}{\operatorname{sd}(\delta_c)},
\]
where $\operatorname{sd}(\delta_c)$ is the (unbiased, $n{-}1$ denominator) standard deviation of the
per-city differences; by convention $|d_z|\approx 0.2,\ 0.5,\ 0.8$ denote small, medium and large
effects. A contrast is deemed significant when its $95\%$ bootstrap CI excludes zero and the permutation
$p<0.05$. }

\subsubsection{Comparison with the Neutral Prompt}

To assess how different prompts influence the model’s safety classifications, we compare each prompt’s tendency to label images as \emph{Unsafe} against the \emph{Neutral prompt} baseline. Specifically, we measure the change in the proportion of images classified as \emph{Unsafe} for each city under a given prompt.

Let \(\mathrm{UNSAFE\_mean}(p,c)\) be the average fraction (\%) of images classified as \emph{Unsafe} under prompt \(p\). Likewise, let \(\mathrm{UNSAFE\_mean}(\mathrm{Neutral}, c)\) be that city’s unsafe classification rate under the \emph{Neutral} prompt. We define the city-level difference as follows:
\begin{equation}
  \Delta_{\mathrm{Unsafe}}(p,c)
  \;=\;
  \mathrm{UNSAFE\_mean}(p,c)
  \;-\;
  \mathrm{UNSAFE\_mean}(\mathrm{Neutral}, c).
  \label{eq:delta_unsafe_city}
\end{equation}
A positive \(\Delta_{\mathrm{Unsafe}}(p,c)\) indicates that prompt \(p\) yields a higher unsafe rate than the neutral baseline in city \(c\).

We further compute a single summary value for prompt \(p\) by averaging across all \(N\) cities:
\begin{equation}
  \Delta_{\mathrm{Unsafe}}(p)
  \;=\;
  \frac{1}{N}
  \sum_{c=1}^{N}
  \Delta_{\mathrm{Unsafe}}(p,c),
  \label{eq:delta_unsafe_agg}
\end{equation}

Further, to assess the alignment of socio-demographic prompts \(p\) with the \emph{Neutral} baseline, for each city $c$, we compute the classification agreement between persona $p$
and the Neutral prompt as

\[
A(p,c)=
\frac{1}{|I_c|}
\sum_{i\in I_c}
\mathbb{1}
\left(
cls_p(i)=cls_{\mathrm{Neutral}}(i)
\right),
\]

where $I_c$ is the set of images from city $c$ evaluated under both prompts.
We then obtain the city-macro agreement by averaging equally across the $N$ cities:

\[
A(p)=\frac{1}{N}\sum_{c=1}^{N}A(p,c).
\]

The Neutral output is used as a behavioural reference, not as ground truth.

\subsubsection{Keyword Network Construction and Community Detection}
\label{sec:keyword_network}

\paragraph{Keyword Extraction and Normalization.}
We construct separate keyword co-occurrence networks for images classified as
\emph{Safe} and \emph{Unsafe}. For each such image, we collect the set of descriptive keywords \( \{k_1, k_2, \ldots, k_n\} \). To standardize these keywords, we apply a normalization procedure that removes punctuation, converts text to lowercase, and maps known synonyms (e.g., \emph{“vehicle traffic”} to \emph{“traffic”}) to a single canonical form. We define the normalized version of a raw keyword \( k \) as:

\[
\widehat{k} = \mathcal{N}(k),
\]

where \(\mathcal{N}(\cdot)\) represents the normalization function. This transformation ensures that semantically equivalent terms are treated consistently throughout the analysis.

\paragraph{Co-occurrence Matrix.}
To build the network, we select the top \(N\) most frequent keywords in the dataset of Unsafe images. Denote this reduced set by 
\(\mathcal{K} = \{\kappa_1, \kappa_2, \ldots, \kappa_N\}\).
For each image \(i\), let \(\mathcal{K}_i \subseteq \mathcal{K}\) be the subset of top-\(N\) normalized keywords present in that image. We then construct an \(N \times N\) co-occurrence matrix \(\mathbf{A}\), where
\[
    A_{mn}
    \;=\;
    \sum_{i=1}^{M}
    \mathbf{1}\bigl( \{\kappa_m,\;\kappa_n\} \subseteq \mathcal{K}_i \bigr),
\]
for \(1 \le m,n \le N\). Here, \(\mathbf{1}(\cdot)\) is an indicator function that is 1 if both keywords \(\kappa_m\) and \(\kappa_n\) appear together in image \(i\), and 0 otherwise. Consequently, \(A_{mn}\) measures the number of images where the two keywords co-occur.

\paragraph{Network Representation.}
We interpret \(\mathbf{A}\) as the adjacency matrix of an undirected graph \(G = (V,E)\), in which \(V\) is the set of keywords (nodes) and \(E\) the set of edges. An edge between \(\kappa_m\) and \(\kappa_n\) exists if \(A_{mn} > 0\), with the edge weight \(w_{mn} = A_{mn}\).

\paragraph{Community Detection via Louvain.}
To identify thematic groupings of keywords, we apply the \emph{Louvain} method to detect communities in the network \(G\) \cite{newman2010networks}. The Louvain algorithm optimizes the modularity \(Q\), defined as
\[
  Q
  = \frac{1}{2W}
  \sum_{(m,n) \,\in\, E}
  \Bigl[
      w_{mn} 
      - \frac{d_m \, d_n}{2W}
  \Bigr]
  \delta(\gamma_m, \gamma_n),
\]
where \(d_m = \sum_{n} w_{mn}\) is the weighted degree of node \(m\), \(W = \tfrac12\sum_{m,n} w_{mn}\) is the sum of all edge weights, \(\delta(\gamma_m,\gamma_n)\) is 1 if nodes \(m\) and \(n\) share the same community label, and 0 otherwise.

\paragraph{Degree Centrality.}
Once each community is identified, we compute a node’s \emph{degree centrality} relative to its own community. For a community \(C \subseteq V\) and a node \(v \in C\), define:
\[
  \mathrm{DC}_{C}(v)
  = \frac{\deg_C(v)}{|C| - 1},
\]
where \(\deg_C(v)\) is the degree of \(v\) within \(C\). Nodes with higher \(\mathrm{DC}_{C}(v)\) in a community appear most frequently in co-occurrence with other community members, suggesting they represent key thematic elements.

\subsection*{Prompt}
\label{Prompt_Section}
In this study, we employed a series of carefully structured prompts to test the model's assessment of
urban safety, with the \emph{Neutral prompt} serving as the foundational baseline. This baseline
scenario, shown in Table~\ref{tab:neutral_prompt}, avoids specifying any socio-demographic context and
instructs the model to classify scenes as \emph{Safe} or \emph{Unsafe} solely based on visible
evidence. The required output is a JSON object containing the
classification and \textcolor{black}{three descriptive keywords} that justify the decision.
 
\begin{table}[ht]
\caption{\textcolor{black}{ Neutral prompt used as the baseline in this study. It specifies no
socio-demographic perspective and does not bias the model toward either class in ambiguous scenes.}}
\label{tab:neutral_prompt}
\centering
\begin{tabular}{p{0.95\linewidth}}
\toprule
\textbf{\textcolor{black}{Neutral Prompt }}\\
\midrule
\textcolor{black}{Your task is to classify the area in the image as ``Safe'' or ``Unsafe'' based on
visible indicators of security, safety, and livability from an individual perspective.}\\[2pt]
\textcolor{black}{Use exactly 3 concise keywords that accurately reflect the visible evidence justifying
your classification.}\\[2pt]
\textcolor{black}{Respond ONLY with a valid JSON object with exactly these keys:}\\
\textcolor{black}{\{}\\
\textcolor{black}{\quad ``Classification'': ``Safe'' or ``Unsafe'',}\\
\textcolor{black}{\quad ``Keywords'': [``keyword1'', ``keyword2'', ``keyword3'']}\\
\textcolor{black}{\}}\\
\bottomrule
\end{tabular}
\end{table}
 
Building on this baseline, we introduced a variety of prompts that adopt specific socio-demographic
perspectives. The core structure of all prompts is kept identical to maximise the comparability of the
results: the only variation is the opening sentence, which specifies the perspective the model is
instructed to adopt (e.g.\ \textcolor{black}{``Assume the role of a person who identifies as \dots''} for
race/ethnicity, ``Assume the role of a \{male, female\} person'' for gender, and ``Assume the role of a
\{young, middle-aged, elderly\} person'' for age). \textcolor{black}{The main text focuses on three
identity axes, comprising 11 persona prompts in total.} The full set of prompts is provided in
Appendix~\ref{Prompts_Comparison_SI}. The persona axes are:
\begin{itemize}
    \item \textbf{\textcolor{black}{Race/ethnicity prompts}}: \textcolor{black}{six categories --- \emph{White/Caucasian}, \emph{Asian/Pacific Islander},
    \emph{Hispanic/Latino}, \emph{Middle Eastern/North African}, \emph{Black/African American}, and
    \emph{Native American/Alaska Native} --- letting us examine how racialised identity shapes the
    model's simulated safety perception.}
    \item \textbf{Gender prompts}: comprising \emph{Male} and \emph{Female}, these prompts let the model
    consider how safety perceptions may vary based on gender-specific experiences.
    \item \textbf{Age prompts}: divided into \emph{Young}, \emph{Middle-aged}, and \emph{Elderly}, these
    prompts adjust the model to consider safety from the viewpoint of different life stages.
\end{itemize}
 By keeping the output requirements consistent and the semantic structure uniform across all
prompts, we can compare variations in the model's perception of urban safety with precision.

\section{Abbreviations}
For clarity, we summarise below the list of abbreviations used throughout this paper:
\begin{itemize}
\item \textbf{AI}: Artificial Intelligence
\item \textbf{CI}: Confidence Interval
\item \textbf{FN}: False Negative
\item \textbf{FP}: False Positive
\item \textbf{MLLM}: Multimodal Large Language Model
\item \textbf{LVM}: Large Vision Model
\item \textbf{SCM}: Stereotype Content Model
\item \textbf{SVI}: Street View Imagery
\item \textbf{TN}: True Negative
\item \textbf{TP}: True Positive
\end{itemize}

\section*{Declarations}

\subsection*{Ethics approval and consent to participate}
Not applicable.

\subsection*{Consent for publication}
All authors have read and approved the final manuscript and consent to its publication.
\subsection*{Availability of data and material}
The dataset analysed in this work is publicly available via \url{https://centerforcollectivelearning.org/urbanperception}.

\subsection*{Competing interests}
The authors declare that they have no competing interests.

\subsection*{Funding}
B.L. has been supported by PRECRISIS (PRotECting public spaces thRough Integrated Smarter Innovative Security), funded by the European Union Internal Security Fund (ISFP-2022-TFI-AG-PROTECT-02-101100539). B.L. and M.L. have been supported by the PNRR ICSC National Research Centre for High Performance Computing, Big Data and Quantum Computing (CN00000013), under the NRRP MUR program funded by the NextGenerationEU. B.L. also acknowledges the support of the PNRR project FAIR - Future AI Research (PE00000013), under the NRRP MUR program funded by the NextGenerationEU. B.L. and M.L. also acknowledge the support of the European Union’s Horizon Europe research and innovation programme under grant agreement No. 101120237 (ELIAS). B.L. and M.L. were
supported by the Horizon Europe Programme No. 101120763 (TANGO).

C.B. acknowledges ISCRA for awarding the project IsCd8 access to
the LEONARDO supercomputer, owned by the EuroHPC
Joint Undertaking, hosted by CINECA (Italy).

\subsection*{Authors' contributions}
C.B. designed and developed the experiments. M.L. designed the study and coordinated the writing of the paper. C.B., B.L., and M.L. contributed to the interpretation of the results and to writing the manuscript. All authors read and approved the final manuscript.

\subsection*{Acknowledgements}
Not applicable.

+

\newpage
\clearpage
\begin{appendices}




\clearpage

\section{Keyword Network Analysis}
\label{keyword_network_SI}

\begin{figure}[p]
  \centering
  \includegraphics[width=\textwidth,height=0.90\textheight,keepaspectratio]{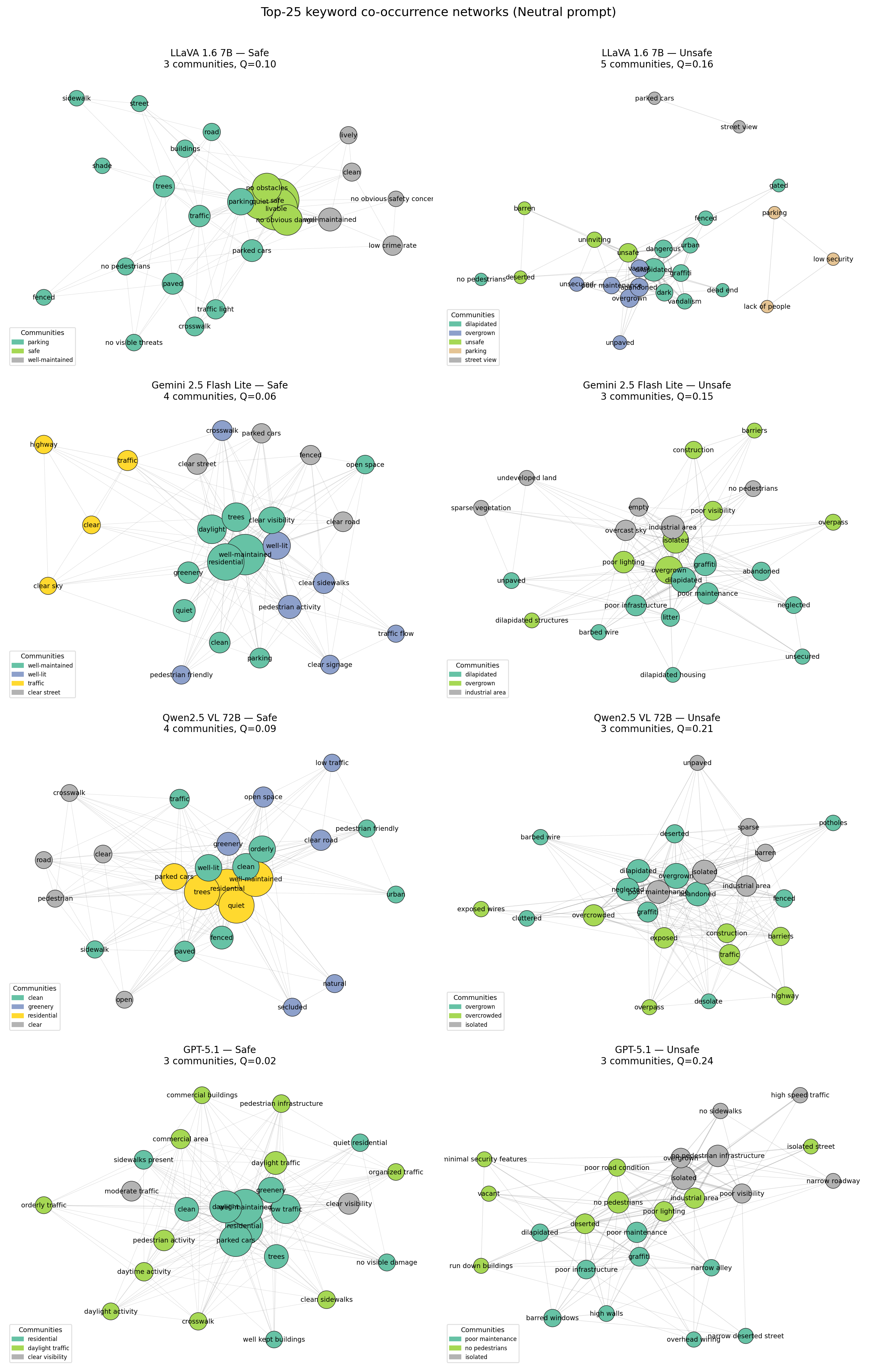}
  \caption{\footnotesize\textcolor{black}{\textbf{Keyword co-occurrence networks across the four MLLMs.}
  Each row corresponds to one model (top to bottom: LLaVA~1.6~7B, Gemini~2.5~Flash~Lite,
  Qwen2.5~VL~72B, GPT-5.1) and shows, side by side, the networks of the 25 most frequent
  keywords for images classified as \emph{Safe} (left) and \emph{Unsafe} (right) under the
  neutral prompt. Nodes are keywords (after canonical-vocabulary normalisation); node size is
  proportional to degree centrality and edge thickness to co-occurrence frequency. Node colours
  denote the communities recovered by the Louvain algorithm, and the number of communities and
  the modularity $Q$ are reported above each panel. The per-community terms and centralities are
  listed in Tables~\ref{tab:safe_llava}--\ref{tab:unsafe_gpt51}.}}
  \label{fig:network_keywords_SI}
\end{figure}

To understand the underlying reasons driving safety classifications, we analyzed the most frequently occurring keywords in the responses. \textcolor{black}{Free-text keywords were first normalised (lower-cased, separators unified, punctuation removed) and mapped to a canonical vocabulary that merges surface variants of the same concept (e.g., ``residential street'' and ``residential buildings'' into ``residential'').} Using network analysis and community detection techniques, we identified clusters of related terms associated with \textit{Safe} and \textit{Unsafe} classifications. \textcolor{black}{We repeated this analysis for all four evaluated models --- LLaVA~1.6~7B, Gemini~2.5~Flash~Lite, Qwen2.5~VL~72B and GPT-5.1 --- building, for each model and class, a co-occurrence network from the 25 most frequent keywords and partitioning it with the Louvain algorithm.} Figure~\ref{fig:network_keywords_SI} shows the resulting networks. Sections~\ref{sec:kw_llava}--\ref{sec:kw_gpt51} report, model by model, the identified communities, sorted by degree centrality computed within each community, which quantifies the relative importance of each term. \textcolor{black}{The reproducibility of these networks is assessed in Section~\ref{sec:kw_robustness}.}

\clearpage

\subsection{LLaVA~1.6~7B}
\label{sec:kw_llava}

Tables~\ref{tab:safe_llava} and~\ref{tab:unsafe_llava} report the communities recovered for the \emph{Safe} and \emph{Unsafe} networks of LLaVA~1.6~7B.

\begin{table}[!ht]
  \centering
  \small
  \caption{LLaVA~1.6~7B --- \emph{Safe} network communities, sorted by within-community degree centrality.}
  \label{tab:safe_llava}
  \begin{tabular}{@{}lr@{}}
    \toprule
    \textbf{Term} & \textbf{Degree centrality} \\
    \midrule
    \multicolumn{2}{@{}l}{\textbf{Community 0}} \\
    trees                       & 0.7143 \\
    parking                     & 0.6429 \\
    traffic                     & 0.6429 \\
    paved                       & 0.5714 \\
    parked cars                 & 0.5000 \\
    no pedestrians              & 0.4286 \\
    crosswalk                   & 0.3571 \\
    street                      & 0.3571 \\
    traffic light               & 0.3571 \\
    no visible threats          & 0.2857 \\
    road                        & 0.2857 \\
    buildings                   & 0.2143 \\
    fenced                      & 0.2143 \\
    sidewalk                    & 0.2143 \\
    shade                       & 0.0714 \\
    \addlinespace
    \multicolumn{2}{@{}l}{\textbf{Community 1}} \\
    livable                     & 1.0000 \\
    safe                        & 1.0000 \\
    no obstacles                & 0.5000 \\
    no obvious danger           & 0.5000 \\
    quiet                       & 0.5000 \\
    \addlinespace
    \multicolumn{2}{@{}l}{\textbf{Community 2}} \\
    low crime rate              & 0.7500 \\
    well-maintained             & 0.7500 \\
    clean                       & 0.5000 \\
    lively                      & 0.5000 \\
    no obvious safety concerns  & 0.5000 \\
    \bottomrule
  \end{tabular}
\end{table}

\begin{table}[!ht]
  \centering
  \small
  \caption{LLaVA~1.6~7B --- \emph{Unsafe} network communities, sorted by within-community degree centrality.}
  \label{tab:unsafe_llava}
  \begin{tabular}{@{}lr@{}}
    \toprule
    \textbf{Term} & \textbf{Degree centrality} \\
    \midrule
    \multicolumn{2}{@{}l}{\textbf{Community 0}} \\
    dilapidated       & 1.0000 \\
    dangerous         & 0.5556 \\
    dark              & 0.5556 \\
    graffiti          & 0.5556 \\
    fenced            & 0.4444 \\
    vandalism         & 0.4444 \\
    urban             & 0.3333 \\
    dead end          & 0.2222 \\
    gated             & 0.2222 \\
    no pedestrians    & 0.1111 \\
    \addlinespace
    \multicolumn{2}{@{}l}{\textbf{Community 1}} \\
    abandoned         & 1.0000 \\
    overgrown         & 1.0000 \\
    poor maintenance  & 0.8000 \\
    unsecured         & 0.6000 \\
    vacant            & 0.6000 \\
    unpaved           & 0.4000 \\
    \addlinespace
    \multicolumn{2}{@{}l}{\textbf{Community 2}} \\
    barren            & 1.0000 \\
    uninviting        & 1.0000 \\
    deserted          & 0.6667 \\
    unsafe            & 0.6667 \\
    \addlinespace
    \multicolumn{2}{@{}l}{\textbf{Community 3}} \\
    lack of people    & 1.0000 \\
    low security      & 1.0000 \\
    parking           & 1.0000 \\
    \addlinespace
    \multicolumn{2}{@{}l}{\textbf{Community 4}} \\
    parked cars       & 1.0000 \\
    street view       & 1.0000 \\
    \bottomrule
  \end{tabular}
\end{table}

\clearpage

\subsection{\textcolor{black}{Gemini~2.5~Flash~Lite}}
\label{sec:kw_gemini}

\textcolor{black}{Tables~\ref{tab:safe_gemini} and~\ref{tab:unsafe_gemini} report the communities recovered for the \emph{Safe} and \emph{Unsafe} networks of Gemini~2.5~Flash~Lite.}

\begin{table}[!ht]
  \centering
  \small
  \caption{\textcolor{black}{Gemini~2.5~Flash~Lite --- \emph{Safe} network communities, sorted by within-community degree centrality.}}
  \label{tab:safe_gemini}
  \begin{tabular}{@{}lr@{}}
    \toprule
    \textbf{Term} & \textbf{Degree centrality} \\
    \midrule
    \multicolumn{2}{@{}l}{\textbf{Community 0}} \\
    well-maintained      & 1.0000 \\
    residential          & 0.8889 \\
    clear visibility     & 0.7778 \\
    daylight             & 0.7778 \\
    trees                & 0.7778 \\
    pedestrian friendly  & 0.6667 \\
    greenery             & 0.5556 \\
    open space           & 0.5556 \\
    clean                & 0.3333 \\
    quiet                & 0.3333 \\
    \addlinespace
    \multicolumn{2}{@{}l}{\textbf{Community 1}} \\
    pedestrian activity  & 0.8333 \\
    well-lit             & 0.8333 \\
    clear sidewalks      & 0.5000 \\
    clear signage        & 0.5000 \\
    parking              & 0.5000 \\
    traffic flow         & 0.5000 \\
    crosswalk            & 0.3333 \\
    \addlinespace
    \multicolumn{2}{@{}l}{\textbf{Community 2}} \\
    clear street         & 0.6667 \\
    fenced               & 0.6667 \\
    clear road           & 0.3333 \\
    parked cars          & 0.3333 \\
    \addlinespace
    \multicolumn{2}{@{}l}{\textbf{Community 3}} \\
    highway              & 1.0000 \\
    traffic              & 1.0000 \\
    clear                & 0.6667 \\
    clear sky            & 0.6667 \\
    \bottomrule
  \end{tabular}
\end{table}

\begin{table}[!ht]
  \centering
  \small
  \caption{\textcolor{black}{Gemini~2.5~Flash~Lite --- \emph{Unsafe} network communities, sorted by within-community degree centrality.}}
  \label{tab:unsafe_gemini}
  \begin{tabular}{@{}lr@{}}
    \toprule
    \textbf{Term} & \textbf{Degree centrality} \\
    \midrule
    \multicolumn{2}{@{}l}{\textbf{Community 0}} \\
    dilapidated            & 0.9000 \\
    graffiti               & 0.8000 \\
    poor infrastructure    & 0.8000 \\
    poor maintenance       & 0.8000 \\
    litter                 & 0.7000 \\
    abandoned              & 0.6000 \\
    neglected              & 0.5000 \\
    unsecured              & 0.5000 \\
    unpaved                & 0.4000 \\
    barbed wire            & 0.3000 \\
    dilapidated housing    & 0.3000 \\
    \addlinespace
    \multicolumn{2}{@{}l}{\textbf{Community 1}} \\
    isolated               & 1.0000 \\
    overgrown              & 0.7143 \\
    construction           & 0.5714 \\
    poor lighting          & 0.5714 \\
    poor visibility        & 0.5714 \\
    barriers               & 0.4286 \\
    dilapidated structures & 0.4286 \\
    overpass               & 0.2857 \\
    \addlinespace
    \multicolumn{2}{@{}l}{\textbf{Community 2}} \\
    industrial area        & 1.0000 \\
    overcast sky           & 1.0000 \\
    sparse vegetation      & 0.6000 \\
    undeveloped land       & 0.6000 \\
    empty                  & 0.4000 \\
    no pedestrians         & 0.4000 \\
    \bottomrule
  \end{tabular}
\end{table}

\clearpage

\subsection{\textcolor{black}{Qwen2.5~VL~72B}}
\label{sec:kw_qwen}

\textcolor{black}{Tables~\ref{tab:safe_qwen} and~\ref{tab:unsafe_qwen} report the communities recovered for the \emph{Safe} and \emph{Unsafe} networks of Qwen2.5~VL~72B.}

\begin{table}[!ht]
  \centering
  \small
  \caption{\textcolor{black}{Qwen2.5~VL~72B --- \emph{Safe} network communities, sorted by within-community degree centrality.}}
  \label{tab:safe_qwen}
  \begin{tabular}{@{}lr@{}}
    \toprule
    \textbf{Term} & \textbf{Degree centrality} \\
    \midrule
    \multicolumn{2}{@{}l}{\textbf{Community 0}} \\
    well-lit             & 1.0000 \\
    orderly              & 0.8750 \\
    clean                & 0.7500 \\
    paved                & 0.6250 \\
    fenced               & 0.5000 \\
    pedestrian friendly  & 0.5000 \\
    urban                & 0.5000 \\
    sidewalk             & 0.3750 \\
    traffic              & 0.3750 \\
    \addlinespace
    \multicolumn{2}{@{}l}{\textbf{Community 1}} \\
    greenery             & 1.0000 \\
    open space           & 0.8000 \\
    clear road           & 0.6000 \\
    low traffic          & 0.6000 \\
    secluded             & 0.6000 \\
    natural              & 0.4000 \\
    \addlinespace
    \multicolumn{2}{@{}l}{\textbf{Community 2}} \\
    parked cars          & 1.0000 \\
    quiet                & 1.0000 \\
    residential          & 1.0000 \\
    trees                & 1.0000 \\
    well-maintained      & 1.0000 \\
    \addlinespace
    \multicolumn{2}{@{}l}{\textbf{Community 3}} \\
    road                 & 0.7500 \\
    clear                & 0.5000 \\
    open                 & 0.5000 \\
    pedestrian           & 0.5000 \\
    crosswalk            & 0.2500 \\
    \bottomrule
  \end{tabular}
\end{table}

\begin{table}[!ht]
  \centering
  \small
  \caption{\textcolor{black}{Qwen2.5~VL~72B --- \emph{Unsafe} network communities, sorted by within-community degree centrality.}}
  \label{tab:unsafe_qwen}
  \begin{tabular}{@{}lr@{}}
    \toprule
    \textbf{Term} & \textbf{Degree centrality} \\
    \midrule
    \multicolumn{2}{@{}l}{\textbf{Community 0}} \\
    abandoned         & 1.0000 \\
    overgrown         & 1.0000 \\
    dilapidated       & 0.8000 \\
    graffiti          & 0.8000 \\
    neglected         & 0.8000 \\
    barbed wire       & 0.7000 \\
    deserted          & 0.7000 \\
    fenced            & 0.7000 \\
    desolate          & 0.5000 \\
    cluttered         & 0.4000 \\
    potholes          & 0.4000 \\
    \addlinespace
    \multicolumn{2}{@{}l}{\textbf{Community 1}} \\
    traffic           & 1.0000 \\
    exposed           & 0.8571 \\
    barriers          & 0.7143 \\
    construction      & 0.7143 \\
    overcrowded       & 0.7143 \\
    highway           & 0.5714 \\
    exposed wires     & 0.4286 \\
    overpass          & 0.4286 \\
    \addlinespace
    \multicolumn{2}{@{}l}{\textbf{Community 2}} \\
    industrial area   & 1.0000 \\
    isolated          & 1.0000 \\
    barren            & 0.8000 \\
    poor maintenance  & 0.8000 \\
    sparse            & 0.8000 \\
    unpaved           & 0.8000 \\
    \bottomrule
  \end{tabular}
\end{table}

\clearpage

\subsection{\textcolor{black}{GPT-5.1}}
\label{sec:kw_gpt51}

\textcolor{black}{Tables~\ref{tab:safe_gpt51} and~\ref{tab:unsafe_gpt51} report the communities recovered for the \emph{Safe} and \emph{Unsafe} networks of GPT-5.1.}

\begin{table}[!ht]
  \centering
  \small
  \caption{\textcolor{black}{GPT-5.1 --- \emph{Safe} network communities, sorted by within-community degree centrality.}}
  \label{tab:safe_gpt51}
  \begin{tabular}{@{}lr@{}}
    \toprule
    \textbf{Term} & \textbf{Degree centrality} \\
    \midrule
    \multicolumn{2}{@{}l}{\textbf{Community 0}} \\
    daylight                  & 1.0000 \\
    parked cars               & 1.0000 \\
    residential               & 0.9091 \\
    well-maintained           & 0.9091 \\
    low traffic               & 0.8182 \\
    greenery                  & 0.7273 \\
    trees                     & 0.7273 \\
    quiet residential         & 0.6364 \\
    no visible damage         & 0.5455 \\
    sidewalks present         & 0.5455 \\
    clean sidewalks           & 0.4545 \\
    well kept buildings       & 0.4545 \\
    \addlinespace
    \multicolumn{2}{@{}l}{\textbf{Community 1}} \\
    clean                     & 0.9000 \\
    commercial area           & 0.7000 \\
    daytime activity          & 0.7000 \\
    organized traffic         & 0.7000 \\
    crosswalk                 & 0.6000 \\
    orderly traffic           & 0.6000 \\
    pedestrian activity       & 0.6000 \\
    commercial buildings      & 0.5000 \\
    daylight activity         & 0.5000 \\
    daylight traffic          & 0.5000 \\
    pedestrian infrastructure & 0.5000 \\
    \addlinespace
    \multicolumn{2}{@{}l}{\textbf{Community 2}} \\
    clear visibility          & 1.0000 \\
    moderate traffic          & 1.0000 \\
    \bottomrule
  \end{tabular}
\end{table}

\begin{table}[!ht]
  \centering
  \small
  \caption{\textcolor{black}{GPT-5.1 --- \emph{Unsafe} network communities, sorted by within-community degree centrality.}}
  \label{tab:unsafe_gpt51}
  \begin{tabular}{@{}lr@{}}
    \toprule
    \textbf{Term} & \textbf{Degree centrality} \\
    \midrule
    \multicolumn{2}{@{}l}{\textbf{Community 0}} \\
    graffiti                     & 0.9286 \\
    deserted                     & 0.7857 \\
    poor infrastructure          & 0.7143 \\
    poor maintenance             & 0.7143 \\
    dilapidated                  & 0.6429 \\
    high walls                   & 0.6429 \\
    no pedestrians               & 0.6429 \\
    narrow alley                 & 0.5714 \\
    poor lighting                & 0.5714 \\
    barred windows               & 0.5000 \\
    overhead wiring              & 0.4286 \\
    run down buildings           & 0.4286 \\
    minimal security features    & 0.3571 \\
    vacant                       & 0.3571 \\
    narrow deserted street       & 0.2857 \\
    \addlinespace
    \multicolumn{2}{@{}l}{\textbf{Community 1}} \\
    isolated                     & 0.8333 \\
    no pedestrian infrastructure & 0.8333 \\
    overgrown                    & 0.8333 \\
    poor visibility              & 0.8333 \\
    no sidewalks                 & 0.6667 \\
    high speed traffic           & 0.5000 \\
    narrow roadway               & 0.5000 \\
    \addlinespace
    \multicolumn{2}{@{}l}{\textbf{Community 2}} \\
    industrial area              & 1.0000 \\
    isolated street              & 0.5000 \\
    poor road condition          & 0.5000 \\
    \bottomrule
  \end{tabular}
\end{table}

\clearpage

\subsection{\textcolor{black}{Robustness Analysis}}
\label{sec:kw_robustness}

\textcolor{black}{Because a co-occurrence matrix can be dominated by a few high-frequency terms, we assessed the reproducibility of the networks. Table~\ref{tab:kw_freq_stab} reports the run-to-run stability of the keyword frequencies at several cutoffs; Table~\ref{tab:kw_net_stab} reports the stability of the network edge weights and of the recovered community partition (adjusted Rand index, ARI); and Table~\ref{tab:kw_cutoff} reports the sensitivity of the community partition to the choice of keyword cutoff. The \emph{safe} network is highly reproducible (edge $\rho=0.90$, ARI $=1.00$, cutoff-invariant), whereas the \emph{unsafe} network, built on far fewer judgements, is noisier: its keyword frequencies remain strongly correlated ($r=0.98$) but its exact top-25 membership and community partition are only moderately stable (ARI $\approx0.57$). We therefore interpret the unsafe communities as indicative thematic groupings rather than exact partitions.}

\begin{table}[!ht]
  \centering
  \small
  \caption{\textcolor{black}{Run-to-run stability of the top-$N$ keyword frequencies (LLaVA~1.6~7B, two neutral runs). Overlap and Jaccard index of the top-$N$ sets, and Spearman/Pearson correlation of the keyword counts.}}
  \label{tab:kw_freq_stab}
  \begin{tabular}{@{}llrrrr@{}}
    \toprule
    \textbf{Class} & \textbf{top-$N$} & \textbf{Overlap} & \textbf{Jaccard} & \textbf{Spearman} & \textbf{Pearson} \\
    \midrule
    Safe   & 10 & 9/10  & 0.818 & 0.916 & 0.996 \\
    Safe   & 20 & 19/20 & 0.905 & 0.938 & 0.996 \\
    Safe   & 25 & 20/25 & 0.667 & 0.889 & 0.997 \\
    Safe   & 30 & 27/30 & 0.818 & 0.872 & 0.997 \\
    \addlinespace
    Unsafe & 10 & 9/10  & 0.818 & 0.830 & 0.963 \\
    Unsafe & 20 & 16/20 & 0.667 & 0.881 & 0.973 \\
    Unsafe & 25 & 16/25 & 0.471 & 0.748 & 0.975 \\
    Unsafe & 30 & 16/30 & 0.364 & 0.646 & 0.976 \\
    \bottomrule
  \end{tabular}
\end{table}

\begin{table}[!ht]
  \centering
  \small
  \caption{\textcolor{black}{Run-to-run stability of the top-25 networks (LLaVA~1.6~7B). Edge-weight correlation over shared nodes and community agreement (ARI).}}
  \label{tab:kw_net_stab}
  \begin{tabular}{@{}lrrrr@{}}
    \toprule
    \textbf{Class} & \textbf{Shared nodes} & \textbf{Edge Spearman} & \textbf{Edge Pearson} & \textbf{Community ARI} \\
    \midrule
    Safe   & 20 & 0.903 & 0.992 & 1.000 \\
    Unsafe & 16 & 0.754 & 0.837 & 0.573 \\
    \bottomrule
  \end{tabular}
\end{table}

\begin{table}[t]
  \centering
  \caption{\textcolor{black}{Sensitivity of the community partition to the keyword cutoff, measured by the ARI between partitions obtained at different top-$N$.}}
  \label{tab:kw_cutoff}
  \begin{tabular}{@{}llr@{}}
    \toprule
    \textbf{Class} & \textbf{Comparison} & \textbf{ARI} \\
    \midrule
    Safe   & top-20 vs.\ top-25 & 1.000 \\
    Safe   & top-25 vs.\ top-30 & 1.000 \\
    Safe   & top-20 vs.\ top-30 & 1.000 \\
    \addlinespace
    Unsafe & top-20 vs.\ top-25 & 0.594 \\
    Unsafe & top-25 vs.\ top-30 & 0.579 \\
    Unsafe & top-20 vs.\ top-30 & 0.332 \\
    \bottomrule
  \end{tabular}
\end{table}

\clearpage

\subsection{Neutral-Prompt Performance in Detail}
\label{Neutral_Prompt_SI}

\textcolor{black}{This appendix expands on the Neutral-prompt results reported in
Section~\ref{sec:neutral_results}. We first define the reported metrics and give the full per-city
breakdown, then analyse where the four models agree and disagree, and finally verify that the results are
robust to the exclusion of cities with few evaluated images. All figures are averaged over the two runs
of each model.}

\textcolor{black}{Throughout, performance is summarised by the \emph{Safe}-class F1 score computed
\emph{within each city} and then \emph{macro-averaged across the 56 cities}, so that every city
contributes equally regardless of how many images it contains. Precision and recall are likewise defined
for the \emph{Safe} class; the reported ``\% Unsafe'' is the share of a model's predictions that fall in
the \emph{Unsafe} class. Cities are assigned to the Global North or Global South following the
classification used throughout the paper (45 North, 11 South).}

\subsubsection{Per-City Metrics}

\textcolor{black}{Table~\ref{app:per_city_all_models} reports, for every city, the number of evaluated images, and the city-level \emph{Safe}-F1 of each model. The city-macro
averages of these columns recover the headline numbers of Table~\ref{tab:neutral_perf} (65.4\%, 69.4\%,
68.9\% and 68.8\% for LLaVA, Gemini, Qwen and GPT-5.1, respectively). The table makes the geographic
pattern explicit: the highest scores occur in affluent, highly-rated Global-North cities such as
Amsterdam, Copenhagen, Sydney and Singapore, while the lowest occur in Global-South cities with low
ground-truth \emph{Safe} rates, most extremely Mexico City, whose images are labelled almost entirely
\emph{Unsafe}. Because per-city \emph{Safe}-F1 is bounded by the availability of truly-safe scenes, it
falls sharply wherever the ground-truth \emph{Safe} rate is low, which is the mechanism discussed in the
main text.}

\begingroup
\scriptsize
\setlength{\tabcolsep}{3pt}
\begin{longtable}{@{}lr rrr rrr rrr rrr@{}}
\caption{Per-city \emph{Safe}-class precision (P), recall (R) and F1 (\%) for all four models,
averaged over two runs and ordered by number of evaluated images $n$. Per-city values are rounded
to the nearest percentage point; the bold row is the city-macro mean.}
\label{app:per_city_all_models}\\
\toprule
& & \multicolumn{3}{c}{\textbf{LLaVA~1.6}} & \multicolumn{3}{c}{\textbf{Gemini}} & \multicolumn{3}{c}{\textbf{Qwen}} & \multicolumn{3}{c}{\textbf{GPT-5.1}} \\
\cmidrule(lr){3-5}\cmidrule(lr){6-8}\cmidrule(lr){9-11}\cmidrule(lr){12-14}
\textbf{City} & \textbf{$n$} & P & R & F1 & P & R & F1 & P & R & F1 & P & R & F1 \\
\midrule
\endfirsthead
\multicolumn{14}{c}{\tablename\ \thetable\ -- \textit{continued from previous page}}\\
\toprule
& & \multicolumn{3}{c}{\textbf{LLaVA~1.6}} & \multicolumn{3}{c}{\textbf{Gemini}} & \multicolumn{3}{c}{\textbf{Qwen}} & \multicolumn{3}{c}{\textbf{GPT-5.1}} \\
\cmidrule(lr){3-5}\cmidrule(lr){6-8}\cmidrule(lr){9-11}\cmidrule(lr){12-14}
\textbf{City} & \textbf{$n$} & P & R & F1 & P & R & F1 & P & R & F1 & P & R & F1 \\
\midrule
\endhead
\midrule \multicolumn{14}{r}{\textit{continued on next page}}\\ \endfoot
\endlastfoot
NewYork & 60 & 56 & 100 & 72 & 65 & 94 & 77 & 66 & 97 & 79 & 65 & 97 & 78 \\
Tokyo & 57 & 37 & 100 & 54 & 33 & 100 & 49 & 34 & 100 & 51 & 33 & 100 & 50 \\
Toronto & 57 & 62 & 100 & 77 & 63 & 94 & 75 & 60 & 95 & 74 & 63 & 100 & 77 \\
Atlanta & 55 & 72 & 100 & 84 & 76 & 92 & 83 & 75 & 100 & 85 & 74 & 97 & 84 \\
Santiago & 55 & 32 & 100 & 49 & 42 & 100 & 59 & 39 & 93 & 55 & 41 & 100 & 58 \\
Chicago & 53 & 68 & 100 & 81 & 77 & 89 & 82 & 74 & 97 & 84 & 72 & 97 & 83 \\
Warsaw & 47 & 49 & 100 & 66 & 54 & 91 & 68 & 52 & 96 & 68 & 53 & 100 & 70 \\
Berlin & 46 & 65 & 98 & 78 & 64 & 89 & 75 & 65 & 93 & 76 & 63 & 93 & 75 \\
SaoPaulo & 46 & 31 & 100 & 47 & 37 & 95 & 53 & 30 & 82 & 44 & 35 & 82 & 49 \\
Houston & 45 & 52 & 100 & 68 & 63 & 85 & 72 & 55 & 85 & 67 & 58 & 92 & 71 \\
Sydney & 44 & 84 & 100 & 91 & 85 & 95 & 90 & 88 & 97 & 92 & 86 & 97 & 91 \\
RioDeJaneiro & 43 & 30 & 100 & 46 & 50 & 88 & 64 & 44 & 88 & 58 & 41 & 88 & 56 \\
Philadelphia & 42 & 49 & 100 & 66 & 54 & 55 & 55 & 64 & 95 & 77 & 60 & 79 & 68 \\
Paris & 39 & 46 & 100 & 63 & 59 & 100 & 74 & 59 & 94 & 73 & 53 & 94 & 68 \\
CapeTown & 39 & 41 & 100 & 58 & 54 & 93 & 68 & 56 & 100 & 71 & 52 & 93 & 67 \\
Moscow & 39 & 51 & 97 & 67 & 59 & 84 & 70 & 58 & 84 & 69 & 57 & 89 & 69 \\
Denver & 37 & 68 & 100 & 81 & 74 & 87 & 80 & 77 & 100 & 87 & 76 & 98 & 86 \\
Singapore & 37 & 81 & 100 & 90 & 83 & 100 & 91 & 83 & 100 & 91 & 81 & 100 & 90 \\
Melbourne & 37 & 76 & 100 & 86 & 77 & 96 & 86 & 76 & 100 & 86 & 77 & 96 & 86 \\
Montreal & 32 & 47 & 100 & 64 & 57 & 93 & 71 & 56 & 93 & 70 & 58 & 100 & 73 \\
Lisbon & 29 & 31 & 100 & 47 & 35 & 100 & 52 & 29 & 86 & 43 & 23 & 71 & 34 \\
Madrid & 28 & 63 & 100 & 77 & 74 & 100 & 85 & 72 & 100 & 84 & 69 & 100 & 82 \\
London & 28 & 67 & 100 & 80 & 78 & 100 & 88 & 69 & 100 & 82 & 69 & 100 & 82 \\
Johannesburg & 27 & 38 & 100 & 56 & 43 & 90 & 58 & 42 & 100 & 59 & 45 & 100 & 62 \\
BeloHorizonte & 27 & 11 & 100 & 21 & 22 & 100 & 36 & 29 & 100 & 44 & 25 & 100 & 40 \\
Milan & 26 & 58 & 100 & 73 & 68 & 100 & 81 & 58 & 100 & 74 & 58 & 100 & 74 \\
Rome & 26 & 58 & 100 & 73 & 68 & 87 & 76 & 61 & 100 & 76 & 66 & 83 & 74 \\
Bucharest & 25 & 38 & 100 & 55 & 41 & 88 & 56 & 47 & 88 & 61 & 41 & 88 & 56 \\
Prague & 25 & 50 & 100 & 67 & 57 & 100 & 73 & 50 & 100 & 67 & 51 & 100 & 68 \\
Bangkok & 25 & 21 & 90 & 35 & 24 & 60 & 34 & 25 & 60 & 35 & 31 & 90 & 46 \\
SanFrancisco & 25 & 48 & 100 & 65 & 51 & 100 & 68 & 52 & 100 & 69 & 52 & 100 & 69 \\
Portland & 22 & 59 & 100 & 74 & 71 & 92 & 80 & 67 & 100 & 80 & 68 & 100 & 81 \\
Stockholm & 20 & 80 & 100 & 89 & 80 & 100 & 89 & 79 & 94 & 86 & 80 & 100 & 89 \\
Dublin & 20 & 72 & 100 & 84 & 78 & 100 & 88 & 70 & 100 & 82 & 70 & 100 & 82 \\
Boston & 20 & 70 & 100 & 82 & 78 & 100 & 88 & 80 & 100 & 89 & 78 & 100 & 88 \\
Seattle & 19 & 47 & 100 & 64 & 44 & 100 & 62 & 47 & 100 & 64 & 47 & 100 & 64 \\
Guadalajara & 19 & 32 & 100 & 48 & 48 & 100 & 65 & 48 & 100 & 65 & 41 & 100 & 59 \\
Munich & 19 & 74 & 100 & 85 & 80 & 100 & 89 & 78 & 100 & 88 & 78 & 100 & 88 \\
Barcelona & 15 & 40 & 100 & 57 & 50 & 83 & 62 & 45 & 83 & 59 & 50 & 100 & 67 \\
WashingtonDC & 14 & 71 & 100 & 83 & 77 & 100 & 87 & 77 & 100 & 87 & 71 & 100 & 83 \\
Taipei & 14 & 50 & 100 & 67 & 71 & 83 & 77 & 77 & 83 & 80 & 45 & 83 & 59 \\
HongKong & 13 & 33 & 100 & 50 & 40 & 100 & 57 & 40 & 100 & 57 & 35 & 100 & 52 \\
Minneapolis & 12 & 75 & 100 & 86 & 78 & 78 & 78 & 80 & 89 & 84 & 82 & 100 & 90 \\
Helsinki & 12 & 58 & 100 & 74 & 58 & 100 & 74 & 62 & 93 & 74 & 58 & 100 & 74 \\
LosAngeles & 12 & 61 & 100 & 76 & 75 & 86 & 80 & 78 & 100 & 88 & 78 & 100 & 88 \\
Valparaiso & 11 & 44 & 100 & 62 & 80 & 100 & 89 & 73 & 100 & 84 & 67 & 100 & 80 \\
Kyoto & 11 & 18 & 100 & 31 & 18 & 100 & 31 & 18 & 100 & 31 & 18 & 100 & 31 \\
Zagreb & 10 & 50 & 100 & 67 & 53 & 100 & 69 & 50 & 100 & 67 & 50 & 100 & 67 \\
Kiev & 9 & 33 & 100 & 50 & 60 & 100 & 75 & 32 & 50 & 39 & 50 & 100 & 67 \\
TelAviv & 8 & 29 & 100 & 44 & 40 & 100 & 57 & 50 & 100 & 67 & 40 & 100 & 57 \\
Gaborone & 8 & 15 & 100 & 27 & 25 & 100 & 40 & 25 & 100 & 40 & 25 & 100 & 40 \\
Glasgow & 7 & 57 & 100 & 73 & 67 & 100 & 80 & 70 & 88 & 78 & 57 & 100 & 73 \\
Bratislava & 7 & 29 & 100 & 44 & 25 & 50 & 33 & 25 & 50 & 33 & 33 & 100 & 50 \\
Copenhagen & 5 & 100 & 100 & 100 & 100 & 100 & 100 & 100 & 100 & 100 & 100 & 90 & 94 \\
Amsterdam & 5 & 100 & 100 & 100 & 100 & 80 & 89 & 100 & 80 & 89 & 100 & 100 & 100 \\
\midrule\addlinespace
\textbf{City-macro mean} & & \textbf{51.5} & \textbf{99.7} & \textbf{65.4} & \textbf{58.6} & \textbf{91.0} & \textbf{69.4} & \textbf{57.4} & \textbf{91.6} & \textbf{68.9} & \textbf{56.3} & \textbf{94.6} & \textbf{68.8} \\
\bottomrule
\end{longtable}
\endgroup

\textcolor{black}{Because all four models are evaluated on the same 1{,}530 images, we can ask not only how
well each performs but also whether their errors are independent or shared. Computing agreement
\emph{within each run} (so that no tie-breaking is required) and averaging over the two runs, the four
models reach a unanimous decision on $1{,}140 \pm 11$ images (74.5\%), and their mean pairwise agreement
is 86.1\%. Agreement is strongly informative: as the number of models voting \emph{Unsafe} rises from
zero to four, the share of images that are truly \emph{Unsafe} increases monotonically (35\%, 75\%, 79\%,
93\% and 98\%), confirming that model consensus tracks the ground truth rather than shared noise.}

\begin{figure}[t]
\centering
\includegraphics[width=\textwidth]{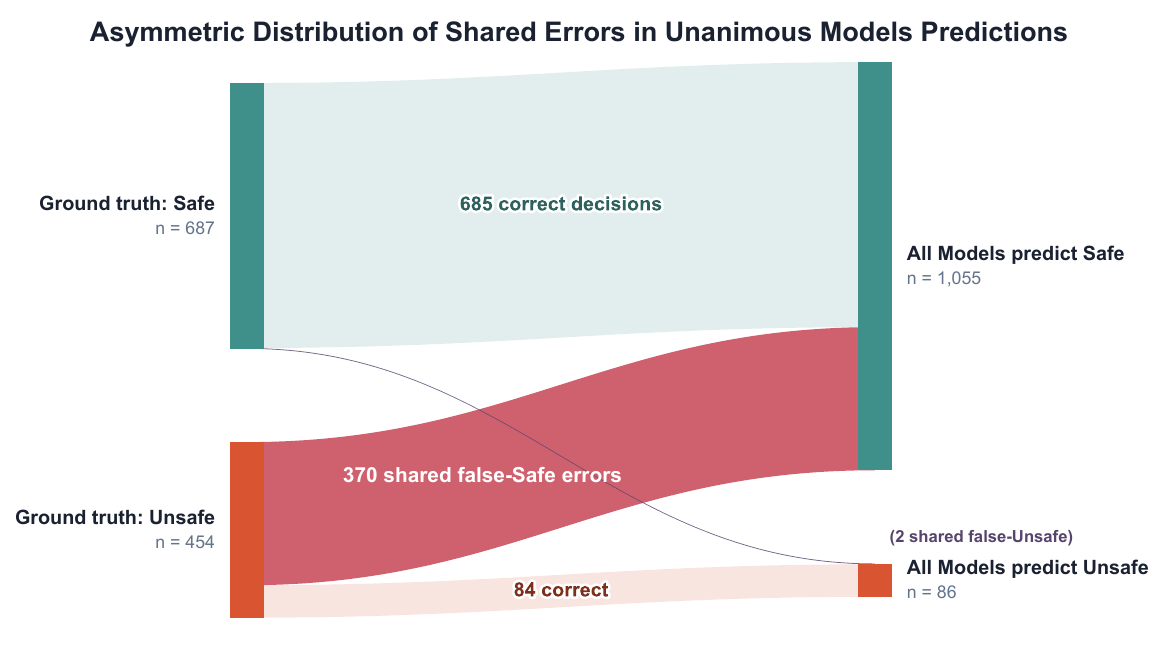}
\caption{\textcolor{black}{Cross-model agreement on the Neutral prompt (counts are means over the two
runs). Where the models unanimously vote \emph{Safe}, $\sim$370 scenes are in fact \emph{Unsafe}; where
they unanimously vote \emph{Unsafe}, essentially none are in fact \emph{Safe}. The shared errors are
almost entirely one-directional, revealing a common Safe-bias.}}
\label{fig:shared_bias}
\end{figure}

\textcolor{black}{The shared errors, however, are highly asymmetric, and this is the clearest signature of
the common Safe-bias. Of the $\sim$1{,}055 images that all four models classify \emph{Safe}, $370 \pm 4$
are in fact labelled \emph{Unsafe} in the ground truth; by contrast, of the $\sim$85 images all four
classify \emph{Unsafe}, fewer than two on average (0--3 across runs) are truly \emph{Safe}
(Figure~\ref{fig:shared_bias}). The models thus fail together almost exclusively in one direction ---
confidently declaring genuinely unsafe scenes safe --- which would be extremely unlikely if their
optimism were merely model-specific noise, and instead points to a Safe-bias intrinsic to how
contemporary MLLMs approach the task.}

\begin{figure}[t]
\centering
\includegraphics[width=\textwidth]{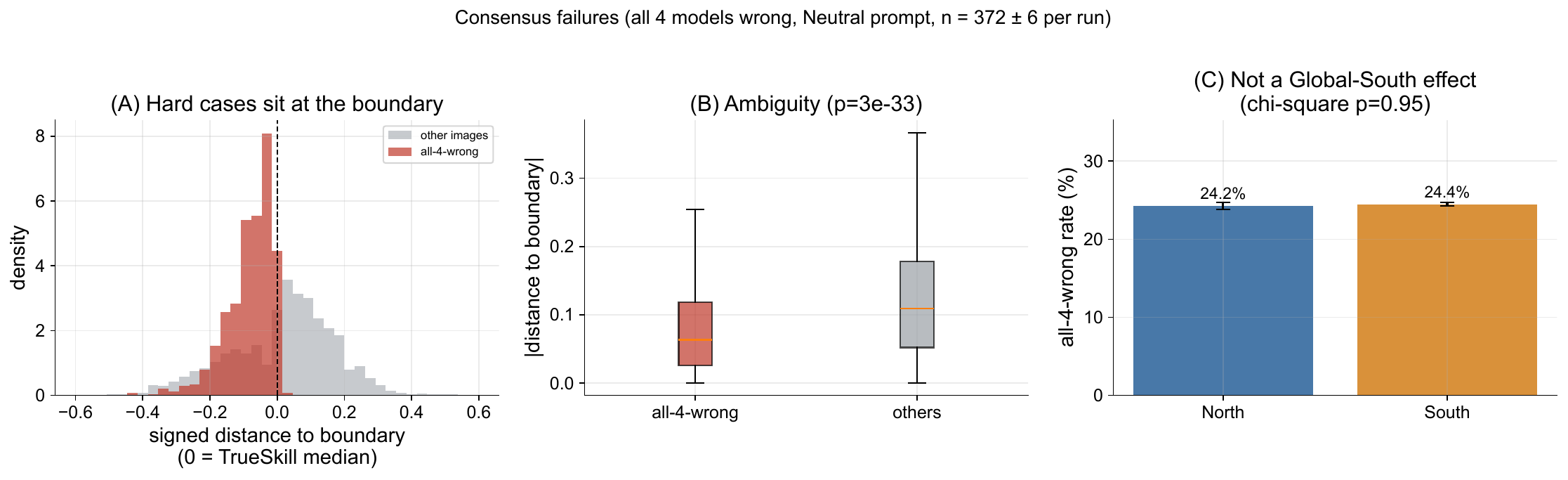}
\caption{\textcolor{black}{Where the models fail together (the $\sim$372 images, on average, that all four
misclassify under the Neutral prompt). \textbf{(A)} Signed distance of each image's TrueSkill score to the
global median (0~$=$~boundary): consensus failures (red) cluster just below the boundary, i.e.\ they are
truly marginally-\emph{Unsafe} scenes. \textbf{(B)} Absolute distance to the boundary is significantly
smaller for consensus failures than for other images, confirming they are more ambiguous. \textbf{(C)} The
consensus-failure rate is statistically indistinguishable between the Global North and Global South,
showing that shared failures are driven by scene ambiguity rather than region.}}
\label{fig:consensus_failures}
\end{figure}

\textcolor{black}{It is worth asking \emph{which} scenes trigger these shared failures. Characterising the
$\sim$372 images (on average, $372 \pm 6$ across the two runs) that all four models misclassify
(Figure~\ref{fig:consensus_failures}), we find they are not arbitrary but concentrated at the decision
boundary: almost all are truly-\emph{Unsafe} scenes whose TrueSkill score sits just below the global
median (panel A), and their absolute distance to that boundary is significantly smaller than for the
remaining images (median $\approx 0.06$ vs.\ $0.11$; Mann--Whitney $p \ll 10^{-10}$, panel B). In other
words, the models fail together precisely on the most ambiguous, borderline cases --- exactly where even
human raters are least decisive. Critically, these consensus failures are \emph{not} a Global-South
phenomenon: their rate is essentially identical in the Global North (24.2\%) and Global South (24.4\%),
with no significant association between region and consensus failure ($\chi^2$ not significant, panel C).
This reinforces the interpretation from the main text: the North--South performance gap arises from the
interaction of the Safe-bias with regional base rates, not from a region-specific breakdown in the
models' judgement, and the shared errors reflect genuine scene ambiguity rather than a failure to
interpret Global-South environments.}

\subsubsection{Robustness to Low-Sample Cities}
 
\textcolor{black}{Several cities are represented by only a handful of images, which could in principle make
the city-macro average unstable. To rule this out, we repeat the analysis on the 40 cities that have at
least 15 evaluated images. As Table~\ref{tab:robust_ge15} shows, both the aggregate performance and the
geographic gap are essentially unchanged: the city-macro \emph{Safe}-F1 moves by at most half a point for
every model, and the North--South gap remains large (26--32 percentage points). Neither the overall
competence nor the regional disparity is therefore an artefact of small, noisy cities.}
 
\begin{table}[ht]
\centering
\caption{\textcolor{black}{Robustness of the Neutral-prompt results to low-sample cities. City-macro
\emph{Safe}-F1 and the North--South gap ($\Delta$) computed on all 56 cities and on the 40 cities with at
least 15 evaluated images.}}
\label{tab:robust_ge15}
\begin{tabular}{lcccc}
\toprule
& \multicolumn{2}{c}{\textbf{Safe-F1 (city-macro)}} & \multicolumn{2}{c}{\textbf{North$-$South gap}} \\
\cmidrule(lr){2-3}\cmidrule(lr){4-5}
\textbf{Model} & \textbf{All (56)} & \textbf{$n\geq 15$ (40)} & \textbf{All (56)} & \textbf{$n\geq 15$ (40)} \\
\midrule
LLaVA~1.6 7B          & 65.4 & 65.8 & 29.4 & 31.7 \\
Gemini~2.5 Flash Lite & 69.4 & 69.3 & 22.2 & 26.6 \\
Qwen2.5-VL-72B        & 68.9 & 69.0 & 22.8 & 27.1 \\
GPT-5.1               & 68.8 & 68.8 & 22.7 & 26.2 \\
\bottomrule
\end{tabular}
\end{table}

\clearpage
\section{Prompts Comparison}
\label{Prompts_Comparison_SI}

\subsection{Overview}
Understanding how urban safety perception varies across different socio-demographic groups requires a well-structured prompting strategy. In this section, we detail three distinct prompts used to evaluate safety classification from different perspectives: ethnicity, gender, and age.

The prompts are designed to:
\begin{itemize}
    \item Encourage the model to consider cultural, gender-based, and age-related factors in safety perception.
    \item Ensure consistency by using a structured JSON output format and minor changes across prompts.
\end{itemize}

The detailed prompt structures are presented in Tables \ref{tab:race_prompt}, \ref{tab:gender_prompt}, and \ref{tab:age_prompt}.

\subsection{Prompts Structure}
\label{Prompts_Comparison_SI}
\textcolor{black}{Every persona prompt shares the balanced structure of the Neutral prompt
(Table~\ref{tab:neutral_prompt}); only the opening role sentence and the ``consider how someone \dots''
line change to specify the perspective, and the required output is always a JSON object with a
\texttt{Classification} and three \texttt{Keywords}. Tables~\ref{tab:race_prompt}--\ref{tab:age_prompt}
give the race/ethnicity, gender and age prompts, with the bracketed placeholder substituted by each
category in turn.}
 
\begin{table}[!htbp]
\centering
\caption{\textcolor{black}{Race/ethnicity prompt. The \texttt{<Race/Ethnicity>} placeholder is
substituted by each of the six broad race/ethnicity categories (White/Caucasian, Asian/Pacific Islander,
Hispanic/Latino, Middle Eastern/North African, Black/African American, Native American/Alaska Native).}}
\label{tab:race_prompt}
\begin{tabular}{p{0.95\linewidth}}
\toprule
\textbf{\textcolor{black}{Race/Ethnicity Prompt}} \\
\midrule
\ttfamily
Assume the role of a person who identifies as <Race/Ethnicity>. Your task is to classify the area in the image as ``Safe'' or ``Unsafe'' based on visible indicators of security, safety, and livability from the perspective of someone with this race or ethnicity background.
 
Consider how someone with this race or ethnicity background might perceive the area.
 
Use exactly 3 concise keywords that accurately reflect the observed evidence justifying your choice.
 
Respond ONLY with a valid JSON object with exactly these keys:\\
\{\\
\quad ``Classification'': ``Safe'' or ``Unsafe'',\\
\quad ``Keywords'': [``keyword1'', ``keyword2'', ``keyword3'']\\
\}\\
\normalfont\\
\bottomrule
\end{tabular}
\end{table}
 
\begin{table}[!htbp]
\centering
\caption{\textcolor{black}{Gender prompt, instructing the model to adopt a male or female perspective.}}
\label{tab:gender_prompt}
\begin{tabular}{p{0.95\linewidth}}
\toprule
\textbf{Gender Prompt} \\
\midrule
\ttfamily
Assume the role of a <male/female> person. Your task is to classify the area in the image as ``Safe'' or ``Unsafe'' based on visible indicators of security, safety, and livability from the perspective of someone of your gender.
 
Consider how someone of your gender would perceive the area.
 
Use exactly 3 concise keywords that accurately reflect the observed evidence justifying your choice.
 
Respond ONLY with a valid JSON object with exactly these keys:\\
\{\\
\quad ``Classification'': ``Safe'' or ``Unsafe'',\\
\quad ``Keywords'': [``keyword1'', ``keyword2'', ``keyword3'']\\
\}\\
\normalfont\\
\bottomrule
\end{tabular}
\end{table}
 
\begin{table}[!htbp]
\centering
\caption{\textcolor{black}{Age prompt, instructing the model to adopt a young, middle-aged, or elderly
perspective.}}
\label{tab:age_prompt}
\begin{tabular}{p{0.95\linewidth}}
\toprule
\textbf{Age Prompt} \\
\midrule
\ttfamily
Assume the role of a <young/middle-aged/elderly> person. Your task is to classify the area in the image as ``Safe'' or ``Unsafe'' based on visible indicators of security, safety, and livability from the perspective of someone of your age.
 
Consider how someone of your age would perceive the area.
 
Use exactly 3 concise keywords that accurately reflect the observed evidence justifying your choice.
 
Respond ONLY with a valid JSON object with exactly these keys:\\
\{\\
\quad ``Classification'': ``Safe'' or ``Unsafe'',\\
\quad ``Keywords'': [``keyword1'', ``keyword2'', ``keyword3'']\\
\}\\
\normalfont\\
\bottomrule
\end{tabular}
\end{table}
 
\clearpage
\section{\textcolor{black}{Persona-Prompt Robustness and Illustrative Outputs}}
\label{Disparities}

\textcolor{black}{This appendix complements the aggregate persona analyses reported
in the main text. For each socio-demographic axis, we first present selected
model outputs for identical images and then report city-level comparisons with
the Neutral baseline. The examples are included to illustrate how classifications
and generated keywords can change when only the persona prompt is modified; they
are not intended to represent the full distribution of outputs. Moreover, the
keywords are model-generated justifications rather than direct measurements of
visual attention, internal reasoning, or the lived perceptions of the populations
named in the prompts.}


\subsection{\textcolor{black}{Race and ethnicity}}

\textcolor{black}{We evaluate six broad race/ethnicity persona prompts:
\emph{White/Caucasian}, \emph{Asian/Pacific Islander},
\emph{Hispanic/Latino}, \emph{Middle Eastern/North African},
\emph{Black/African American}, and
\emph{Native American/Alaska Native}. Each of the four MLLMs evaluates the
same retained image set under every prompt, with two runs per configuration.
The image, task instructions, and output format remain fixed; only the
persona-related wording changes. These labels are treated as analytical prompt
conditions and should not be interpreted as exhaustive or internally homogeneous
demographic categories.}

\subsubsection*{\textcolor{black}{Illustrative outputs}}

\textcolor{black}{Tables~\ref{tab:Appendix_full_ethnicity_toronto}
and~\ref{tab:Appendix_full_ethnicity_bangkok} reproduce the outputs obtained
for two selected images. They illustrate both persona-induced variation and
substantial heterogeneity across models. For some model--image combinations,
persona prompting changes the classification from \emph{Safe} to \emph{Unsafe}
or in the opposite direction; for others, the classification remains stable
while the generated keywords change. The cited terms range from maintenance,
daylight, and visibility to isolation, deterioration, and limited pedestrian
activity. A small number of LLaVA outputs repeat persona labels rather than
providing image-grounded visual cues; these outputs are retained verbatim because
they document a failure of the requested keyword-generation format.}

\begin{table}[!htbp]
\centering
\caption{\textcolor{black}{\textbf{Model outputs across race/ethnicity prompts
for one Toronto image.}
The table reports the classification and keywords generated by each model for
image ID \texttt{50f42eabfdc9f065f00022de} under the Neutral prompt and the
six race/ethnicity personas. The visual input is identical across all conditions.
Outputs are reproduced verbatim after formatting; an em dash indicates that no
valid keyword list was returned.}}
\label{tab:Appendix_full_ethnicity_toronto}
\footnotesize

\begin{tabular}{lllp{7.2cm}}
\toprule
\textbf{Model} & \textbf{Race/Ethnicity} & \textbf{Class.} & \textbf{Keywords} \\
\midrule

\multirow{7}{*}{LLaVA 1.6 7B}
& Neutral & Safe & Safe, Livable, No Obstacles \\
& White/Caucasian & Safe & Safe, Livable, No Obvious Safety Concerns \\
& Asian/Pacific Islander & Safe & Safe, Livable, Clean \\
& Hispanic/Latino & Safe & Safe, Livable, Clean \\
& Middle Eastern/N.\,African & Safe & Safe, Livable, No Obvious Tensions \\
& Black/African American & Unsafe & dilapidated, vacant, unmaintained \\
& Native American & Unsafe & Native American, Alaska Native, Unsafe \\

\midrule

\multirow{7}{*}{Gemini 2.5 FL}
& Neutral & Unsafe & industrial area, isolated, limited visibility \\
& White/Caucasian & Safe & industrial area, well-maintained, clear visibility \\
& Asian/Pacific Islander & Unsafe & isolated, industrial, empty \\
& Hispanic/Latino & Safe & well-maintained, clear road, daylight \\
& Middle Eastern/N.\,African & Safe & --- \\
& Black/African American & Safe & industrial, well-maintained, open \\
& Native American & Unsafe & isolated, industrial, empty \\

\midrule

\multirow{7}{*}{Qwen2.5-VL-72B}
& Neutral & Safe & Well-maintained, Open-space, Light-traffic \\
& White/Caucasian & Safe & Well-maintained, Industrial, Quiet \\
& Asian/Pacific Islander & Safe & Well-maintained, Industrial, Secluded \\
& Hispanic/Latino & Unsafe & Industrial, Isolation, Barren \\
& Middle Eastern/N.\,African & Safe & Well-maintained, Industrial, Open \\
& Black/African American & Unsafe & isolated, industrial, barren \\
& Native American & Unsafe & Industrial, Isolation, Barren \\

\midrule

\multirow{7}{*}{GPT-5.1}
& Neutral & Unsafe & isolated, industrial, minimal pedestrian activity \\
& White/Caucasian & Safe & industrial-area, low-traffic, maintained-road \\
& Asian/Pacific Islander & Unsafe & isolated, industrial, minimal-activity \\
& Hispanic/Latino & Unsafe & isolated, industrial, no pedestrians \\
& Middle Eastern/N.\,African & Unsafe & isolated-industrial, lack-of-pedestrians, minimal-surveillance \\
& Black/African American & Unsafe & isolated, industrial, no pedestrians \\
& Native American & Unsafe & industrial isolation, lack of pedestrians, minimal surveillance \\

\bottomrule
\end{tabular}
\end{table}

\begin{table}[!htbp]
\centering
\caption{\textcolor{black}{\textbf{Model outputs across race/ethnicity prompts
for one Bangkok image.}
The table reports the classification and keywords generated by each model for
image ID \texttt{5140c9adfdc9f049260025df} under the Neutral prompt and the
six race/ethnicity personas. The visual input is identical across all conditions.
The example illustrates that persona sensitivity varies substantially across
models: Gemini retains the same classification in all conditions, whereas the
other models change their classification under selected prompts.}}
\label{tab:Appendix_full_ethnicity_bangkok}
\footnotesize

\begin{tabular}{lllp{7.2cm}}
\toprule
\textbf{Model} & \textbf{Race/Ethnicity} & \textbf{Class.} & \textbf{Keywords} \\
\midrule

\multirow{7}{*}{LLaVA 1.6 7B}
& Neutral & Safe & urban area, bridge, traffic light \\
& White/Caucasian & Safe & urban, modern, developed \\
& Asian/Pacific Islander & Safe & Asian-friendly, Safety, Livable \\
& Hispanic/Latino & Unsafe & Lack of diversity, Limited public spaces, Insufficient safety measures \\
& Middle Eastern/N.\,African & Unsafe & Crowded, Dense, Poor infrastructure \\
& Black/African American & Unsafe & dilapidated, unsafe, uninviting \\
& Native American & Unsafe & Native American, Alaska Native, Safety Concerns \\

\midrule

\multirow{7}{*}{Gemini 2.5 FL}
& Neutral & Unsafe & overpass, traffic, urban decay \\
& White/Caucasian & Unsafe & poor infrastructure, unmaintained buildings, overcast sky \\
& Asian/Pacific Islander & Unsafe & poor infrastructure, overcast sky, dense urban \\
& Hispanic/Latino & Unsafe & poor infrastructure, overcrowded, unmaintained \\
& Middle Eastern/N.\,African & Unsafe & poor infrastructure, overcrowded, unmaintained \\
& Black/African American & Unsafe & poor infrastructure, overcrowded, neglected \\
& Native American & Unsafe & overpass, industrial, traffic \\

\midrule

\multirow{7}{*}{Qwen2.5-VL-72B}
& Neutral & Unsafe & Overcrowded, Exposed, Unmaintained \\
& White/Caucasian & Safe & well-maintained, organized, infrastructure \\
& Asian/Pacific Islander & Safe & Well-maintained, Clear-roads, Urban-infrastructure \\
& Hispanic/Latino & Unsafe & Overcrowded, Infrastructure, Maintenance \\
& Middle Eastern/N.\,African & Unsafe & Overcrowded, Infrastructure, Maintenance \\
& Black/African American & Unsafe & isolated, industrial, neglected \\
& Native American & Unsafe & Isolated, Infrastructure, Urban \\

\midrule

\multirow{7}{*}{GPT-5.1}
& Neutral & Unsafe & elevated highway, no pedestrian infrastructure, heavy traffic \\
& White/Caucasian & Safe & elevated highway, commercial area, daytime traffic \\
& Asian/Pacific Islander & Safe & busy arterial road, commercial area, daytime visibility \\
& Hispanic/Latino & Unsafe & elevated highway, industrial surroundings, lack of pedestrian space \\
& Middle Eastern/N.\,African & Safe & busy arterial road, commercial area, daylight visibility \\
& Black/African American & Unsafe & isolated-highway, lack-of-pedestrians, limited-surveillance \\
& Native American & Unsafe & elevated highway, heavy traffic, lack of pedestrian space \\

\bottomrule
\end{tabular}
\end{table}

\subsubsection*{\textcolor{black}{City-level comparisons with Neutral}}

\textcolor{black}{We compare each race/ethnicity persona with the Neutral baseline
using a paired city-level design with $n=56$ cities. For each city, we compute
the persona-induced change in unsafe-classification rate after averaging the
two experimental runs. We report the mean paired difference, a 95\% bootstrap
confidence interval based on 10,000 city-level resamples, a two-sided sign-flip
permutation test based on 10,000 permutations, and Cohen's $d_z$ for paired
observations. The resulting $p$-values are unadjusted for multiple comparisons.}

\textcolor{black}{Native American prompting produces a statistically significant
increase relative to Neutral in all four models, ranging from $+6.35$ percentage
points for GPT-5.1 to $+38.00$ for LLaVA ($p<0.001$ in every model;
$d_z=0.65$--$2.04$). Black/African American prompting increases the unsafe rate
in LLaVA ($+21.65$ pp), Qwen ($+22.38$ pp), and GPT-5.1 ($+10.41$ pp), with
$p<0.001$ in all three cases. Gemini shows a different direction: its
Black/African American persona produces a small but statistically significant
decrease relative to Neutral ($\Delta=-2.15$ pp, $p=0.007$, $d_z=-0.37$).}

\textcolor{black}{The White/Caucasian persona remains close to Neutral in effect
magnitude across all four models ($|d_z|\leq0.36$). Nevertheless, small
differences are statistically detectable for LLaVA
($\Delta=+1.32$ pp, $p=0.006$), Gemini
($\Delta=-2.34$ pp, $p=0.042$), and GPT-5.1
($\Delta=+2.54$ pp, $p=0.044$), while Qwen shows no difference
($\Delta=+0.03$ pp, $p=0.978$). Among the intermediate personas, all three
produce significant positive shifts in LLaVA and Qwen. In GPT-5.1,
Asian/Pacific Islander and Hispanic/Latino prompting produce significant
positive shifts, whereas Middle Eastern/North African does not. None of the
three intermediate personas differs significantly from Neutral in Gemini.}

\textcolor{black}{These results demonstrate that race- and ethnicity-conditioned
language can systematically change the models' classification criteria for
identical images. However, the direction and magnitude of the effects are not
uniform across models. The analysis establishes persona sensitivity within the
evaluated prompting design, not demographic fidelity or authentic
perspective-taking.}

\begin{table}[!htbp]
    \centering
    \caption{\textcolor{black}{\textbf{Race/ethnicity personas compared with the
    Neutral baseline.}
    For each model and persona, the table reports the mean paired city-level
    difference in unsafe-classification rate
    ($\Delta$, percentage points), its 95\% bootstrap confidence interval, the
    two-sided sign-flip permutation $p$-value, and Cohen's $d_z$
    ($n=56$ paired cities). Positive values indicate a higher unsafe rate than
    Neutral, whereas negative values indicate a lower rate. Statistically
    significant differences at $p<0.05$ are shown in bold. The reported
    $p$-values are unadjusted for multiple comparisons.}}
    \label{tab:Appendix_race_significance}
    \small
    \begin{tabular}{llrrr}
        \toprule
        \textbf{Model} & \textbf{Race/Ethnicity} & \textbf{$\Delta$ (pp)} &
        \textbf{95\% CI} & \textbf{$p$ \; ($d_z$)} \\
        \midrule
        \multirow{6}{*}{LLaVA 1.6 7B}
          & White/Caucasian            & $+1.32$  & $[0.45,\ 2.35]$   & \textbf{0.006}\;($0.36$) \\
          & Asian/Pacific Isl.         & $+4.87$  & $[3.22,\ 6.95]$   & \textbf{$<$0.001}\;($0.66$) \\
          & Hispanic/Latino            & $+7.73$  & $[5.59,\ 10.41]$  & \textbf{$<$0.001}\;($0.80$) \\
          & Middle Eastern/N.\,Afr.    & $+10.00$ & $[7.50,\ 13.06]$  & \textbf{$<$0.001}\;($0.91$) \\
          & Black/African Am.          & $+21.65$ & $[17.99,\ 25.63]$ & \textbf{$<$0.001}\;($1.45$) \\
          & Native American            & $+38.00$ & $[33.36,\ 42.89]$ & \textbf{$<$0.001}\;($2.04$) \\
        \midrule
        \multirow{6}{*}{Gemini 2.5 FL}
          & White/Caucasian            & $-2.34$  & $[-4.38,\ -0.08]$ & \textbf{0.042}\;($-0.28$) \\
          & Asian/Pacific Isl.         & $+0.63$  & $[-1.16,\ 2.39]$  & 0.507\;($0.09$) \\
          & Hispanic/Latino            & $-0.36$  & $[-1.90,\ 1.20]$  & 0.666\;($-0.06$) \\
          & Middle Eastern/N.\,Afr.    & $-1.28$  & $[-3.15,\ 0.69]$  & 0.211\;($-0.17$) \\
          & Black/African Am.          & $-2.15$  & $[-3.65,\ -0.62]$ & \textbf{0.007}\;($-0.37$) \\
          & Native American            & $+17.11$ & $[14.57,\ 19.70]$ & \textbf{$<$0.001}\;($1.70$) \\
        \midrule
        \multirow{6}{*}{Qwen2.5-VL-72B}
          & White/Caucasian            & $+0.03$  & $[-1.56,\ 1.75]$  & 0.978\;($0.00$) \\
          & Asian/Pacific Isl.         & $+5.50$  & $[3.03,\ 8.17]$   & \textbf{$<$0.001}\;($0.56$) \\
          & Hispanic/Latino            & $+7.58$  & $[5.49,\ 9.79]$   & \textbf{$<$0.001}\;($0.91$) \\
          & Middle Eastern/N.\,Afr.    & $+3.30$  & $[1.53,\ 5.13]$   & \textbf{$<$0.001}\;($0.48$) \\
          & Black/African Am.          & $+22.38$ & $[18.88,\ 25.82]$ & \textbf{$<$0.001}\;($1.67$) \\
          & Native American            & $+18.77$ & $[15.59,\ 22.22]$ & \textbf{$<$0.001}\;($1.45$) \\
        \midrule
        \multirow{6}{*}{GPT-5.1}
          & White/Caucasian            & $+2.54$  & $[0.28,\ 5.13]$   & \textbf{0.044}\;($0.27$) \\
          & Asian/Pacific Isl.         & $+3.97$  & $[1.86,\ 6.45]$   & \textbf{$<$0.001}\;($0.44$) \\
          & Hispanic/Latino            & $+4.79$  & $[3.08,\ 6.83]$   & \textbf{$<$0.001}\;($0.66$) \\
          & Middle Eastern/N.\,Afr.    & $+1.04$  & $[-0.31,\ 2.54]$  & 0.163\;($0.19$) \\
          & Black/African Am.          & $+10.41$ & $[7.89,\ 13.25]$  & \textbf{$<$0.001}\;($1.00$) \\
          & Native American            & $+6.35$  & $[3.91,\ 8.95]$   & \textbf{$<$0.001}\;($0.65$) \\
        \bottomrule
    \end{tabular}
\end{table}


\clearpage
\subsection{\textcolor{black}{Gender}}

\textcolor{black}{Gender prompting produces the most directionally consistent
persona effect in the study: the Female persona yields a higher
unsafe-classification rate than the Male persona in all four models. The direct
Female--Male comparisons are reported in the main text. Here, we provide
selected qualitative outputs and compare each gender persona separately with
the Neutral baseline.}

\subsubsection*{\textcolor{black}{Illustrative outputs}}

\textcolor{black}{Figure~\ref{fig:keyword_gender} examines only images for which
the Male and Female prompts produce different classifications. Within this
subset, Female-prompt outputs more frequently cite isolation, poor lighting,
limited visibility, and the absence of pedestrians, whereas Male-prompt outputs
more frequently cite residential character, maintenance, daylight, and visual
order. These associations describe differences in the generated explanations
and should not be interpreted as causal evidence about the internal features
used by the models.}

\textcolor{black}{Tables~\ref{tab:Appendix_full_gender_1}
and~\ref{tab:Appendix_full_gender_2} provide two image-level examples. For the
Tel Aviv image, the Female prompt changes the classification to \emph{Unsafe}
for LLaVA, Gemini, and GPT-5.1, while Qwen already classifies the image as
\emph{Unsafe} under all three conditions. For the San Francisco image, Gemini,
Qwen, and GPT-5.1 change from \emph{Safe} under the Male prompt to
\emph{Unsafe} under the Female prompt, while LLaVA remains \emph{Safe}. These
examples illustrate the effect but are not used to estimate its prevalence.}

\begin{table}[!htbp]
  \centering
  \caption{\textcolor{black}{\textbf{Model outputs across gender prompts for one
  Tel Aviv image.}
  Classifications and keywords generated for image ID
  \texttt{50f3981ffdc9f065f0000dfa} under the Neutral, Male, and Female prompts.
  The visual input is identical across all conditions.}}
  \label{tab:Appendix_full_gender_1}
  \footnotesize
  \begin{tabular}{lllp{8.8cm}}
    \toprule
    \textbf{Model} & \textbf{Persona} & \textbf{Class.} & \textbf{Keywords} \\
    \midrule
    \multirow{3}{*}{LLaVA 1.6 7B}
     & Neutral & Safe & Safe, Livable, Parking Available \\
     & Male & Safe & Parked cars, No visible threats, Low activity \\
     & Female & Unsafe & Crowded, Parked Vehicles, Motorcycles \\
    \midrule
    \multirow{3}{*}{Gemini 2.5 FL}
     & Neutral & Safe & well-maintained buildings, street parking, pedestrian activity \\
     & Male & Safe & well-maintained, active, residential \\
     & Female & Unsafe & poor lighting, dilapidated buildings, lack of pedestrian presence \\
    \midrule
    \multirow{3}{*}{Qwen2.5-VL-72B}
     & Neutral & Unsafe & Dilapidated, Overcrowded, Neglected \\
     & Male & Unsafe & Dilapidated, Crowded, Unmaintained \\
     & Female & Unsafe & Dilapidated, Crowded, Unmaintained \\
    \midrule
    \multirow{3}{*}{GPT-5.1}
     & Neutral & Safe & residential buildings, parked vehicles, daylight visibility \\
     & Male & Safe & residential-street, parked-vehicles, daytime-visibility \\
     & Female & Unsafe & narrow-deserted-street, visible-disrepair, cluttered-parking \\
    \bottomrule
  \end{tabular}
\end{table}

\begin{table}[!htbp]
  \centering
  \caption{\textcolor{black}{\textbf{Model outputs across gender prompts for one
  San Francisco image.}
  Classifications and keywords generated for image ID
  \texttt{50f5e8a8fdc9f065f000746e} under the Neutral, Male, and Female prompts.
  The visual input is identical across all conditions.}}
  \label{tab:Appendix_full_gender_2}
  \footnotesize
  \begin{tabular}{lllp{8.8cm}}
    \toprule
    \textbf{Model} & \textbf{Persona} & \textbf{Class.} & \textbf{Keywords} \\
    \midrule
    \multirow{3}{*}{LLaVA 1.6 7B}
     & Neutral & Safe & Safe, Livable, No Obstacles \\
     & Male & Safe & Safe, Livable, No Obstacles \\
     & Female & Safe & Safe, Livable, No Obstacles \\
    \midrule
    \multirow{3}{*}{Gemini 2.5 FL}
     & Neutral & Safe & industrial area, daylight, clear visibility \\
     & Male & Safe & industrial area, daylight, clear visibility \\
     & Female & Unsafe & isolated, industrial, limited visibility \\
    \midrule
    \multirow{3}{*}{Qwen2.5-VL-72B}
     & Neutral & Safe & well-maintained, clear-roads, industrial-area \\
     & Male & Safe & well-maintained, low-traffic, visible-business \\
     & Female & Unsafe & Overgrown, Deserted, Industrial \\
    \midrule
    \multirow{3}{*}{GPT-5.1}
     & Neutral & Safe & daylight, industrial area, light traffic \\
     & Male & Safe & daylight, industrial area, low pedestrian activity \\
     & Female & Unsafe & isolated road, limited pedestrian presence, poor natural surveillance \\
    \bottomrule
  \end{tabular}
\end{table}

\subsubsection*{\textcolor{black}{City-level comparisons with Neutral}}

\textcolor{black}{Relative to the Neutral baseline, the Female persona produces
a statistically significant positive shift in every model, ranging from
$+7.14$ percentage points for LLaVA to $+27.88$ for GPT-5.1
($p<0.001$ throughout; $d_z=0.84$--$2.31$). The Male persona also produces a
positive shift in LLaVA ($+3.45$ pp, $p<0.001$), Gemini
($+1.83$ pp, $p=0.015$), and GPT-5.1
($+5.81$ pp, $p<0.001$), but not in Qwen
($+0.93$ pp, $p=0.210$). In every model, the Female shift is substantially
larger than the Male shift. The displacement from Neutral is therefore driven
primarily by the Female prompt increasing the unsafe-classification rate, not
by the Male prompt lowering it.}

\begin{table}[!htbp]
  \centering
  \caption{\textcolor{black}{\textbf{Gender personas compared with the Neutral
  baseline.}
  Mean paired city-level difference in unsafe-classification rate
  ($\Delta$, percentage points), 95\% bootstrap confidence interval,
  two-sided sign-flip permutation $p$-value, and Cohen's $d_z$
  ($n=56$ paired cities). Positive values indicate a higher unsafe rate than
  Neutral. Statistically significant differences at $p<0.05$ are shown in
  bold.}}
  \label{tab:Appendix_gender_significance}
  \small
  \begin{tabular}{llrrr}
    \toprule
    \textbf{Model} & \textbf{Persona} & \textbf{$\Delta$ (pp)} &
    \textbf{95\% CI} & \textbf{$p$ \; ($d_z$)} \\
    \midrule
    \multirow{2}{*}{LLaVA 1.6 7B}
      & Male & $+3.45$ & $[+2.32,\ +4.75]$ & \textbf{$<$0.001}\;($+0.73$) \\
      & Female & $+7.14$ & $[+5.11,\ +9.45]$ & \textbf{$<$0.001}\;($+0.84$) \\
    \midrule
    \multirow{2}{*}{Gemini 2.5 FL}
      & Male & $+1.83$ & $[+0.42,\ +3.26]$ & \textbf{0.015}\;($+0.33$) \\
      & Female & $+18.47$ & $[+16.02,\ +20.98]$ & \textbf{$<$0.001}\;($+1.95$) \\
    \midrule
    \multirow{2}{*}{Qwen2.5-VL-72B}
      & Male & $+0.93$ & $[-0.51,\ +2.35]$ & 0.210\;($+0.17$) \\
      & Female & $+8.50$ & $[+6.73,\ +10.37]$ & \textbf{$<$0.001}\;($+1.19$) \\
    \midrule
    \multirow{2}{*}{GPT-5.1}
      & Male & $+5.81$ & $[+4.39,\ +7.39]$ & \textbf{$<$0.001}\;($+1.01$) \\
      & Female & $+27.88$ & $[+24.88,\ +31.06]$ & \textbf{$<$0.001}\;($+2.31$) \\
    \bottomrule
  \end{tabular}
\end{table}


\clearpage
\subsection{\textcolor{black}{Age}}
\label{disparities_age}

\textcolor{black}{Age prompting produces a less uniform pattern than gender
prompting. The relative position of the Young, Middle-aged, and Elderly personas
varies across models, and the selected image-level examples should therefore not
be interpreted as evidence of a universal age ordering.}

\subsubsection*{\textcolor{black}{Illustrative outputs}}

\textcolor{black}{Tables~\ref{tab:Appendix_full_age_1}
and~\ref{tab:Appendix_full_age_2} report outputs for two images classified as
\emph{Safe} under the Neutral prompt by all four models. In both examples, the
Middle-aged persona retains the Neutral classification for every model.
Gemini and GPT-5.1 change both the Young and Elderly classifications to
\emph{Unsafe}; Qwen changes only the Elderly classification; and LLaVA remains
\emph{Safe} under all three age prompts. The risk-related keywords generated
under changed classifications include isolation, heavy or fast-moving traffic,
poor lighting, missing crossings, and limited pedestrian infrastructure. These
examples illustrate model-dependent response changes but do not establish their
aggregate frequency.}

\begin{table}[!htbp]
  \centering
  \caption{\textcolor{black}{\textbf{Model outputs across age prompts for one
  Tokyo image.}
  Classifications and keywords generated for image ID
  \texttt{513e6d9bfdc9f0358700c308} under the Neutral, Young, Middle-aged,
  and Elderly prompts. The visual input is identical across all conditions.}}
  \label{tab:Appendix_full_age_1}
  \footnotesize
  \begin{tabular}{lllp{8.8cm}}
    \toprule
    \textbf{Model} & \textbf{Persona} & \textbf{Class.} & \textbf{Keywords} \\
    \midrule
    \multirow{4}{*}{LLaVA 1.6 7B}
     & Neutral & Safe & Safe, Livable, No Obvious Danger \\
     & Young & Safe & Safe, Lively, Clean \\
     & Middle-aged & Safe & Safe, Lively, Well-Maintained \\
     & Elderly & Safe & Paved road, No visible obstacles, No people on the street \\
    \midrule
    \multirow{4}{*}{Gemini 2.5 FL}
     & Neutral & Safe & well-lit, maintained road, active traffic \\
     & Young & Unsafe & overpass, dark, isolated \\
     & Middle-aged & Safe & well-lit, maintained, residential \\
     & Elderly & Unsafe & overpass shadow, busy road, poor lighting \\
    \midrule
    \multirow{4}{*}{Qwen2.5-VL-72B}
     & Neutral & Safe & well-maintained, organized, infrastructure \\
     & Young & Safe & well-lit, organized, infrastructure \\
     & Middle-aged & Safe & well-maintained, organized, infrastructure \\
     & Elderly & Unsafe & Overpass, Traffic, Exposed \\
    \midrule
    \multirow{4}{*}{GPT-5.1}
     & Neutral & Safe & well-maintained road, organized traffic, urban infrastructure \\
     & Young & Unsafe & heavy traffic, no sidewalks, underpass area \\
     & Middle-aged & Safe & well-maintained road, active traffic, commercial buildings \\
     & Elderly & Unsafe & heavy-traffic, no-sidewalks, underpass-shadow \\
    \bottomrule
  \end{tabular}
\end{table}

\begin{table}[!htbp]
  \centering
  \caption{\textcolor{black}{\textbf{Model outputs across age prompts for one
  Bucharest image.}
  Classifications and keywords generated for image ID
  \texttt{5140d2a9fdc9f0492600330f} under the Neutral, Young, Middle-aged,
  and Elderly prompts. The visual input is identical across all conditions.}}
  \label{tab:Appendix_full_age_2}
  \footnotesize
  \begin{tabular}{lllp{8.8cm}}
    \toprule
    \textbf{Model} & \textbf{Persona} & \textbf{Class.} & \textbf{Keywords} \\
    \midrule
    \multirow{4}{*}{LLaVA 1.6 7B}
     & Neutral & Safe & Safe, Livable, No Obstruction \\
     & Young & Safe & Low Traffic, Parked Cars, No Obvious Danger \\
     & Middle-aged & Safe & Safe, Livable, No Obvious Danger \\
     & Elderly & Safe & Low traffic, Clear road, No visible hazards \\
    \midrule
    \multirow{4}{*}{Gemini 2.5 FL}
     & Neutral & Safe & well-maintained road, traffic lights, clear visibility \\
     & Young & Unsafe & busy road, no pedestrian, isolated \\
     & Middle-aged & Safe & well-maintained road, traffic lights, clear visibility \\
     & Elderly & Unsafe & heavy traffic, no pedestrian, fast cars \\
    \midrule
    \multirow{4}{*}{Qwen2.5-VL-72B}
     & Neutral & Safe & Roadway, Trees, Vehicles \\
     & Young & Safe & well-lit, organized, traffic \\
     & Middle-aged & Safe & well-lit, organized, traffic \\
     & Elderly & Unsafe & Roadway, Overgrown, Isolated \\
    \midrule
    \multirow{4}{*}{GPT-5.1}
     & Neutral & Safe & multiple lanes, daylight traffic, sidewalk present \\
     & Young & Unsafe & no pedestrian crossings, fast traffic, isolated sidewalk \\
     & Middle-aged & Safe & busy road, daylight visibility, maintained sidewalk \\
     & Elderly & Unsafe & fast-traffic, no-crosswalks, isolated-sidewalk \\
    \bottomrule
  \end{tabular}
\end{table}

\subsubsection*{\textcolor{black}{City-level comparisons with Neutral}}

\textcolor{black}{All three age personas produce statistically significant positive
shifts relative to Neutral in every model ($p<0.001$ throughout), but their
relative magnitudes differ. In LLaVA, the Young persona produces the largest
shift ($+16.13$ pp, $d_z=1.22$), while the Elderly persona produces the smallest
($+3.45$ pp, $d_z=0.74$); the Middle-aged shift lies between them
($+5.14$ pp, $d_z=0.76$).}

\textcolor{black}{In Gemini, the Young and Elderly shifts are nearly identical:
$+18.26$ and $+18.14$ percentage points, respectively. The Young value is
numerically larger by $0.12$ pp, but the present table does not test the direct
Young--Elderly difference. The Middle-aged persona produces a substantially
smaller shift of $+6.60$ pp. In Qwen and GPT-5.1, the Elderly persona produces
the largest shift, reaching $+10.96$ pp for Qwen and $+22.65$ pp for GPT-5.1.
The Middle-aged persona produces the smallest shift in both models:
$+3.09$ and $+5.69$ pp, respectively.}

\textcolor{black}{The Middle-aged persona is therefore closest to Neutral in
Gemini, Qwen, and GPT-5.1, but not in LLaVA, where the Elderly persona is
closest. Similarly, the largest age-related departure is associated with the
Young persona in LLaVA, is effectively shared by the Young and Elderly personas
in Gemini, and is associated with the Elderly persona in Qwen and GPT-5.1.
These results do not support a single model-independent ordering of perceived
unsafety across age personas.}

\begin{table}[!htbp]
  \centering
  \caption{\textcolor{black}{\textbf{Age personas compared with the Neutral
  baseline.}
  Mean paired city-level difference in unsafe-classification rate
  ($\Delta$, percentage points), 95\% bootstrap confidence interval,
  two-sided sign-flip permutation $p$-value, and Cohen's $d_z$
  ($n=56$ paired cities). All reported differences are positive and statistically
  significant at $p<0.001$. The Middle-aged shift is the smallest in Gemini,
  Qwen, and GPT-5.1; in LLaVA, the Elderly shift is slightly smaller than the
  Middle-aged shift. Young and Elderly prompting produces nearly identical
  shifts in Gemini, with the Young value numerically higher by $0.12$ percentage
  points.}}
  \label{tab:Appendix_age_significance}
  \small
  \begin{tabular}{llrrr}
    \toprule
    \textbf{Model} & \textbf{Persona} & \textbf{$\Delta$ (pp)} &
    \textbf{95\% CI} & \textbf{$p$ \; ($d_z$)} \\
    \midrule
    \multirow{3}{*}{LLaVA 1.6 7B}
      & Young & $+16.13$ & $[+12.88,\ +19.70]$ & \textbf{$<$0.001}\;($+1.22$) \\
      & Middle-aged & $+5.14$ & $[+3.54,\ +6.99]$ & \textbf{$<$0.001}\;($+0.76$) \\
      & Elderly & $+3.45$ & $[+2.36,\ +4.73]$ & \textbf{$<$0.001}\;($+0.74$) \\
    \midrule
    \multirow{3}{*}{Gemini 2.5 FL}
      & Young & $+18.26$ & $[+15.71,\ +20.84]$ & \textbf{$<$0.001}\;($+1.88$) \\
      & Middle-aged & $+6.60$ & $[+5.01,\ +8.35]$ & \textbf{$<$0.001}\;($+1.03$) \\
      & Elderly & $+18.14$ & $[+15.47,\ +20.93]$ & \textbf{$<$0.001}\;($+1.74$) \\
    \midrule
    \multirow{3}{*}{Qwen2.5-VL-72B}
      & Young & $+3.92$ & $[+2.75,\ +5.08]$ & \textbf{$<$0.001}\;($+0.88$) \\
      & Middle-aged & $+3.09$ & $[+1.68,\ +4.45]$ & \textbf{$<$0.001}\;($+0.57$) \\
      & Elderly & $+10.96$ & $[+9.31,\ +12.65]$ & \textbf{$<$0.001}\;($+1.70$) \\
    \midrule
    \multirow{3}{*}{GPT-5.1}
      & Young & $+10.83$ & $[+8.63,\ +13.27]$ & \textbf{$<$0.001}\;($+1.21$) \\
      & Middle-aged & $+5.69$ & $[+4.03,\ +7.51]$ & \textbf{$<$0.001}\;($+0.85$) \\
      & Elderly & $+22.65$ & $[+19.99,\ +25.46]$ & \textbf{$<$0.001}\;($+2.13$) \\
    \bottomrule
  \end{tabular}
\end{table}

\clearpage
\section{Dataset Statistics}
\label{Dataset_SI}
\begin{figure}[!htbp]
\centering
\includegraphics[width=\linewidth]{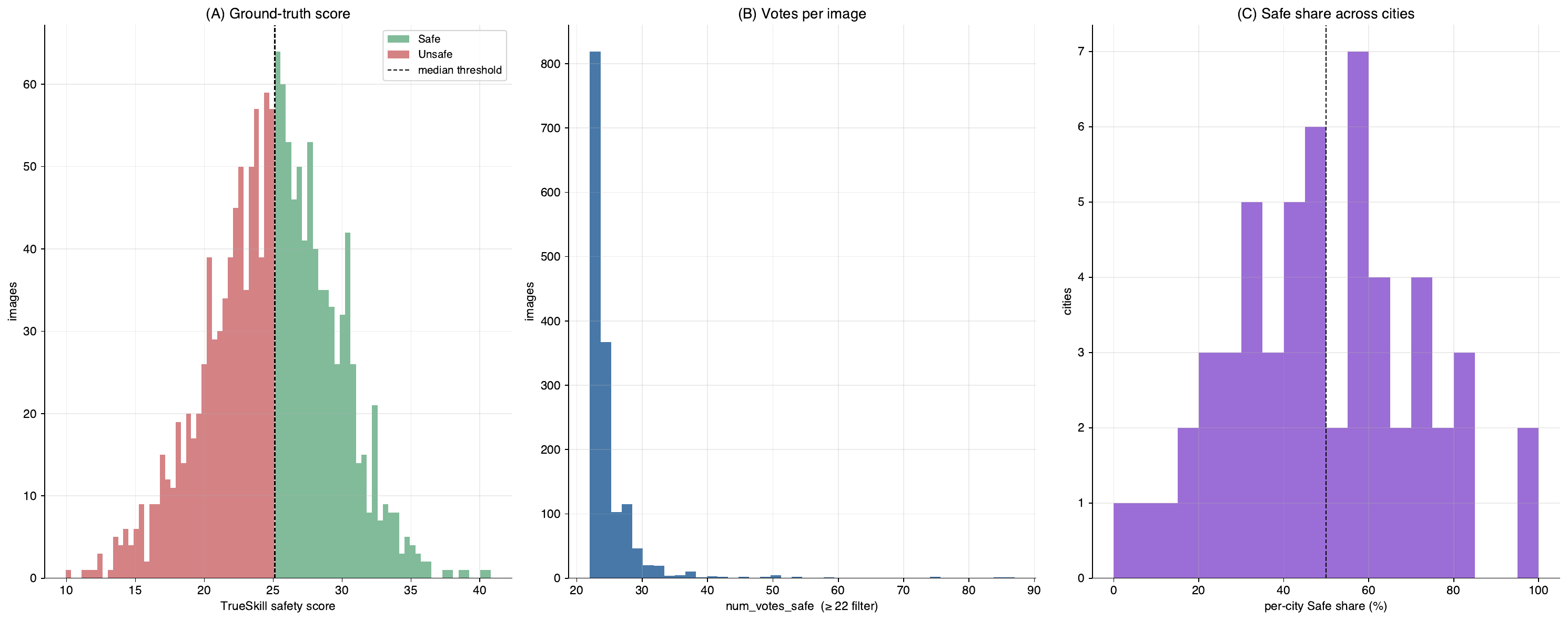}
\caption{\textcolor{black}{Distributions in the threshold-22 evaluation set. (A) Image-level TrueSkill
safety scores, split \emph{Safe}/\emph{Unsafe} at the global median (dashed). (B) Safety votes per
image (all $\ge$22 by construction; the mode is 22--25). (C) Per-city \emph{Safe} share, with the
50\% line dashed.}}
\label{fig:dataset_dist}
\end{figure}

\textcolor{black}{This appendix characterises the \emph{threshold-22} evaluation set used throughout the
paper (Place Pulse 2.0 images retained only if they have at least 22 safety votes, following
Gu et al.~\cite{gu2025designing}; see the Dataset section). The set contains \textbf{1{,}530 images
across 56 cities}, binarised at the global-median TrueSkill score into 751 \emph{Safe} (49.1\%) and
779 \emph{Unsafe} (50.9\%) images. Figure~\ref{fig:dataset_dist} shows the score, vote-count and
per-city Safe-share distributions, and Table~\ref{tab:summary_by_city} lists the per-city composition.}

\small
\begin{longtable}{l c r r r r r}
\caption{\textcolor{black}{Per-city composition of the threshold-22 evaluation set: region (N = Global
North, S = Global South), number of retained images, share labelled \emph{Safe} (global-median split),
mean and standard deviation of the TrueSkill safety score, and median vote count.}}
\label{tab:summary_by_city} \\
\toprule
City & Reg. & Images & Safe\% & TS mean & TS std & Med. votes \\
\midrule
\endfirsthead
\multicolumn{7}{c}{\tablename\ \thetable\ -- continued} \\
\toprule
City & Reg. & Images & Safe\% & TS mean & TS std & Med. votes \\
\midrule
\endhead
\bottomrule
\endfoot
Amsterdam & N & 5 & 100 & 28.1 & 2.7 & 22 \\
Atlanta & N & 55 & 69 & 27.1 & 4.1 & 23 \\
Bangkok & S & 25 & 20 & 21.8 & 4.4 & 23 \\
Barcelona & N & 15 & 40 & 24.0 & 4.8 & 24 \\
Belo Horizonte & S & 27 & 7 & 20.4 & 4.0 & 24 \\
Berlin & N & 46 & 61 & 25.8 & 3.3 & 24 \\
Boston & N & 20 & 70 & 26.2 & 4.8 & 23 \\
Bratislava & N & 7 & 29 & 23.1 & 2.0 & 24 \\
Bucharest & N & 25 & 32 & 23.3 & 4.2 & 24 \\
Cape Town & S & 39 & 38 & 24.5 & 3.6 & 24 \\
Chicago & N & 53 & 66 & 27.3 & 5.2 & 23 \\
Copenhagen & N & 5 & 100 & 27.5 & 0.8 & 25 \\
Denver & N & 37 & 62 & 26.4 & 4.4 & 23 \\
Dublin & N & 20 & 70 & 26.3 & 3.4 & 23 \\
Gaborone & S & 8 & 12 & 21.0 & 3.2 & 24 \\
Glasgow & N & 7 & 57 & 24.6 & 2.0 & 22 \\
Guadalajara & S & 19 & 32 & 22.8 & 4.9 & 24 \\
Helsinki & N & 12 & 58 & 26.4 & 3.5 & 23 \\
Hong Kong & N & 13 & 31 & 23.3 & 3.6 & 23 \\
Houston & N & 45 & 44 & 25.4 & 4.2 & 23 \\
Johannesburg & S & 27 & 37 & 24.6 & 3.0 & 23 \\
Kiev & N & 9 & 33 & 23.0 & 7.4 & 24 \\
Kyoto & N & 11 & 18 & 22.2 & 2.9 & 23 \\
Lisbon & N & 29 & 24 & 23.6 & 3.8 & 23 \\
London & N & 28 & 64 & 26.1 & 4.1 & 23 \\
Los Angeles & N & 12 & 58 & 23.5 & 5.6 & 23 \\
Madrid & N & 28 & 61 & 25.2 & 3.1 & 23 \\
Melbourne & N & 37 & 76 & 28.3 & 3.8 & 23 \\
Mexico City & S & 17 & 0 & 19.8 & 4.0 & 23 \\
Milan & N & 26 & 54 & 24.8 & 4.6 & 23 \\
Minneapolis & N & 12 & 75 & 26.2 & 3.1 & 24 \\
Montreal & N & 32 & 47 & 25.1 & 5.2 & 24 \\
Moscow & N & 39 & 49 & 24.5 & 4.0 & 24 \\
Munich & N & 19 & 74 & 27.5 & 3.3 & 23 \\
New York & N & 60 & 55 & 26.0 & 4.1 & 23 \\
Paris & N & 39 & 44 & 24.0 & 4.5 & 23 \\
Philadelphia & N & 42 & 45 & 23.8 & 4.5 & 23 \\
Portland & N & 22 & 59 & 26.5 & 2.8 & 24 \\
Prague & N & 25 & 48 & 24.7 & 4.4 & 23 \\
Rio de Janeiro & S & 43 & 19 & 20.9 & 3.7 & 23 \\
Rome & N & 26 & 58 & 24.8 & 3.6 & 23 \\
San Francisco & N & 25 & 48 & 25.0 & 3.3 & 24 \\
Santiago & S & 55 & 27 & 22.9 & 4.4 & 23 \\
Sao Paulo & S & 46 & 24 & 22.2 & 5.0 & 23 \\
Seattle & N & 19 & 42 & 25.6 & 3.6 & 23 \\
Singapore & N & 37 & 81 & 27.7 & 4.1 & 23 \\
Stockholm & N & 20 & 80 & 27.3 & 3.6 & 23 \\
Sydney & N & 44 & 84 & 27.6 & 2.7 & 24 \\
Taipei & N & 14 & 43 & 22.5 & 5.6 & 24 \\
Tel Aviv & N & 8 & 25 & 22.0 & 5.9 & 23 \\
Tokyo & N & 57 & 32 & 23.8 & 3.9 & 23 \\
Toronto & N & 57 & 56 & 25.8 & 4.5 & 24 \\
Valparaiso & S & 11 & 36 & 22.1 & 4.4 & 23 \\
Warsaw & N & 47 & 49 & 25.7 & 3.3 & 23 \\
Washington DC & N & 14 & 71 & 27.1 & 5.0 & 22 \\
Zagreb & N & 10 & 50 & 25.9 & 3.3 & 23 \\
\bottomrule
\end{longtable}

\subsection{Effect of Model Size}
\label{Model_Size_SI}

\textcolor{black}{We evaluate three scales of the same family --- LLaVA~1.6 \textbf{7B}, \textbf{13B},
\textbf{34B} --- under one identical setup (same 4-bit pipeline, same prompts, two seeds each).
Table~\ref{tab:model_size} reports, for every persona, the city-macro unsafe-classification rate
(mean of two seeds; race/ethnicity personas form a separate batch, hence their own Neutral).}

\begin{table}[ht]
  \centering
  \caption{\textcolor{black}{Effect of model size (LLaVA~1.6 7B/13B/34B): city-macro Safe-F1 and
  unsafe-classification rate (\%) for every persona, mean of two seeds.}}
  \label{tab:model_size}
  \small
  \begin{tabular}{lccc}
    \toprule
     & \textbf{7B} & \textbf{13B} & \textbf{34B} \\
    \midrule
    \textit{Safe-F1 (\%)} & 65.4 & 66.1 & 68.1 \\
    \midrule
    \multicolumn{4}{l}{\textit{Unsafe rate (\%) --- neutral}}\\
        Neutral            & \phantom{0}5.5 & 39.2 & 14.9 \\
        \midrule

    \multicolumn{4}{l}{\textit{Unsafe rate (\%) --- gender/age batch}}\\

    Male               & \phantom{0}8.9 & 62.6 & \phantom{0}9.0 \\
    Female             & 12.6 & 70.2 & 12.3 \\
    Young              & 21.6 & 63.2 & 13.6 \\
    Middle-aged        & 10.6 & 28.4 & 11.0 \\
    Elderly            & \phantom{0}8.9 & 21.3 & 11.7 \\
    \midrule
    \multicolumn{4}{l}{\textit{Unsafe rate (\%) --- race/ethnicity batch}}\\
    White / Caucasian           & \phantom{0}7.2 & 15.2 & \phantom{0}9.5 \\
    Asian / Pacific Islander     & 10.8 & 76.5 & \phantom{0}6.7 \\
    Hispanic / Latino           & 13.6 & 81.2 & \phantom{0}8.5 \\
    Middle Eastern / N. African & 15.9 & 89.6 & \phantom{0}5.8 \\
    Black / African American    & 27.6 & 94.2 & 12.0 \\
    Native American             & 43.9 & 99.1 & 19.2 \\
    \bottomrule
  \end{tabular}
\end{table}

\textcolor{black}{Performance improves modestly with scale (Safe-F1 65.4\%~$\rightarrow$~66.1\%~$\rightarrow$~68.1\%),
while the Safe-lean persists at every size (the Unsafe share stays well below the $\sim$51\% ground-truth
rate). Crucially, the \emph{directions} of the socio-demographic biases are stable across the whole
family: Female~$>$~Male and Young~$>$~Elderly at every size, and Native American (highest) with
Black/African American (second-highest) lead the racial ordering at every size. Absolute rates are
non-monotonic in size --- the 13B is uniformly the most Unsafe-prone --- so model size modulates the
\emph{magnitude} of the Safe-lean but not the \emph{direction} of the biases. Together with the
four-model evidence in the main text, this shows the bias structure is robust across scales as well as
across architectures.}

\clearpage

\section{Intersectional Prompt Effects}
\label{app:intersectional_bias}

\textcolor{black}{To examine whether the pipeline can condition the models on multiple
socio-demographic attributes simultaneously, we conducted an additional
factorial experiment combining six race/ethnicity descriptors, two gender
descriptors, and three age groups. Table~\ref{tab:intersectional_threeway}
reports the percentage of images classified as Unsafe under all 36
race/ethnicity $\times$ gender $\times$ age conditions.}

\begingroup
\scriptsize
\setlength{\tabcolsep}{2pt}
\renewcommand{\arraystretch}{1.08}
\setlength{\LTleft}{\fill}
\setlength{\LTright}{\fill}

\begin{longtable}{
  @{}
  >{\raggedright\arraybackslash}p{0.46\linewidth}
  *{4}{>{\centering\arraybackslash}p{0.105\linewidth}}
  @{}
}

\caption{\textcolor{black}{Unsafe-classification rates for the 36 three-way intersectional
persona prompts. Values report the percentage of valid classifications assigned
to the Unsafe category. Every prompt was evaluated on the same 1,530 images.
The results derive from one dedicated factorial run per model and should be
interpreted as descriptive prompt-conditioned model behaviour rather than as
estimates of how demographic groups perceive urban environments.}}
\label{tab:intersectional_threeway}
\\

\toprule
\textbf{Persona combination}
&
\shortstack{\textbf{LLaVA}\\\textbf{1.6 7B}}
&
\shortstack{\textbf{Gemini}\\\textbf{2.5 FL}}
&
\shortstack{\textbf{Qwen2.5}\\\textbf{VL-72B}}
&
\shortstack{\textbf{GPT}\\\textbf{5.1}}
\\
\midrule
\endfirsthead

\multicolumn{5}{c}{
  \tablename\ \thetable{} -- continued
}
\\
\toprule
\textbf{Persona combination}
&
\shortstack{\textbf{LLaVA}\\\textbf{1.6 7B}}
&
\shortstack{\textbf{Gemini}\\\textbf{2.5 FL}}
&
\shortstack{\textbf{Qwen2.5}\\\textbf{VL-72B}}
&
\shortstack{\textbf{GPT}\\\textbf{5.1}}
\\
\midrule
\endhead

\midrule
\multicolumn{5}{r}{\scriptsize Continued on next page}
\\
\endfoot

\bottomrule
\endlastfoot

Asian/Pacific Islander $\times$ Male $\times$ Young
& 2.42 & 25.88 & 24.51 & 25.42 \\

Asian/Pacific Islander $\times$ Female $\times$ Young
& 2.42 & 28.50 & 28.37 & 35.95 \\

Asian/Pacific Islander $\times$ Male $\times$ Middle-aged
& 1.57 & 22.68 & 22.22 & 24.44 \\

Asian/Pacific Islander $\times$ Female $\times$ Middle-aged
& 1.70 & 27.91 & 25.16 & 29.48 \\

Asian/Pacific Islander $\times$ Male $\times$ Elderly
& 1.31 & 25.23 & 27.06 & 29.93 \\

Asian/Pacific Islander $\times$ Female $\times$ Elderly
& 1.37 & 30.20 & 31.11 & 33.79 \\

\addlinespace[3pt]

Black/African American $\times$ Male $\times$ Young
& 3.73 & 23.07 & 37.19 & 30.78 \\

Black/African American $\times$ Female $\times$ Young
& 3.27 & 32.22 & 43.01 & 38.37 \\

Black/African American $\times$ Male $\times$ Middle-aged
& 2.35 & 25.56 & 32.94 & 27.71 \\

Black/African American $\times$ Female $\times$ Middle-aged
& 2.29 & 32.88 & 37.32 & 33.53 \\

Black/African American $\times$ Male $\times$ Elderly
& 2.03 & 24.25 & 37.06 & 31.83 \\

Black/African American $\times$ Female $\times$ Elderly
& 1.90 & 32.81 & 41.11 & 36.60 \\

\addlinespace[3pt]

Hispanic/Latino $\times$ Male $\times$ Young
& 3.53 & 27.45 & 26.14 & 26.54 \\

Hispanic/Latino $\times$ Female $\times$ Young
& 3.46 & 30.98 & 28.24 & 34.38 \\

Hispanic/Latino $\times$ Male $\times$ Middle-aged
& 2.29 & 24.05 & 24.71 & 24.31 \\

Hispanic/Latino $\times$ Female $\times$ Middle-aged
& 2.42 & 31.20 & 26.54 & 30.59 \\

Hispanic/Latino $\times$ Male $\times$ Elderly
& 2.29 & 27.06 & 27.52 & 29.67 \\

Hispanic/Latino $\times$ Female $\times$ Elderly
& 2.22 & 35.10 & 30.46 & 34.18 \\

\addlinespace[3pt]

Middle Eastern/North African $\times$ Male $\times$ Young
& 1.63 & 23.68 & 23.40 & 23.01 \\

Middle Eastern/North African $\times$ Female $\times$ Young
& 1.44 & 27.60 & 26.86 & 33.27 \\

Middle Eastern/North African $\times$ Male $\times$ Middle-aged
& 0.98 & 23.48 & 24.05 & 23.07 \\

Middle Eastern/North African $\times$ Female $\times$ Middle-aged
& 0.85 & 25.57 & 25.95 & 28.30 \\

Middle Eastern/North African $\times$ Male $\times$ Elderly
& 0.65 & 25.88 & 25.49 & 24.44 \\

Middle Eastern/North African $\times$ Female $\times$ Elderly
& 0.65 & 30.35 & 29.80 & 31.18 \\

\addlinespace[3pt]

Native American/Alaska Native $\times$ Male $\times$ Young
& 3.20 & 36.60 & 32.94 & 27.39 \\

Native American/Alaska Native $\times$ Female $\times$ Young
& 3.20 & 42.88 & 36.21 & 34.71 \\

Native American/Alaska Native $\times$ Male $\times$ Middle-aged
& 1.70 & 35.78 & 29.15 & 25.42 \\

Native American/Alaska Native $\times$ Female $\times$ Middle-aged
& 1.96 & 44.02 & 34.44 & 31.50 \\

Native American/Alaska Native $\times$ Male $\times$ Elderly
& 1.18 & 36.17 & 34.05 & 28.89 \\

Native American/Alaska Native $\times$ Female $\times$ Elderly
& 1.11 & 46.34 & 41.11 & 36.67 \\

\addlinespace[3pt]

White/Caucasian $\times$ Male $\times$ Young
& 1.44 & 23.01 & 20.65 & 23.40 \\

White/Caucasian $\times$ Female $\times$ Young
& 1.44 & 27.52 & 22.35 & 30.78 \\

White/Caucasian $\times$ Male $\times$ Middle-aged
& 1.05 & 26.47 & 22.22 & 23.59 \\

White/Caucasian $\times$ Female $\times$ Middle-aged
& 1.11 & 30.59 & 23.53 & 28.24 \\

White/Caucasian $\times$ Male $\times$ Elderly
& 0.59 & 29.45 & 24.05 & 25.56 \\

White/Caucasian $\times$ Female $\times$ Elderly
& 0.59 & 34.77 & 26.67 & 33.07 \\

\end{longtable}

\endgroup

\end{appendices}



\begin{thebibliography}{70}
\ifx \bisbn   \undefined \def \bisbn  #1{ISBN #1}\fi
\ifx \binits  \undefined \def \binits#1{#1}\fi
\ifx \bauthor  \undefined \def \bauthor#1{#1}\fi
\ifx \batitle  \undefined \def \batitle#1{#1}\fi
\ifx \bjtitle  \undefined \def \bjtitle#1{#1}\fi
\ifx \bvolume  \undefined \def \bvolume#1{\textbf{#1}}\fi
\ifx \byear  \undefined \def \byear#1{#1}\fi
\ifx \bissue  \undefined \def \bissue#1{#1}\fi
\ifx \bfpage  \undefined \def \bfpage#1{#1}\fi
\ifx \blpage  \undefined \def \blpage #1{#1}\fi
\ifx \burl  \undefined \def \burl#1{\textsf{#1}}\fi
\ifx \doiurl  \undefined \def \doiurl#1{\url{https://doi.org/#1}}\fi
\ifx \betal  \undefined \def \betal{\textit{et al.}}\fi
\ifx \binstitute  \undefined \def \binstitute#1{#1}\fi
\ifx \binstitutionaled  \undefined \def \binstitutionaled#1{#1}\fi
\ifx \bctitle  \undefined \def \bctitle#1{#1}\fi
\ifx \beditor  \undefined \def \beditor#1{#1}\fi
\ifx \bpublisher  \undefined \def \bpublisher#1{#1}\fi
\ifx \bbtitle  \undefined \def \bbtitle#1{#1}\fi
\ifx \bedition  \undefined \def \bedition#1{#1}\fi
\ifx \bseriesno  \undefined \def \bseriesno#1{#1}\fi
\ifx \blocation  \undefined \def \blocation#1{#1}\fi
\ifx \bsertitle  \undefined \def \bsertitle#1{#1}\fi
\ifx \bsnm \undefined \def \bsnm#1{#1}\fi
\ifx \bsuffix \undefined \def \bsuffix#1{#1}\fi
\ifx \bparticle \undefined \def \bparticle#1{#1}\fi
\ifx \barticle \undefined \def \barticle#1{#1}\fi
\bibcommenthead
\ifx \bconfdate \undefined \def \bconfdate #1{#1}\fi
\ifx \botherref \undefined \def \botherref #1{#1}\fi
\ifx \url \undefined \def \url#1{\textsf{#1}}\fi
\ifx \bchapter \undefined \def \bchapter#1{#1}\fi
\ifx \bbook \undefined \def \bbook#1{#1}\fi
\ifx \bcomment \undefined \def \bcomment#1{#1}\fi
\ifx \oauthor \undefined \def \oauthor#1{#1}\fi
\ifx \citeauthoryear \undefined \def \citeauthoryear#1{#1}\fi
\ifx \endbibitem  \undefined \def \endbibitem {}\fi
\ifx \bconflocation  \undefined \def \bconflocation#1{#1}\fi
\ifx \arxivurl  \undefined \def \arxivurl#1{\textsf{#1}}\fi
\csname PreBibitemsHook\endcsname

\bibitem[\protect\citeauthoryear{Carmona}{2010}]{carmona2010contemporary}
\begin{barticle}
\bauthor{\bsnm{Carmona}, \binits{M.}}:
\batitle{Contemporary public space: Critique and classification, part one: Critique}.
\bjtitle{Journal of urban design}
\bvolume{15}(\bissue{1}),
\bfpage{123}--\blpage{148}
(\byear{2010})
\end{barticle}
\endbibitem

\bibitem[\protect\citeauthoryear{Luca et~al.}{2024}]{luca2024towards}
\begin{botherref}
\oauthor{\bsnm{Luca}, \binits{M.}},
\oauthor{\bsnm{Lepri}, \binits{B.}},
\oauthor{\bsnm{Gallotti}, \binits{R.}},
\oauthor{\bsnm{Paolazzi}, \binits{S.}},
\oauthor{\bsnm{Bigi}, \binits{M.}},
\oauthor{\bsnm{Pistore}, \binits{M.}}:
Towards civic digital twins: Co-design the citizen-centric future of bologna.
arXiv preprint arXiv:2412.06328
(2024)
\end{botherref}
\endbibitem

\bibitem[\protect\citeauthoryear{Ahn et~al.}{2025}]{Ahn2025}
\begin{barticle}
\bauthor{\bsnm{Ahn}, \binits{C.}},
\bauthor{\bsnm{Lotfi-Jam}, \binits{F.}},
\bauthor{\bsnm{Graham}, \binits{C.}},
\bauthor{\bsnm{Bunnell}, \binits{T.}},
\bauthor{\bsnm{Marvin}, \binits{S.}}:
\batitle{Critical urban informatics for urban digital twin models}.
\bjtitle{Nature Cities}
(\byear{2025})
\doiurl{10.1038/s44284-024-00171-0}
\end{barticle}
\endbibitem

\bibitem[\protect\citeauthoryear{Lee et~al.}{2024}]{lee2024co}
\begin{barticle}
\bauthor{\bsnm{Lee}, \binits{D.}},
\bauthor{\bsnm{Feiertag}, \binits{P.}},
\bauthor{\bsnm{Unger}, \binits{L.}}:
\batitle{Co-production, co-creation or co-design of public space? a systematic review}.
\bjtitle{Cities}
\bvolume{154},
\bfpage{105372}
(\byear{2024})
\end{barticle}
\endbibitem

\bibitem[\protect\citeauthoryear{Salesses et~al.}{2013}]{salesses2013collaborative}
\begin{barticle}
\bauthor{\bsnm{Salesses}, \binits{P.}},
\bauthor{\bsnm{Schechtner}, \binits{K.}},
\bauthor{\bsnm{Hidalgo}, \binits{C.A.}}:
\batitle{The collaborative image of the city: mapping the inequality of urban perception}.
\bjtitle{PloS one}
\bvolume{8}(\bissue{7}),
\bfpage{68400}
(\byear{2013})
\end{barticle}
\endbibitem

\bibitem[\protect\citeauthoryear{Zamanifard et~al.}{2019}]{zamanifard2019measuring}
\begin{barticle}
\bauthor{\bsnm{Zamanifard}, \binits{H.}},
\bauthor{\bsnm{Alizadeh}, \binits{T.}},
\bauthor{\bsnm{Bosman}, \binits{C.}},
\bauthor{\bsnm{Coiacetto}, \binits{E.}}:
\batitle{Measuring experiential qualities of urban public spaces: users’ perspective}.
\bjtitle{Journal of Urban Design}
\bvolume{24}(\bissue{3}),
\bfpage{340}--\blpage{364}
(\byear{2019})
\end{barticle}
\endbibitem

\bibitem[\protect\citeauthoryear{Mehta}{2014}]{mehta2014evaluating}
\begin{barticle}
\bauthor{\bsnm{Mehta}, \binits{V.}}:
\batitle{Evaluating public space}.
\bjtitle{Journal of Urban design}
\bvolume{19}(\bissue{1}),
\bfpage{53}--\blpage{88}
(\byear{2014})
\end{barticle}
\endbibitem

\bibitem[\protect\citeauthoryear{Gu et~al.}{2025}]{gu2025designing}
\begin{barticle}
\bauthor{\bsnm{Gu}, \binits{Y.}},
\bauthor{\bsnm{Quintana}, \binits{M.}},
\bauthor{\bsnm{Liang}, \binits{X.}},
\bauthor{\bsnm{Ito}, \binits{K.}},
\bauthor{\bsnm{Yap}, \binits{W.}},
\bauthor{\bsnm{Biljecki}, \binits{F.}}:
\batitle{Designing effective image-based surveys for urban visual perception}.
\bjtitle{Landscape and Urban Planning}
\bvolume{260},
\bfpage{105368}
(\byear{2025})
\end{barticle}
\endbibitem

\bibitem[\protect\citeauthoryear{Westwood}{2025}]{westwood2025potential}
\begin{barticle}
\bauthor{\bsnm{Westwood}, \binits{S.J.}}:
\batitle{The potential existential threat of large language models to online survey research}.
\bjtitle{Proceedings of the National Academy of Sciences}
\bvolume{122}(\bissue{47}),
\bfpage{2518075122}
(\byear{2025})
\end{barticle}
\endbibitem

\bibitem[\protect\citeauthoryear{Simini et~al.}{2021}]{simini2021deep}
\begin{barticle}
\bauthor{\bsnm{Simini}, \binits{F.}},
\bauthor{\bsnm{Barlacchi}, \binits{G.}},
\bauthor{\bsnm{Luca}, \binits{M.}},
\bauthor{\bsnm{Pappalardo}, \binits{L.}}:
\batitle{A deep gravity model for mobility flows generation}.
\bjtitle{Nature communications}
\bvolume{12}(\bissue{1}),
\bfpage{6576}
(\byear{2021})
\end{barticle}
\endbibitem

\bibitem[\protect\citeauthoryear{Bahrami et~al.}{2022}]{bahrami2022using}
\begin{barticle}
\bauthor{\bsnm{Bahrami}, \binits{M.}},
\bauthor{\bsnm{Xu}, \binits{Y.}},
\bauthor{\bsnm{Tweed}, \binits{M.}},
\bauthor{\bsnm{Bozkaya}, \binits{B.}}, \betal:
\batitle{Using gravity model to make store closing decisions: A data driven approach}.
\bjtitle{Expert systems with applications}
\bvolume{205},
\bfpage{117703}
(\byear{2022})
\end{barticle}
\endbibitem

\bibitem[\protect\citeauthoryear{Suhara et~al.}{2021}]{suhara2021validating}
\begin{barticle}
\bauthor{\bsnm{Suhara}, \binits{Y.}},
\bauthor{\bsnm{Bahrami}, \binits{M.}},
\bauthor{\bsnm{Bozkaya}, \binits{B.}},
\bauthor{\bsnm{Pentland}, \binits{A.S.}}:
\batitle{Validating gravity-based market share models using large-scale transactional data}.
\bjtitle{Big Data}
\bvolume{9}(\bissue{3}),
\bfpage{188}--\blpage{202}
(\byear{2021})
\end{barticle}
\endbibitem

\bibitem[\protect\citeauthoryear{Gebru et~al.}{2017}]{Gebru_2017}
\begin{barticle}
\bauthor{\bsnm{Gebru}, \binits{T.}},
\bauthor{\bsnm{Krause}, \binits{J.}},
\bauthor{\bsnm{Wang}, \binits{Y.}},
\bauthor{\bsnm{Chen}, \binits{D.}},
\bauthor{\bsnm{Deng}, \binits{J.}},
\bauthor{\bsnm{Aiden}, \binits{E.L.}},
\bauthor{\bsnm{Fei-Fei}, \binits{L.}}:
\batitle{Using deep learning and google street view to estimate the demographic makeup of neighborhoods across the united states}.
\bjtitle{Proceedings of the National Academy of Sciences}
\bvolume{114},
\bfpage{13108}--\blpage{13113}
(\byear{2017})
\end{barticle}
\endbibitem

\bibitem[\protect\citeauthoryear{Helbich et~al.}{2019}]{HELBICH2019107}
\begin{barticle}
\bauthor{\bsnm{Helbich}, \binits{M.}},
\bauthor{\bsnm{Yao}, \binits{Y.}},
\bauthor{\bsnm{Liu}, \binits{Y.}},
\bauthor{\bsnm{Zhang}, \binits{J.}},
\bauthor{\bsnm{Liu}, \binits{P.}},
\bauthor{\bsnm{Wang}, \binits{R.}}:
\batitle{Using deep learning to examine street view green and blue spaces and their associations with geriatric depression in beijing, china}.
\bjtitle{Environment International}
\bvolume{126},
\bfpage{107}--\blpage{117}
(\byear{2019})
\doiurl{10.1016/j.envint.2019.02.013}
\end{barticle}
\endbibitem

\bibitem[\protect\citeauthoryear{Ho et~al.}{2024}]{ho2024elev}
\begin{botherref}
\oauthor{\bsnm{Ho}, \binits{Y.-H.}},
\oauthor{\bsnm{Li}, \binits{L.}},
\oauthor{\bsnm{Mostafavi}, \binits{A.}}:
ELEV-VISION-SAM: Integrated Vision Language and Foundation Model for Automated Estimation of Building Lowest Floor Elevation
(2024).
\url{https://arxiv.org/abs/2404.12606}
\end{botherref}
\endbibitem

\bibitem[\protect\citeauthoryear{Naik et~al.}{2014}]{naik2014streetscore}
\begin{bchapter}
\bauthor{\bsnm{Naik}, \binits{N.}},
\bauthor{\bsnm{Philipoom}, \binits{J.}},
\bauthor{\bsnm{Raskar}, \binits{R.}},
\bauthor{\bsnm{Hidalgo}, \binits{C.}}:
\bctitle{Streetscore-predicting the perceived safety of one million streetscapes}.
In: \bbtitle{Proceedings of the IEEE Conference on Computer Vision and Pattern Recognition Workshops},
pp. \bfpage{779}--\blpage{785}
(\byear{2014})
\end{bchapter}
\endbibitem

\bibitem[\protect\citeauthoryear{De~Nadai et~al.}{2016}]{de2016safer}
\begin{bchapter}
\bauthor{\bsnm{De~Nadai}, \binits{M.}},
\bauthor{\bsnm{Vieriu}, \binits{R.L.}},
\bauthor{\bsnm{Zen}, \binits{G.}},
\bauthor{\bsnm{Dragicevic}, \binits{S.}},
\bauthor{\bsnm{Naik}, \binits{N.}},
\bauthor{\bsnm{Caraviello}, \binits{M.}},
\bauthor{\bsnm{Hidalgo}, \binits{C.A.}},
\bauthor{\bsnm{Sebe}, \binits{N.}},
\bauthor{\bsnm{Lepri}, \binits{B.}}:
\bctitle{Are safer looking neighborhoods more lively? a multimodal investigation into urban life}.
In: \bbtitle{Proceedings of the 24th ACM International Conference on Multimedia},
pp. \bfpage{1127}--\blpage{1135}
(\byear{2016})
\end{bchapter}
\endbibitem

\bibitem[\protect\citeauthoryear{Dubey et~al.}{2016}]{dubey2016deep}
\begin{bchapter}
\bauthor{\bsnm{Dubey}, \binits{A.}},
\bauthor{\bsnm{Naik}, \binits{N.}},
\bauthor{\bsnm{Parikh}, \binits{D.}},
\bauthor{\bsnm{Raskar}, \binits{R.}},
\bauthor{\bsnm{Hidalgo}, \binits{C.A.}}:
\bctitle{Deep learning the city: Quantifying urban perception at a global scale}.
In: \bbtitle{Computer Vision--ECCV 2016: 14th European Conference, Amsterdam, The Netherlands, October 11--14, 2016, Proceedings, Part I 14},
pp. \bfpage{196}--\blpage{212}
(\byear{2016}).
\bcomment{Springer}
\end{bchapter}
\endbibitem

\bibitem[\protect\citeauthoryear{Redi et~al.}{2018}]{redi2018spirit}
\begin{barticle}
\bauthor{\bsnm{Redi}, \binits{M.}},
\bauthor{\bsnm{Aiello}, \binits{L.M.}},
\bauthor{\bsnm{Schifanella}, \binits{R.}},
\bauthor{\bsnm{Quercia}, \binits{D.}}:
\batitle{The spirit of the city: Using social media to capture neighborhood ambiance}.
\bjtitle{Proceedings of the ACM on human-computer interaction}
\bvolume{2}(\bissue{CSCW}),
\bfpage{1}--\blpage{18}
(\byear{2018})
\end{barticle}
\endbibitem

\bibitem[\protect\citeauthoryear{Quercia et~al.}{2014}]{quercia2014aesthetic}
\begin{bchapter}
\bauthor{\bsnm{Quercia}, \binits{D.}},
\bauthor{\bsnm{O'Hare}, \binits{N.K.}},
\bauthor{\bsnm{Cramer}, \binits{H.}}:
\bctitle{Aesthetic capital: what makes london look beautiful, quiet, and happy?}
In: \bbtitle{Proceedings of the 17th ACM Conference on Computer Supported Cooperative Work \& Social Computing},
pp. \bfpage{945}--\blpage{955}
(\byear{2014})
\end{bchapter}
\endbibitem

\bibitem[\protect\citeauthoryear{Hou et~al.}{2024}]{hou2024global}
\begin{barticle}
\bauthor{\bsnm{Hou}, \binits{Y.}},
\bauthor{\bsnm{Quintana}, \binits{M.}},
\bauthor{\bsnm{Khomiakov}, \binits{M.}},
\bauthor{\bsnm{Yap}, \binits{W.}},
\bauthor{\bsnm{Ouyang}, \binits{J.}},
\bauthor{\bsnm{Ito}, \binits{K.}},
\bauthor{\bsnm{Wang}, \binits{Z.}},
\bauthor{\bsnm{Zhao}, \binits{T.}},
\bauthor{\bsnm{Biljecki}, \binits{F.}}:
\batitle{Global streetscapes—a comprehensive dataset of 10 million street-level images across 688 cities for urban science and analytics}.
\bjtitle{ISPRS Journal of Photogrammetry and Remote Sensing}
\bvolume{215},
\bfpage{216}--\blpage{238}
(\byear{2024})
\end{barticle}
\endbibitem

\bibitem[\protect\citeauthoryear{Gustafsod}{1998}]{gustafsod1998gender}
\begin{barticle}
\bauthor{\bsnm{Gustafsod}, \binits{P.E.}}:
\batitle{Gender differences in risk perception: Theoretical and methodological erspectives}.
\bjtitle{Risk analysis}
\bvolume{18}(\bissue{6}),
\bfpage{805}--\blpage{811}
(\byear{1998})
\end{barticle}
\endbibitem

\bibitem[\protect\citeauthoryear{Quintana et~al.}{2025}]{quintana2025global}
\begin{botherref}
\oauthor{\bsnm{Quintana}, \binits{M.}},
\oauthor{\bsnm{Gu}, \binits{Y.}},
\oauthor{\bsnm{Liang}, \binits{X.}},
\oauthor{\bsnm{Hou}, \binits{Y.}},
\oauthor{\bsnm{Ito}, \binits{K.}},
\oauthor{\bsnm{Zhu}, \binits{Y.}},
\oauthor{\bsnm{Abdelrahman}, \binits{M.}},
\oauthor{\bsnm{Biljecki}, \binits{F.}}:
Global urban visual perception varies across demographics and personalities.
Nature Cities,
1--15
(2025)
\end{botherref}
\endbibitem

\bibitem[\protect\citeauthoryear{Kang et~al.}{2023}]{kang2023assessing}
\begin{barticle}
\bauthor{\bsnm{Kang}, \binits{Y.}},
\bauthor{\bsnm{Abraham}, \binits{J.}},
\bauthor{\bsnm{Ceccato}, \binits{V.}},
\bauthor{\bsnm{Duarte}, \binits{F.}},
\bauthor{\bsnm{Gao}, \binits{S.}},
\bauthor{\bsnm{Ljungqvist}, \binits{L.}},
\bauthor{\bsnm{Zhang}, \binits{F.}},
\bauthor{\bsnm{N{\"a}sman}, \binits{P.}},
\bauthor{\bsnm{Ratti}, \binits{C.}}:
\batitle{Assessing differences in safety perceptions using geoai and survey across neighbourhoods in stockholm, sweden}.
\bjtitle{Landscape and Urban Planning}
\bvolume{236},
\bfpage{104768}
(\byear{2023})
\end{barticle}
\endbibitem

\bibitem[\protect\citeauthoryear{Pereira and Rebelo}{2024}]{pereira2024women}
\begin{barticle}
\bauthor{\bsnm{Pereira}, \binits{A.}},
\bauthor{\bsnm{Rebelo}, \binits{E.M.}}:
\batitle{Women in public spaces: Perceptions and initiatives to promote gender equality}.
\bjtitle{Cities}
\bvolume{154},
\bfpage{105346}
(\byear{2024})
\end{barticle}
\endbibitem

\bibitem[\protect\citeauthoryear{Navarrete-Hernandez et~al.}{2021}]{navarrete2021building}
\begin{barticle}
\bauthor{\bsnm{Navarrete-Hernandez}, \binits{P.}},
\bauthor{\bsnm{Vetro}, \binits{A.}},
\bauthor{\bsnm{Concha}, \binits{P.}}:
\batitle{Building safer public spaces: Exploring gender difference in the perception of safety in public space through urban design interventions}.
\bjtitle{Landscape and Urban Planning}
\bvolume{214},
\bfpage{104180}
(\byear{2021})
\end{barticle}
\endbibitem

\bibitem[\protect\citeauthoryear{Van~Hoof et~al.}{2021}]{van2021ten}
\begin{barticle}
\bauthor{\bsnm{Van~Hoof}, \binits{J.}},
\bauthor{\bsnm{Marston}, \binits{H.R.}},
\bauthor{\bsnm{Kazak}, \binits{J.K.}},
\bauthor{\bsnm{Buffel}, \binits{T.}}:
\batitle{Ten questions concerning age-friendly cities and communities and the built environment}.
\bjtitle{Building and environment}
\bvolume{199},
\bfpage{107922}
(\byear{2021})
\end{barticle}
\endbibitem

\bibitem[\protect\citeauthoryear{Figueiredo et~al.}{2023}]{figueiredo2023older}
\begin{barticle}
\bauthor{\bsnm{Figueiredo}, \binits{M.}},
\bauthor{\bsnm{Eloy}, \binits{S.}},
\bauthor{\bsnm{Marques}, \binits{S.}},
\bauthor{\bsnm{Dias}, \binits{L.}}:
\batitle{Older people perceptions on the built environment: a scoping review}.
\bjtitle{Applied ergonomics}
\bvolume{108},
\bfpage{103951}
(\byear{2023})
\end{barticle}
\endbibitem

\bibitem[\protect\citeauthoryear{Raudsepp}{2001}]{raudsepp2001some}
\begin{barticle}
\bauthor{\bsnm{Raudsepp}, \binits{M.}}:
\batitle{Some socio-demographic and socio-psychological predictors of environmentalism}.
\bjtitle{Trames}
\bvolume{5}(\bissue{55/50}),
\bfpage{355}--\blpage{367}
(\byear{2001})
\end{barticle}
\endbibitem

\bibitem[\protect\citeauthoryear{Christian et~al.}{2021}]{christian2021households}
\begin{barticle}
\bauthor{\bsnm{Christian}, \binits{A.K.}},
\bauthor{\bsnm{Dovie}, \binits{B.D.}},
\bauthor{\bsnm{Akpalu}, \binits{W.}},
\bauthor{\bsnm{Codjoe}, \binits{S.N.A.}}:
\batitle{Households' socio-demographic characteristics, perceived and underestimated vulnerability to floods and related risk reduction in ghana}.
\bjtitle{Urban Climate}
\bvolume{35},
\bfpage{100759}
(\byear{2021})
\end{barticle}
\endbibitem

\bibitem[\protect\citeauthoryear{Sharma and Gursoy}{2015}]{sharma2015examination}
\begin{barticle}
\bauthor{\bsnm{Sharma}, \binits{B.}},
\bauthor{\bsnm{Gursoy}, \binits{D.}}:
\batitle{An examination of changes in residents' perceptions of tourism impacts over time: The impact of residents' socio-demographic characteristics}.
\bjtitle{Asia Pacific Journal of Tourism Research}
\bvolume{20}(\bissue{12}),
\bfpage{1332}--\blpage{1352}
(\byear{2015})
\end{barticle}
\endbibitem

\bibitem[\protect\citeauthoryear{Finucane et~al.}{2013}]{finucane2013gender}
\begin{bchapter}
\bauthor{\bsnm{Finucane}, \binits{M.L.}},
\bauthor{\bsnm{Slovic}, \binits{P.}},
\bauthor{\bsnm{Mertz}, \binits{C.K.}},
\bauthor{\bsnm{Flynn}, \binits{J.}},
\bauthor{\bsnm{Satterfield}, \binits{T.}}:
\bctitle{Gender, race and perceived risk: The ‘white-male'effect}.
In: \bbtitle{The Feeling of Risk},
pp. \bfpage{125}--\blpage{139}.
\bpublisher{Routledge}, \blocation{???}
(\byear{2013})
\end{bchapter}
\endbibitem

\bibitem[\protect\citeauthoryear{Beneduce et~al.}{2026}]{beneduce_2026_20287027}
\begin{botherref}
\oauthor{\bsnm{Beneduce}, \binits{C.}},
\oauthor{\bsnm{Luca}, \binits{M.}},
\oauthor{\bsnm{Lepri}, \binits{B.}}:
AI's Blind Spots: Geographic Knowledge and Diversity Deficit in Generated Urban Scenario.
Zenodo
(2026).
\doiurl{10.48550/arXiv.2506.16898} .
\url{https://doi.org/10.48550/arXiv.2506.16898}
\end{botherref}
\endbibitem

\bibitem[\protect\citeauthoryear{Luca et~al.}{2025}]{luca2025llm}
\begin{botherref}
\oauthor{\bsnm{Luca}, \binits{M.}},
\oauthor{\bsnm{Beneduce}, \binits{C.}},
\oauthor{\bsnm{Lepri}, \binits{B.}},
\oauthor{\bsnm{Staiano}, \binits{J.}}:
The llm wears prada: Analysing gender bias and stereotypes through online shopping data.
arXiv preprint arXiv:2504.01951
(2025)
\end{botherref}
\endbibitem

\bibitem[\protect\citeauthoryear{Wang et~al.}{2025}]{wang2025uncovering}
\begin{barticle}
\bauthor{\bsnm{Wang}, \binits{C.}},
\bauthor{\bsnm{Tang}, \binits{H.}},
\bauthor{\bsnm{Yang}, \binits{X.}},
\bauthor{\bsnm{Xie}, \binits{Y.}},
\bauthor{\bsnm{Suh}, \binits{J.}},
\bauthor{\bsnm{Sitaram}, \binits{S.}},
\bauthor{\bsnm{Huang}, \binits{J.}},
\bauthor{\bsnm{Xie}, \binits{Y.}},
\bauthor{\bsnm{Zhao}, \binits{P.}},
\bauthor{\bsnm{Gong}, \binits{Z.}}, \betal:
\batitle{Uncovering inequalities in new knowledge learning by large language models across different languages}.
\bjtitle{Proceedings of the National Academy of Sciences}
\bvolume{122}(\bissue{51}),
\bfpage{2514626122}
(\byear{2025})
\end{barticle}
\endbibitem

\bibitem[\protect\citeauthoryear{Peng et~al.}{2025}]{peng2025vision}
\begin{botherref}
\oauthor{\bsnm{Peng}, \binits{Z.}},
\oauthor{\bsnm{Lu}, \binits{W.}},
\oauthor{\bsnm{An}, \binits{H.}},
\oauthor{\bsnm{Xia}, \binits{X.}},
\oauthor{\bsnm{Zhang}, \binits{Y.}},
\oauthor{\bsnm{Xue}, \binits{F.}},
\oauthor{\bsnm{Chen}, \binits{J.}}:
Vision language model (vlm)-enabled street view analytics: a systematic literature review.
Engineering, Construction and Architectural Management,
1--19
(2025)
\end{botherref}
\endbibitem

\bibitem[\protect\citeauthoryear{Wu et~al.}{2025}]{wu2025exploring}
\begin{botherref}
\oauthor{\bsnm{Wu}, \binits{S.}},
\oauthor{\bsnm{Wang}, \binits{W.}},
\oauthor{\bsnm{Zhang}, \binits{C.}},
\oauthor{\bsnm{Zeng}, \binits{Q.}},
\oauthor{\bsnm{Wang}, \binits{H.}},
\oauthor{\bsnm{Lin}, \binits{J.}},
\oauthor{\bsnm{Li}, \binits{S.}}:
Exploring multimodal large language models’ potential in simulating human perception of urban cycling environments: A street-view perspective.
Information Geography,
100037
(2025)
\end{botherref}
\endbibitem

\bibitem[\protect\citeauthoryear{Hou et~al.}{2025}]{hou2025transferring}
\begin{botherref}
\oauthor{\bsnm{Hou}, \binits{C.}},
\oauthor{\bsnm{Zhang}, \binits{F.}},
\oauthor{\bsnm{Kang}, \binits{Y.}},
\oauthor{\bsnm{Fan}, \binits{Z.}},
\oauthor{\bsnm{Li}, \binits{S.}}:
Transferring population group knowledge from multimodal large language model to small model: using urban safety perception evaluation as case study
(2025)
\end{botherref}
\endbibitem

\bibitem[\protect\citeauthoryear{Zhang et~al.}{2025}]{zhang2025urban}
\begin{barticle}
\bauthor{\bsnm{Zhang}, \binits{J.}},
\bauthor{\bsnm{Li}, \binits{Y.}},
\bauthor{\bsnm{Fukuda}, \binits{T.}},
\bauthor{\bsnm{Wang}, \binits{B.}}:
\batitle{Urban safety perception assessments via integrating multimodal large language models with street view images}.
\bjtitle{Cities}
\bvolume{165},
\bfpage{106122}
(\byear{2025})
\end{barticle}
\endbibitem

\bibitem[\protect\citeauthoryear{Malekzadeh et~al.}{2024}]{malekzadeh2024urban}
\begin{botherref}
\oauthor{\bsnm{Malekzadeh}, \binits{M.}},
\oauthor{\bsnm{Willberg}, \binits{E.}},
\oauthor{\bsnm{Torkko}, \binits{J.}},
\oauthor{\bsnm{Toivonen}, \binits{T.}}:
Urban visual appeal according to chatgpt: contrasting ai and human insights.
arXiv preprint arXiv:2407.14268
(2024)
\end{botherref}
\endbibitem

\bibitem[\protect\citeauthoryear{H{\"a}m{\"a}l{\"a}inen et~al.}{2023}]{hamalainen2023evaluating}
\begin{bchapter}
\bauthor{\bsnm{H{\"a}m{\"a}l{\"a}inen}, \binits{P.}},
\bauthor{\bsnm{Tavast}, \binits{M.}},
\bauthor{\bsnm{Kunnari}, \binits{A.}}:
\bctitle{Evaluating large language models in generating synthetic hci research data: a case study}.
In: \bbtitle{Proceedings of the 2023 CHI Conference on Human Factors in Computing Systems},
pp. \bfpage{1}--\blpage{19}
(\byear{2023})
\end{bchapter}
\endbibitem

\bibitem[\protect\citeauthoryear{Park et~al.}{2023}]{park2023generative}
\begin{bchapter}
\bauthor{\bsnm{Park}, \binits{J.S.}},
\bauthor{\bsnm{O'Brien}, \binits{J.}},
\bauthor{\bsnm{Cai}, \binits{C.J.}},
\bauthor{\bsnm{Morris}, \binits{M.R.}},
\bauthor{\bsnm{Liang}, \binits{P.}},
\bauthor{\bsnm{Bernstein}, \binits{M.S.}}:
\bctitle{Generative agents: Interactive simulacra of human behavior}.
In: \bbtitle{Proceedings of the 36th Annual Acm Symposium on User Interface Software and Technology},
pp. \bfpage{1}--\blpage{22}
(\byear{2023})
\end{bchapter}
\endbibitem

\bibitem[\protect\citeauthoryear{Fleisig et~al.}{2023}]{fleisig2023majority}
\begin{botherref}
\oauthor{\bsnm{Fleisig}, \binits{E.}},
\oauthor{\bsnm{Abebe}, \binits{R.}},
\oauthor{\bsnm{Klein}, \binits{D.}}:
When the majority is wrong: Modeling annotator disagreement for subjective tasks.
arXiv preprint arXiv:2305.06626
(2023)
\end{botherref}
\endbibitem

\bibitem[\protect\citeauthoryear{Horton}{2023}]{horton2023large}
\begin{botherref}
\oauthor{\bsnm{Horton}, \binits{J.J.}}:
Large language models as simulated economic agents: What can we learn from homo silicus?
Technical report,
National Bureau of Economic Research
(2023)
\end{botherref}
\endbibitem

\bibitem[\protect\citeauthoryear{Argyle et~al.}{2023}]{argyle2023out}
\begin{barticle}
\bauthor{\bsnm{Argyle}, \binits{L.P.}},
\bauthor{\bsnm{Busby}, \binits{E.C.}},
\bauthor{\bsnm{Fulda}, \binits{N.}},
\bauthor{\bsnm{Gubler}, \binits{J.R.}},
\bauthor{\bsnm{Rytting}, \binits{C.}},
\bauthor{\bsnm{Wingate}, \binits{D.}}:
\batitle{Out of one, many: Using language models to simulate human samples}.
\bjtitle{Political Analysis}
\bvolume{31}(\bissue{3}),
\bfpage{337}--\blpage{351}
(\byear{2023})
\end{barticle}
\endbibitem

\bibitem[\protect\citeauthoryear{T{\"o}rnberg et~al.}{2023}]{tornberg2023simulating}
\begin{botherref}
\oauthor{\bsnm{T{\"o}rnberg}, \binits{P.}},
\oauthor{\bsnm{Valeeva}, \binits{D.}},
\oauthor{\bsnm{Uitermark}, \binits{J.}},
\oauthor{\bsnm{Bail}, \binits{C.}}:
Simulating social media using large language models to evaluate alternative news feed algorithms.
arXiv preprint arXiv:2310.05984
(2023)
\end{botherref}
\endbibitem

\bibitem[\protect\citeauthoryear{Tosato et~al.}{2026}]{tosato2026persistent}
\begin{bchapter}
\bauthor{\bsnm{Tosato}, \binits{T.}},
\bauthor{\bsnm{Helbling}, \binits{S.}},
\bauthor{\bsnm{Mantilla-Ramos}, \binits{Y.-J.}},
\bauthor{\bsnm{Hegazy}, \binits{M.}},
\bauthor{\bsnm{Tosato}, \binits{A.}},
\bauthor{\bsnm{Lemay}, \binits{D.J.}},
\bauthor{\bsnm{Rish}, \binits{I.}},
\bauthor{\bsnm{Dumas}, \binits{G.}}:
\bctitle{Persistent instability in llm’s personality measurements: Effects of scale, reasoning, and conversation history}.
In: \bbtitle{Proceedings of the AAAI Conference on Artificial Intelligence},
vol. \bseriesno{40},
pp. \bfpage{37961}--\blpage{37969}
(\byear{2026})
\end{bchapter}
\endbibitem

\bibitem[\protect\citeauthoryear{Liu et~al.}{2024}]{liu2024improved}
\begin{bchapter}
\bauthor{\bsnm{Liu}, \binits{H.}},
\bauthor{\bsnm{Li}, \binits{C.}},
\bauthor{\bsnm{Li}, \binits{Y.}},
\bauthor{\bsnm{Lee}, \binits{Y.J.}}:
\bctitle{Improved baselines with visual instruction tuning}.
In: \bbtitle{Proceedings of the IEEE/CVF Conference on Computer Vision and Pattern Recognition},
pp. \bfpage{26296}--\blpage{26306}
(\byear{2024})
\end{bchapter}
\endbibitem

\bibitem[\protect\citeauthoryear{Comanici et~al.}{2025}]{comanici2025gemini}
\begin{botherref}
\oauthor{\bsnm{Comanici}, \binits{G.}},
\oauthor{\bsnm{Bieber}, \binits{E.}},
\oauthor{\bsnm{Schaekermann}, \binits{M.}},
\oauthor{\bsnm{Pasupat}, \binits{I.}},
\oauthor{\bsnm{Sachdeva}, \binits{N.}},
\oauthor{\bsnm{Dhillon}, \binits{I.}},
\oauthor{\bsnm{Blistein}, \binits{M.}},
\oauthor{\bsnm{Ram}, \binits{O.}},
\oauthor{\bsnm{Zhang}, \binits{D.}},
\oauthor{\bsnm{Rosen}, \binits{E.}}, et al.:
Gemini 2.5: Pushing the frontier with advanced reasoning, multimodality, long context, and next generation agentic capabilities.
arXiv preprint arXiv:2507.06261
(2025)
\end{botherref}
\endbibitem

\bibitem[\protect\citeauthoryear{Bai et~al.}{2025}]{bai2025qwen3}
\begin{botherref}
\oauthor{\bsnm{Bai}, \binits{S.}},
\oauthor{\bsnm{Cai}, \binits{Y.}},
\oauthor{\bsnm{Chen}, \binits{R.}},
\oauthor{\bsnm{Chen}, \binits{K.}},
\oauthor{\bsnm{Chen}, \binits{X.}},
\oauthor{\bsnm{Cheng}, \binits{Z.}},
\oauthor{\bsnm{Deng}, \binits{L.}},
\oauthor{\bsnm{Ding}, \binits{W.}},
\oauthor{\bsnm{Gao}, \binits{C.}},
\oauthor{\bsnm{Ge}, \binits{C.}}, et al.:
Qwen3-vl technical report.
arXiv preprint arXiv:2511.21631
(2025)
\end{botherref}
\endbibitem

\bibitem[\protect\citeauthoryear{Singh et~al.}{2025}]{singh2025openai}
\begin{botherref}
\oauthor{\bsnm{Singh}, \binits{A.}},
\oauthor{\bsnm{Fry}, \binits{A.}},
\oauthor{\bsnm{Perelman}, \binits{A.}},
\oauthor{\bsnm{Tart}, \binits{A.}},
\oauthor{\bsnm{Ganesh}, \binits{A.}},
\oauthor{\bsnm{El-Kishky}, \binits{A.}},
\oauthor{\bsnm{McLaughlin}, \binits{A.}},
\oauthor{\bsnm{Low}, \binits{A.}},
\oauthor{\bsnm{Ostrow}, \binits{A.}},
\oauthor{\bsnm{Ananthram}, \binits{A.}}, et al.:
Openai gpt-5 system card.
arXiv preprint arXiv:2601.03267
(2025)
\end{botherref}
\endbibitem

\bibitem[\protect\citeauthoryear{Tao et~al.}{2024}]{tao2024cultural}
\begin{barticle}
\bauthor{\bsnm{Tao}, \binits{Y.}},
\bauthor{\bsnm{Viberg}, \binits{O.}},
\bauthor{\bsnm{Baker}, \binits{R.S.}},
\bauthor{\bsnm{Kizilcec}, \binits{R.F.}}:
\batitle{Cultural bias and cultural alignment of large language models}.
\bjtitle{PNAS nexus}
\bvolume{3}(\bissue{9}),
\bfpage{346}
(\byear{2024})
\end{barticle}
\endbibitem

\bibitem[\protect\citeauthoryear{Branco et~al.}{2016}]{branco2016survey}
\begin{barticle}
\bauthor{\bsnm{Branco}, \binits{P.}},
\bauthor{\bsnm{Torgo}, \binits{L.}},
\bauthor{\bsnm{Ribeiro}, \binits{R.P.}}:
\batitle{A survey of predictive modeling on imbalanced domains}.
\bjtitle{ACM computing surveys (CSUR)}
\bvolume{49}(\bissue{2}),
\bfpage{1}--\blpage{50}
(\byear{2016})
\end{barticle}
\endbibitem

\bibitem[\protect\citeauthoryear{D'Inc{\`a} et~al.}{2024}]{d2024openbias}
\begin{bchapter}
\bauthor{\bsnm{D'Inc{\`a}}, \binits{M.}},
\bauthor{\bsnm{Peruzzo}, \binits{E.}},
\bauthor{\bsnm{Mancini}, \binits{M.}},
\bauthor{\bsnm{Xu}, \binits{D.}},
\bauthor{\bsnm{Goel}, \binits{V.}},
\bauthor{\bsnm{Xu}, \binits{X.}},
\bauthor{\bsnm{Wang}, \binits{Z.}},
\bauthor{\bsnm{Shi}, \binits{H.}},
\bauthor{\bsnm{Sebe}, \binits{N.}}:
\bctitle{Openbias: Open-set bias detection in text-to-image generative models}.
In: \bbtitle{Proceedings of the IEEE/CVF Conference on Computer Vision and Pattern Recognition},
pp. \bfpage{12225}--\blpage{12235}
(\byear{2024})
\end{bchapter}
\endbibitem

\bibitem[\protect\citeauthoryear{Kelling and Wilson}{1982}]{kelling1982broken}
\begin{botherref}
\oauthor{\bsnm{Kelling}, \binits{G.L.}},
\oauthor{\bsnm{Wilson}, \binits{J.Q.}}:
Broken windows: The police and neighborhood safety.
The Atlantic
(1982)
\end{botherref}
\endbibitem

\bibitem[\protect\citeauthoryear{Jacobs}{1992}]{jacobs1992death}
\begin{bbook}
\bauthor{\bsnm{Jacobs}, \binits{J.}}:
\bbtitle{The Death and Life of Great American Cities}.
\bpublisher{Knopf Doubleday Publishing Group},
\blocation{New York}
(\byear{1992})
\end{bbook}
\endbibitem

\bibitem[\protect\citeauthoryear{Newman}{1972}]{newman1972defensible}
\begin{bbook}
\bauthor{\bsnm{Newman}, \binits{O.}}:
\bbtitle{Defensible Space: Crime Prevention Through Urban Design}.
\bsertitle{Architecture/Urban Affairs}.
\bpublisher{Macmillan},
\blocation{New York}
(\byear{1972})
\end{bbook}
\endbibitem

\bibitem[\protect\citeauthoryear{Shaw and McKay}{1942}]{shaw1942juvenile}
\begin{bbook}
\bauthor{\bsnm{Shaw}, \binits{C.R.}},
\bauthor{\bsnm{McKay}, \binits{H.D.}}:
\bbtitle{Juvenile Delinquency and Urban Areas: A Study of Rates of Delinquents in Relation to Differential Characteristics of Local Communities in American Cities}.
\bsertitle{Behavior Research Fund Monographs}.
\bpublisher{University of Chicago Press},
\blocation{Chicago}
(\byear{1942})
\end{bbook}
\endbibitem

\bibitem[\protect\citeauthoryear{Sampson and Groves}{1989}]{sampson1989community}
\begin{barticle}
\bauthor{\bsnm{Sampson}, \binits{R.J.}},
\bauthor{\bsnm{Groves}, \binits{W.B.}}:
\batitle{Community structure and crime: Testing social-disorganization theory}.
\bjtitle{American journal of sociology}
\bvolume{94}(\bissue{4}),
\bfpage{774}--\blpage{802}
(\byear{1989})
\end{barticle}
\endbibitem

\bibitem[\protect\citeauthoryear{Beneduce et~al.}{2025}]{beneduce2025large}
\begin{botherref}
\oauthor{\bsnm{Beneduce}, \binits{C.}},
\oauthor{\bsnm{Lepri}, \binits{B.}},
\oauthor{\bsnm{Luca}, \binits{M.}}:
Large language models are zero-shot next location predictors.
IEEE Access
(2025)
\end{botherref}
\endbibitem

\bibitem[\protect\citeauthoryear{Li et~al.}{2023}]{li2023frozenlanguagemodelhelps}
\begin{botherref}
\oauthor{\bsnm{Li}, \binits{J.}},
\oauthor{\bsnm{Liu}, \binits{C.}},
\oauthor{\bsnm{Cheng}, \binits{S.}},
\oauthor{\bsnm{Arcucci}, \binits{R.}},
\oauthor{\bsnm{Hong}, \binits{S.}}:
Frozen Language Model Helps ECG Zero-Shot Learning
(2023).
\url{https://arxiv.org/abs/2303.12311}
\end{botherref}
\endbibitem

\bibitem[\protect\citeauthoryear{Gruver et~al.}{2024}]{gruver2024largelanguagemodelszeroshot}
\begin{botherref}
\oauthor{\bsnm{Gruver}, \binits{N.}},
\oauthor{\bsnm{Finzi}, \binits{M.}},
\oauthor{\bsnm{Qiu}, \binits{S.}},
\oauthor{\bsnm{Wilson}, \binits{A.G.}}:
Large Language Models Are Zero-Shot Time Series Forecasters
(2024).
\url{https://arxiv.org/abs/2310.07820}
\end{botherref}
\endbibitem

\bibitem[\protect\citeauthoryear{Forester}{1999}]{forester1999deliberative}
\begin{bbook}
\bauthor{\bsnm{Forester}, \binits{J.}}:
\bbtitle{The Deliberative Practitioner: Encouraging Participatory Planning Processes}.
\bsertitle{Mit Press}.
\bpublisher{MIT Press},
\blocation{Cambridge, MA}
(\byear{1999})
\end{bbook}
\endbibitem

\bibitem[\protect\citeauthoryear{Healey}{1997}]{healey1997collaborative}
\begin{bbook}
\bauthor{\bsnm{Healey}, \binits{P.}}:
\bbtitle{Collaborative Planning: Shaping Places in Fragmented Societies}.
\bsertitle{Canadian Electronic Library: Books Collection}.
\bpublisher{UBC Press},
\blocation{Vancouver}
(\byear{1997})
\end{bbook}
\endbibitem

\bibitem[\protect\citeauthoryear{Wei et~al.}{2023}]{wei2023chainofthoughtpromptingelicitsreasoning}
\begin{botherref}
\oauthor{\bsnm{Wei}, \binits{J.}},
\oauthor{\bsnm{Wang}, \binits{X.}},
\oauthor{\bsnm{Schuurmans}, \binits{D.}},
\oauthor{\bsnm{Bosma}, \binits{M.}},
\oauthor{\bsnm{Ichter}, \binits{B.}},
\oauthor{\bsnm{Xia}, \binits{F.}},
\oauthor{\bsnm{Chi}, \binits{E.}},
\oauthor{\bsnm{Le}, \binits{Q.}},
\oauthor{\bsnm{Zhou}, \binits{D.}}:
Chain-of-Thought Prompting Elicits Reasoning in Large Language Models
(2023).
\url{https://arxiv.org/abs/2201.11903}
\end{botherref}
\endbibitem

\bibitem[\protect\citeauthoryear{Bommasani et~al.}{2022}]{bommasani2022opportunitiesrisksfoundationmodels}
\begin{botherref}
\oauthor{\bsnm{Bommasani}, \binits{R.}},
\oauthor{\bsnm{Hudson}, \binits{D.A.}},
\oauthor{\bsnm{Adeli}, \binits{E.}},
\oauthor{\bsnm{Altman}, \binits{R.}},
\oauthor{\bsnm{Arora}, \binits{S.}},
\oauthor{\bsnm{Arx}, \binits{S.}},
\oauthor{\bsnm{Bernstein}, \binits{M.S.}},
\oauthor{\bsnm{Bohg}, \binits{J.}},
\oauthor{\bsnm{Bosselut}, \binits{A.}},
\oauthor{\bsnm{Brunskill}, \binits{E.}},
\oauthor{\bsnm{Brynjolfsson}, \binits{E.}},
\oauthor{\bsnm{Buch}, \binits{S.}},
\oauthor{\bsnm{Card}, \binits{D.}},
\oauthor{\bsnm{Castellon}, \binits{R.}},
\oauthor{\bsnm{Chatterji}, \binits{N.}},
\oauthor{\bsnm{Chen}, \binits{A.}},
\oauthor{\bsnm{Creel}, \binits{K.}},
\oauthor{\bsnm{Davis}, \binits{J.Q.}},
\oauthor{\bsnm{Demszky}, \binits{D.}},
\oauthor{\bsnm{Donahue}, \binits{C.}},
\oauthor{\bsnm{Doumbouya}, \binits{M.}},
\oauthor{\bsnm{Durmus}, \binits{E.}},
\oauthor{\bsnm{Ermon}, \binits{S.}},
\oauthor{\bsnm{Etchemendy}, \binits{J.}},
\oauthor{\bsnm{Ethayarajh}, \binits{K.}},
\oauthor{\bsnm{Fei-Fei}, \binits{L.}},
\oauthor{\bsnm{Finn}, \binits{C.}},
\oauthor{\bsnm{Gale}, \binits{T.}},
\oauthor{\bsnm{Gillespie}, \binits{L.}},
\oauthor{\bsnm{Goel}, \binits{K.}},
\oauthor{\bsnm{Goodman}, \binits{N.}},
\oauthor{\bsnm{Grossman}, \binits{S.}},
\oauthor{\bsnm{Guha}, \binits{N.}},
\oauthor{\bsnm{Hashimoto}, \binits{T.}},
\oauthor{\bsnm{Henderson}, \binits{P.}},
\oauthor{\bsnm{Hewitt}, \binits{J.}},
\oauthor{\bsnm{Ho}, \binits{D.E.}},
\oauthor{\bsnm{Hong}, \binits{J.}},
\oauthor{\bsnm{Hsu}, \binits{K.}},
\oauthor{\bsnm{Huang}, \binits{J.}},
\oauthor{\bsnm{Icard}, \binits{T.}},
\oauthor{\bsnm{Jain}, \binits{S.}},
\oauthor{\bsnm{Jurafsky}, \binits{D.}},
\oauthor{\bsnm{Kalluri}, \binits{P.}},
\oauthor{\bsnm{Karamcheti}, \binits{S.}},
\oauthor{\bsnm{Keeling}, \binits{G.}},
\oauthor{\bsnm{Khani}, \binits{F.}},
\oauthor{\bsnm{Khattab}, \binits{O.}},
\oauthor{\bsnm{Koh}, \binits{P.W.}},
\oauthor{\bsnm{Krass}, \binits{M.}},
\oauthor{\bsnm{Krishna}, \binits{R.}},
\oauthor{\bsnm{Kuditipudi}, \binits{R.}},
\oauthor{\bsnm{Kumar}, \binits{A.}},
\oauthor{\bsnm{Ladhak}, \binits{F.}},
\oauthor{\bsnm{Lee}, \binits{M.}},
\oauthor{\bsnm{Lee}, \binits{T.}},
\oauthor{\bsnm{Leskovec}, \binits{J.}},
\oauthor{\bsnm{Levent}, \binits{I.}},
\oauthor{\bsnm{Li}, \binits{X.L.}},
\oauthor{\bsnm{Li}, \binits{X.}},
\oauthor{\bsnm{Ma}, \binits{T.}},
\oauthor{\bsnm{Malik}, \binits{A.}},
\oauthor{\bsnm{Manning}, \binits{C.D.}},
\oauthor{\bsnm{Mirchandani}, \binits{S.}},
\oauthor{\bsnm{Mitchell}, \binits{E.}},
\oauthor{\bsnm{Munyikwa}, \binits{Z.}},
\oauthor{\bsnm{Nair}, \binits{S.}},
\oauthor{\bsnm{Narayan}, \binits{A.}},
\oauthor{\bsnm{Narayanan}, \binits{D.}},
\oauthor{\bsnm{Newman}, \binits{B.}},
\oauthor{\bsnm{Nie}, \binits{A.}},
\oauthor{\bsnm{Niebles}, \binits{J.C.}},
\oauthor{\bsnm{Nilforoshan}, \binits{H.}},
\oauthor{\bsnm{Nyarko}, \binits{J.}},
\oauthor{\bsnm{Ogut}, \binits{G.}},
\oauthor{\bsnm{Orr}, \binits{L.}},
\oauthor{\bsnm{Papadimitriou}, \binits{I.}},
\oauthor{\bsnm{Park}, \binits{J.S.}},
\oauthor{\bsnm{Piech}, \binits{C.}},
\oauthor{\bsnm{Portelance}, \binits{E.}},
\oauthor{\bsnm{Potts}, \binits{C.}},
\oauthor{\bsnm{Raghunathan}, \binits{A.}},
\oauthor{\bsnm{Reich}, \binits{R.}},
\oauthor{\bsnm{Ren}, \binits{H.}},
\oauthor{\bsnm{Rong}, \binits{F.}},
\oauthor{\bsnm{Roohani}, \binits{Y.}},
\oauthor{\bsnm{Ruiz}, \binits{C.}},
\oauthor{\bsnm{Ryan}, \binits{J.}},
\oauthor{\bsnm{Ré}, \binits{C.}},
\oauthor{\bsnm{Sadigh}, \binits{D.}},
\oauthor{\bsnm{Sagawa}, \binits{S.}},
\oauthor{\bsnm{Santhanam}, \binits{K.}},
\oauthor{\bsnm{Shih}, \binits{A.}},
\oauthor{\bsnm{Srinivasan}, \binits{K.}},
\oauthor{\bsnm{Tamkin}, \binits{A.}},
\oauthor{\bsnm{Taori}, \binits{R.}},
\oauthor{\bsnm{Thomas}, \binits{A.W.}},
\oauthor{\bsnm{Tramèr}, \binits{F.}},
\oauthor{\bsnm{Wang}, \binits{R.E.}},
\oauthor{\bsnm{Wang}, \binits{W.}},
\oauthor{\bsnm{Wu}, \binits{B.}},
\oauthor{\bsnm{Wu}, \binits{J.}},
\oauthor{\bsnm{Wu}, \binits{Y.}},
\oauthor{\bsnm{Xie}, \binits{S.M.}},
\oauthor{\bsnm{Yasunaga}, \binits{M.}},
\oauthor{\bsnm{You}, \binits{J.}},
\oauthor{\bsnm{Zaharia}, \binits{M.}},
\oauthor{\bsnm{Zhang}, \binits{M.}},
\oauthor{\bsnm{Zhang}, \binits{T.}},
\oauthor{\bsnm{Zhang}, \binits{X.}},
\oauthor{\bsnm{Zhang}, \binits{Y.}},
\oauthor{\bsnm{Zheng}, \binits{L.}},
\oauthor{\bsnm{Zhou}, \binits{K.}},
\oauthor{\bsnm{Liang}, \binits{P.}}:
On the Opportunities and Risks of Foundation Models
(2022).
\url{https://arxiv.org/abs/2108.07258}
\end{botherref}
\endbibitem

\bibitem[\protect\citeauthoryear{Longpre et~al.}{2024}]{longpre2024bridging}
\begin{botherref}
\oauthor{\bsnm{Longpre}, \binits{S.}},
\oauthor{\bsnm{Singh}, \binits{N.}},
\oauthor{\bsnm{Cherep}, \binits{M.}},
\oauthor{\bsnm{Tiwary}, \binits{K.}},
\oauthor{\bsnm{Materzynska}, \binits{J.}},
\oauthor{\bsnm{Brannon}, \binits{W.}},
\oauthor{\bsnm{Mahari}, \binits{R.}},
\oauthor{\bsnm{Dey}, \binits{M.}},
\oauthor{\bsnm{Hamdy}, \binits{M.}},
\oauthor{\bsnm{Saxena}, \binits{N.}}, et al.:
Bridging the data provenance gap across text, speech and video.
arXiv preprint arXiv:2412.17847
(2024)
\end{botherref}
\endbibitem

\bibitem[\protect\citeauthoryear{Herbrich et~al.}{2006}]{herbrich2006trueskill}
\begin{botherref}
\oauthor{\bsnm{Herbrich}, \binits{R.}},
\oauthor{\bsnm{Minka}, \binits{T.}},
\oauthor{\bsnm{Graepel}, \binits{T.}}:
Trueskill™: a bayesian skill rating system.
Advances in neural information processing systems
\textbf{19}
(2006)
\end{botherref}
\endbibitem

\bibitem[\protect\citeauthoryear{Liu et~al.}{2024}]{liu2024visual}
\begin{botherref}
\oauthor{\bsnm{Liu}, \binits{H.}},
\oauthor{\bsnm{Li}, \binits{C.}},
\oauthor{\bsnm{Wu}, \binits{Q.}},
\oauthor{\bsnm{Lee}, \binits{Y.J.}}:
Visual instruction tuning.
Advances in neural information processing systems
\textbf{36}
(2024)
\end{botherref}
\endbibitem

\bibitem[\protect\citeauthoryear{Newman}{2010}]{newman2010networks}
\begin{bbook}
\bauthor{\bsnm{Newman}, \binits{M.E.J.}}:
\bbtitle{Networks: An Introduction}.
\bpublisher{Oxford University Press},
\blocation{Oxford}
(\byear{2010})
\end{bbook}
\endbibitem

\end{thebibliography}
\end{document}